\documentclass[11pt,a4paper]{article}
\usepackage{jcappub}

\newcommand{\de}{\mathrm d}

\newcommand{\om}{{\Omega_m}}

\newcommand{\odm}{{\Omega_\mathrm{DM}}}

\newcommand{\ho}{{H_0}}

\newcommand{\mdm}{{m_{\rm DM}}}

\newcommand{\gd}{{\Gamma_d}}
\newcommand{\sv}{{\langle\sigma_a v\rangle}}

\newcommand{\g}{$\gamma$}
\newcommand{\ns}{{\sc ns}}
\newcommand{\low}{{\sc low}}
\newcommand{\high}{{\sc high}}

\newcommand{\Fermitenyr}{{\it Fermi-10yr}}
\newcommand{\Fermissimo}{{\it `Fermissimo'}}
\newcommand{\Euclid}{{\it Euclid}}
\newcommand{\DES}{{DES}}
\newcommand{\Fermi}{{\it Fermi}}

\def\bea{\begin{eqnarray}}
\def\eea{\end{eqnarray}}
\def\be{\begin{equation}}
\def\ee{\end{equation}}

\begin{document}

\title{Tomographic-spectral approach for dark matter detection in the cross-correlation between cosmic shear and diffuse \g-ray emission}

\author[a,b]{S. Camera,}
\author[c]{M. Fornasa,}
\author[d,e]{N. Fornengo}
\author[d,e]{\& M. Regis}

\affiliation[a]{Jodrell Bank Centre for Astrophysics, The University of Manchester,\\Alan Turing Building, Oxford Road, Manchester M13 9PL,UK}
\affiliation[b]{CENTRA, Instituto Superior T\'ecnico, Universidade de Lisboa,\\Avenida Rovisco Pais 1, 1049-001 Lisboa, Portugal}
\affiliation[c]{School of Physics \& Astronomy, University of Nottingham,\\University Campus, Nottingham NG7 2RD, UK}
\affiliation[d]{Dipartimento di Fisica, Universit\`a  di Torino,\\Via P. Giuria 1, 10125 Torino, Italy}
\affiliation[e]{INFN, Sezione di Torino,\\Via P. Giuria 1, 10125 Torino, Italy}

\emailAdd{stefano.camera@manchester.ac.uk}
\emailAdd{fornasam@gmail.com}
\emailAdd{fornengo@to.infn.it}
\emailAdd{regis@to.infn.it}

\abstract{We recently proposed to cross-correlate the diffuse extragalactic \g-ray background with the gravitational lensing signal of cosmic shear. This represents a novel and promising strategy to search for annihilating or decaying particle dark matter (DM) candidates. In the present work, we demonstrate the potential of a tomographic-spectral approach: measuring the cross-correlation in separate bins of redshift {\it and} energy significantly improves the sensitivity to a DM signal. Indeed, the technique proposed here takes advantage of the different scaling of the astrophysical and DM components with redshift and, {\it simultaneously}, of their different energy spectra and different angular extensions. The sensitivity to a particle DM signal is extremely promising even when the DM-induced emission is quite faint. We first quantify the prospects of detecting DM by cross-correlating the \Fermi\ Large Area Telescope (LAT) diffuse \g-ray background with the cosmic shear expected from the Dark Energy Survey. Under the hypothesis of a significant subhalo boost, such a measurement can deliver a $5\sigma$ detection of DM, if the DM particle is lighter than 300 GeV and has a thermal annihilation rate. We then forecast the capability of the European Space Agency \Euclid\ satellite (whose launch is planned for 2020), in combination with an hypothetical future \g-ray detector with slightly improved specifications compared to current telescopes. We predict that the cross-correlation of their data will allow a measurement of the DM mass with an uncertainty of a factor of 1.5--2, even for moderate subhalo boosts, for DM masses up to few hundreds of GeV and thermal annihilation rates.}


\maketitle

\section{Introduction}
A tremendous experimental effort is underway with the goal of delivering the first non-gravitational detection of dark matter (DM). Indeed, despite its huge abundance \cite{Ade:2015xua}, very little is known about the true nature of DM, and the community is eager for new and innovative ideas able to finally pin down its properties.

In Ref.~\cite{Camera:2012cj} (Paper I), we demonstrated, for the first time, how is possibile to single out distinctive DM signatures in the cross-correlation between the so-called extragalactic \g-ray background (EGB) and the weak lensing effect of cosmic shear---an unbiased tracer of the matter distribution in the Universe.

The EGB is the residual emission remaining after the contribution of the resolved \g-ray sources (both point-like and extended) and of the Galactic foreground (due to the interaction of cosmic rays with the Galactic interstellar medium and radiation fields) are subtracted from the total \g-ray radiation. The latest measurement of the EGB was performed by the Large Area Telescope (LAT) instrument on the \Fermi satellite over the energy range between 100 MeV and 820 GeV \cite{Ackermann:2014usa}. Unresolved astrophysical sources like blazars \cite{Inoue:2008pk,Abazajian:2010pc,Collaboration:2010gqa,Ajello:2011zi,Harding:2012gk,DiMauro:2013xta,Ajello:2013lka}, misaligned Active Galactic Nuclei (mAGNs) \cite{Inoue:2011bm,DiMauro:2013xta} and star-forming galaxies (SFGs) \cite{Ackermann:2012vca,Lacki:2012si,Tamborra:2014xia} are guaranteed components to the EGB, but \g\ rays produced by DM annihilations or decays in Galactic or extragalactic (sub)haloes may also play a relevant r\^ole (see Refs.~\cite{Ullio:2002pj,Ando:2005xg,Ando:2006cr,SiegalGaskins:2008ge,SiegalGaskins:2009ux,Zavala:2009zr,Fornasa:2009qh,Zavala:2011tt,Fornasa:2012gu,Ando:2013ff}, amongst others and Ref.~\cite{Fornasa:2015qua} for a recent review).

In Ref.~\cite{Ackermann:2012uf}, \Fermi-LAT measured the auto-correlation angular power spectrum (APS) of the EGB, reporting a Poisson-like signal with a significance ranging from 7.1$\sigma$ (between 2 and 5 GeV) to 2.4$\sigma$ (between 10 and 50 GeV). The measurement of the intensity and of the auto-correlation APS of the EGB can be used to reconstruct its composition and constrain the DM-induced component (see e.g. Refs~\cite{Cuoco:2012yf,Ando:2013ff,Gomez-Vargas:2013cna}). The picture that arises suggests that unresolved blazars can explain the whole auto-correlation signal \cite{Cuoco:2012yf,Harding:2012gk}, whereas they are responsible for only $\lesssim25\%$ of the EGB intensity \cite{Ajello:2011zi,Ajello:2013lka,DiMauro:2013xta}. Thus, other contributions are needed to reproduce the whole EGB intensity, while being more isotropic than blazars, in order not to overshoot the auto-correlation APS data.

Cross-correlating the EGB with other observables can provide additional and complementary information (some possibilities are summarised in Ref.~\cite{Fornengo:2013rga}). In Paper I, we demonstrated that the cross-correlation with the cosmic shear is particularly compelling since, in this case, the contribution of unresolved blazars is subdominant. Such a signal is, therefore, more sensitive to components that would be difficult to isolate by the study of the EGB intensity or of its auto-correlation APS.

The cosmic shear is a statistical measure of the distortions of the image of distant galaxies due to the weak gravitational lensing produced by the matter present between the galaxies and the observer \cite{Kaiser:1991qi,Bartelmann:1999yn,Munshi:2006fn,Bartelmann:2010fz}. It is expected to cross-correlate with the EGB since the same distribution of matter acting as gravitational lens is also responsible for the \g-ray emission, either through annihilations (or decays) of DM or because DM structures host astrophysical \g-ray emitters.

In Paper I, we computed the cross-correlation APS between the cosmic shear expected from the Dark Energy Survey (\DES) \cite{Abbott:2005bi} or the \Euclid\ satellite \cite{Laureijs:2011gra,Amendola:2012ys}\footnote{\texttt{http://www.euclid-ec.org/}} and the \g-ray emission produced by different classes of sources, namely DM extragalactic (sub)haloes, blazars and SFGs. Results are extremely promising and they suggest that DM haloes can be responsible for the largest cross-correlation signal (both in the case of annihilating and decaying DM): the analysis of 5 years of \Fermi-LAT data may already be enough to detect DM-induced signatures in the cross-correlation with cosmic shear.

Since Paper I, Ref.~\cite{Shirasaki:2014noa} measured the 2-point correlation function of the \Fermi-LAT data gathered until January 2014 with the cosmic shear detected by the Canada-France-Hawaii Telescope Lensing Survey (CFTHLenS) \cite{Heymans:2012gg}. Their measurement was consistent with no signal and the null detection was used to derive constraints on the DM annihilation cross section. For a light DM particle and for specific annihilation channels, the upper limits of Ref.~\cite{Shirasaki:2014noa} approach the `thermal' value of $3 \times 10^{-26}\,\mathrm{cm^3\,s^{-1}}$ for the annihilation cross-section. This fact additionally proves the potential of the cross-correlation between \g-ray emission and cosmic shear for the indirect detection of DM. 

Along the same line, the cross-correlation of the EGB with other tracers of the cosmic large-scale structure is also a potentially valuable technique to infer the composition of the EGB and to disentangle a DM contribution~\cite{Xia:2011ax,Fornengo:2013rga,Ando:2013xwa,Ando:2014aoa}. In particular, Ref.~\cite{Fornengo:2014cya} found the first evidence of a cross-correlation between the weak lensing of the cosmic microwave background (CMB) and the EGB. Since the kernel of the CMB lensing peaks at a redshift $z \sim 2$, this observable is more suited to constraint astrophysical contributions to the EGB. Indeed, their emission mostly comes from intermediate $z$, while the DM-induced signal peaks at lower redshift, as we shall discuss later. More recently, Ref.~\cite{Xia:2015wka} reported the first detection of a cross-correlation between \g-ray emission and galaxy catalogues, while Ref.~\cite{Regis:2015zka} discussed the implication of such a measurement for DM searches. Despite being very promising to reveal a DM-induced \g-ray emission \cite{Fornengo:2013rga,Ando:2013xwa,Ando:2014aoa}, low-redshift galaxy catalogues trace light and, therefore, are a biased tracer of the distribution of DM \cite{Fornengo:2013rga}. On the other hand, cosmic-shear is a direct measurement of the DM distribution in the Universe.  

In the present work, we investigate the importance of a {\it combined} tomographic and spectral approach in the study of the cross-correlation between cosmic shear and \g-ray emission---where by `tomographic' and `spectral' we mean the exploitation of the redshift and energy dependence of the observables. Moreover, with respect to Paper I, we complement our model of the EGB by including the contribution of unresolved mAGNs. Such a component, together with unresolved SFGs, is associated to a quite large cross-correlation APS. Unresolved SFGs and mAGNs, which are most probably responsible for the majority of the EGB intensity, represent the largest astrophysical backgrounds when looking for DM signatures in the cross-correlation. However, the abundance of those classes of sources as a function of redshift is quite different from that of DM. Tomography takes advantage of this fact, isolating the cross-correlation APS in different redshift bins and studying how the it changes as a function of redshift. This is expected to enhance our capability to dissect the EGB into components and to discriminate DM from astrophysical contributions.

Here, we also investigate how performing the analysis in different bins in energy can further increment the possibility of distinguishing between astrophysical sources (typically characterised by power-law spectra) and the emission expected from DM. As we will show below, the spectral analysis is also crucial to infer the DM microscopic properties, allowing us to lift the degeneracy between its mass and annihilation/decay rate---both entering the computation of the cross-correlation signal.

Our final aim is to fully exploit the complementary information coming from the angular, redshift, and spectral distribution of the extragalactic signal. The analysis will focus on the expected weak lensing observations from \DES\ (whose data and maps will be available in the near future, if not already as in Ref.~\citep[][]{Vikram:2015leg}) and \Euclid, which will be launched in 2020 and will have an optimal sensitivity to the cosmic shear signal. For what concerns \g\ rays, we will consider 10 nominal years (i.e.\ after cuts) of data from \Fermi-LAT (available, approximately, at the end of the expected lifetime of the mission), and a possible near-future instrument similar to \Fermi-LAT but with improved capabilities (dubbed \Fermissimo, just for definiteness). The cross-correlation of 10 years of \Fermi-LAT data with DES represents a scenario available in the near future, while the combination of \Fermissimo\ a \Euclid\ will be possible on a longer timescale.

Apart from computing the sensitivity of the tomographic-spectral approach to the detection of DM in the cross-correlation APS between the EGB and cosmic shear, we will also estimate the precision with which such a measurement can determine the properties of DM, e.g. its mass and its annihilation or decay rates. This will be achieved by means of a statistical analysis based on the Fisher information matrix technique \citep{Tegmark:1996bz}. The method is extremely flexible and it allows for a systematic scrutiny of the main uncertainties pertaining our modelling of the DM distribution and the astrophysical sources contributing to the EGB.

The paper is organised as follows: the main results are presented in Sec.~\ref{sec:results}, where we derive forecasts for DM detection/bounds, whilst Sec.~\ref{sec:tomospec} introduces our tomographic and spectral technique. The Fisher analysis is described in Sec.~\ref{sec:analysis}. For the interested reader, in Sec.~\ref{sec:intensity} we summarise the details of the computation of the mean (all-sky averaged) \g-ray emission associated with the classes of sources considered, as well as of the cosmic shear signal. In Sec.~\ref{sec:auto_and_cross_correlations} we derive the \g-ray auto-correlation APS of those sources, as well as their cross-correlation APS with cosmic shear. We shall take special care in emphasising the assumptions made and in estimating their impact on our results. Finally, conclusions are drawn in Sec.~\ref{sec:conclusions}.

Throughout the paper we assume a flat $\Lambda$CDM cosmology with cosmological parameters as derived by the \textit{Planck} satellite \cite{Ade:2015xua}. We also assume the standard weakly interacting massive particle (WIMP) paradigm in which the DM particle has a mass $\mdm$ of the order of GeV-TeV and its interactions are of the weak scale.

\section{Intensity of \g-ray emission and cosmic shear}
\label{sec:intensity}
In this section, we introduce the formalism adopted when describing the \g-ray emission produced by different classes of sources, as well as the intensity of the cosmic-shear signal. In this common framework, an observable $X(\chi,\hat{\mathbf n})$ can be written as
\begin{equation}
\mathcal I^X(\hat{\mathbf n}) = \int \de\chi \, X(\chi,\hat{\mathbf n}) =
\int \de\chi \, g_X(\chi,\hat{\mathbf n}) \widetilde W^X(\chi).
\end{equation}
Depending on the observable considered, $\mathcal I^{X}(\hat{\mathbf n})$ indicates the \g-ray emission expected from a given class of sources or the intensity of the shear signal from the direction $\hat{\mathbf n}$ in the sky. Here, $\chi=\chi(z)$ is the comoving distance at redshift $z$ and, for a flat Universe, $\de\chi=c/H\,\de z$ being $c$ the speed of light and $H(z)$ the Hubble rate. The observable $X(\chi,\hat{\mathbf n})$ is seeded by the source field $g_X(\chi,\hat{\mathbf n})$, which encodes the dependence on the line-of-sight direction, $\hat{\mathbf n}$. We define the window function $W^X(\chi)$ (also called weight or selection function) as $W^X(\chi)\equiv\langle X\rangle(\chi)=\langle g_X \rangle(\chi) \widetilde W^X(\chi)$, where $\langle \cdot \rangle$ stands for sky average. The mean intensity $\langle \mathcal I^{X} \rangle$ can simply be written as the integral of the window function, viz.
\begin{equation}
\langle \mathcal I^X \rangle = \int \de \chi \, W^X(\chi).
\end{equation}

In the following, we will derive the window functions for the \g-ray emission from different classes of sources (Secs. from \ref{sec:annihilating_DM} to \ref{sec:MAGNs}), as well as the window function for the cosmic shear (Sec.~\ref{sec:cosmic_shear}).

\subsection{Dark matter}
\subsubsection{Annihilating dark matter}
\label{sec:annihilating_DM}
The \g\ rays produced by DM annihilations in (sub)haloes ($X=\gamma_{{\rm DMa}}$) trace the DM density squared $\rho^2_{\rm DM}$, so that $g_{\gamma_{{\rm DMa}}}(\chi,\hat{\mathbf n}) = \rho^2_{\rm DM}(\chi,\hat{\mathbf n})$. Following Ref.~\cite{Fornengo:2013rga} and references therein, the window function for annihilating DM can be written as follows:
\begin{equation}
W^{\gamma_{{\rm DMa}}}(\chi) = \frac{(\odm \rho_c)^2}{4\pi} 
\frac{\sv}{2\mdm^2} \left[1+z(\chi)\right]^3 
\Delta^2(\chi) \int_{\Delta E_\gamma} \de E_\gamma \, \frac{\de N_a}{\de E_\gamma} 
\left[E_\gamma(\chi) \right] e^{-\tau\left[\chi,E_\gamma(\chi)\right]},
\label{eqn:window_annihilating_DM}
\end{equation}
where $\odm$ is the cosmological abundance of DM, $\rho_c$ is the critical density of the Universe and $z(\chi)$ is the redshift corresponding to the comoving distance $\chi$. $\de N_a / \de E_\gamma$ indicates the number of photons produced per annihilation and its dependence upon the energy determines the \g-ray spectrum, whilst $\sv$ is the velocity-averaged annihilation rate, assumed to be the same in all haloes. 

Different \g-ray production mechanisms can contribute to $\de N_a / \de E_\gamma$. At the energies of interest for this paper (i.e.\ from a few to a few hundreds GeV) primary production is probably the most important. In this case, \g\ rays are either produced directly in the annihilations of the two DM particles (and therefore the photon yield is a monochromatic line) or they are generated by hadronisation or radiative emission of the particles produced in the annihilations (such as quarks, leptons, gauge or Higgs bosons). Robust predictions can be obtained for $\de N_a / \de E_\gamma$, based on the results of events generators like PYTHIA \cite{Sjostrand:2006za}, which is used in the present work. Photon yields originated from the hadronisation of quarks or gauge bosons share a quite common spectral shape, with a cut-off at energies equal to $\mdm$. Hadronisations of $\tau^\pm$ leptons are associated with hard spectra. Possible bumps, localised close to the cut-off, may occur due to final-state radiation or internal bremsstrahlung (for simplicity, we shall only consider the former, since it is model independent). The photon yield in Eq.~\eqref{eqn:window_annihilating_DM} is integrated in an energy window $\Delta E_\gamma$ that will be established later balancing the necessity of large statistics and the desire to consider small energy bins. The factor $\exp\{-\tau[\chi,E_\gamma(\chi)]\}$ accounts for the absorption due to the extragalactic background light and we model $\tau$ as in Ref.~\cite{Franceschini:2008tp}. Radiative emissions (most notably, inverse Compton scattering) also contribute to the total DM-induced \g-ray production. However, for the sake of simplicity, they will be neglected here, since they are more model dependent than prompt emission, and they typically affect predictions only for multi-TeV DM annihilating into light leptons.

The last quantity in Eq.~\eqref{eqn:window_annihilating_DM} is $\Delta^2(\chi)$, the so-called clumping factor. It measures how much the \g-ray flux increases due to the clumpiness of the DM distribution. It reads
\begin{equation}
\Delta^2(\chi) \equiv 
\frac{\langle \rho^2_{\rm DM} \rangle}{{\bar \rho}^2_{\rm DM}} =
\int_{M_{\rm min}}^{M_{\rm max}} \de M \frac{\de n}{\de M} 
\int \de^3 \mathbf{x} \, 
\frac{\rho^2_h({\mathbf{x}|M,\chi)}}{{\bar \rho}^2_{\rm DM}}.
\label{eqn:clumping}
\end{equation}
Eq.~\eqref{eqn:clumping} shows that $\Delta^2(\chi)$ is computed integrating the halo number density $\de n/\de M$ above the minimal halo mass $M_{\rm min}$ after multiplying by the total number of annihilations produced a generic halo of mass $M$ at distance $\chi$ with density profile $\rho_h(\mathbf{x}|M,\chi)$. The exact value of $M_{\rm min}$ is unknown and it depends on the free-streaming length of the DM particle, with reasonable values ranging between $10^{-12} M_\odot$ and $10^{-3} M_\odot$ \cite{Profumo:2006bv,Bringmann:2009vf}. We consider a reference value of $10^{-6} M_\odot$, but we also report our results for a much more conservative value of $10^7 M_\odot$. This represents the smallest mass scale for which we have an indirect evidence of DM haloes from the study of the stellar dynamics in the dwarf spheroidal galaxies of the Milky Way. $M_{\rm min}=10^7 M_\odot$ is the minimum mass assumed for one of the three benchmarks that will be considered later for $\Delta^2(\chi)$. We refer to this as the \ns\ scenario and we further assumes that, in this case, subhaloes provide a negligible contribution to the DM-induced \g-ray emission. The \ns\ model is, thus, a very conservative scenario. The maximal halo mass $M_{\rm max}$ will always be set to $10^{18} M_\odot$ and its value has only a minor impact on the results. 

Many quantities in Eq.~\eqref{eqn:clumping} can be derived from $N$-body simulations: we take the halo mass function from Ref.~\cite{Sheth:1999mn} and we assume that haloes are characterised by the Navarro-Frenk-White (NFW) universal density profile \citep{Navarro:1995iw}. The profile is completely determined by the total mass of the halo and by its size. We express the latter in terms of the concentration parameter $c(M)$, taken as a function of the halo mass $M$ from Ref.~\cite{MunozCuartas:2010ig}. Even though $N$-body simulations provide valuable information for massive DM haloes, unfortunately they do not cover the mass range considered in Eq.~\eqref{eqn:clumping}. Below their mass resolution our knowledge is poor and one has to resort to debatable assumptions. In the following sections, we shall assume that haloes with a mass below $10^{10} M_\odot$ are still well described by a NFW profile with a mass function from Ref.~\cite{Sheth:1999mn}, but we adopt the $c(M)$ relation presented in Ref.~\cite{Bullock:1999he}, renormalised so that it gives the same concentration as the model from Ref.~\cite{MunozCuartas:2010ig} at $10^{10} M_\odot$. The resulting $c(M)$ at low masses is within the uncertainty band of Fig. 1 in Ref.~\cite{SanchezConde:2013yxa}.

One additional source of uncertainty is the amount of subhaloes hosted by main haloes. We consider two scenarios---dubbed \low\ and \high---as extreme cases bracketing the effect of subhaloes. The \low\ scenario follows the procedure introduced in Ref.~\cite{Kamionkowski:2010mi} and encodes the results of Ref.~\cite{SanchezConde:2011ap}. The \high\ scenario stems from the results of Ref.~\cite{Gao:2011rf} and is implemented by means of the semi-analytical approach described in Ref.~\cite{Ando:2013ff}. Note that, when subhaloes are included in the computation of the DM-induced \g-ray emission, the last factor in Eq.~\eqref{eqn:clumping} has to be changed into $\int \de V \rho^2_h(\mathbf{x}|M,\chi) [1+B(M,\chi)] /{\bar \rho}^2_{\rm DM}$, where $B(M,\chi)$ is the subhalo boost factor. Again, for \low\ and \high\ we take $M_{\rm min}=10^{-6} M_\odot$.

\begin{figure}
\centering
\hspace{-4mm}
\includegraphics[width=0.45\textwidth]{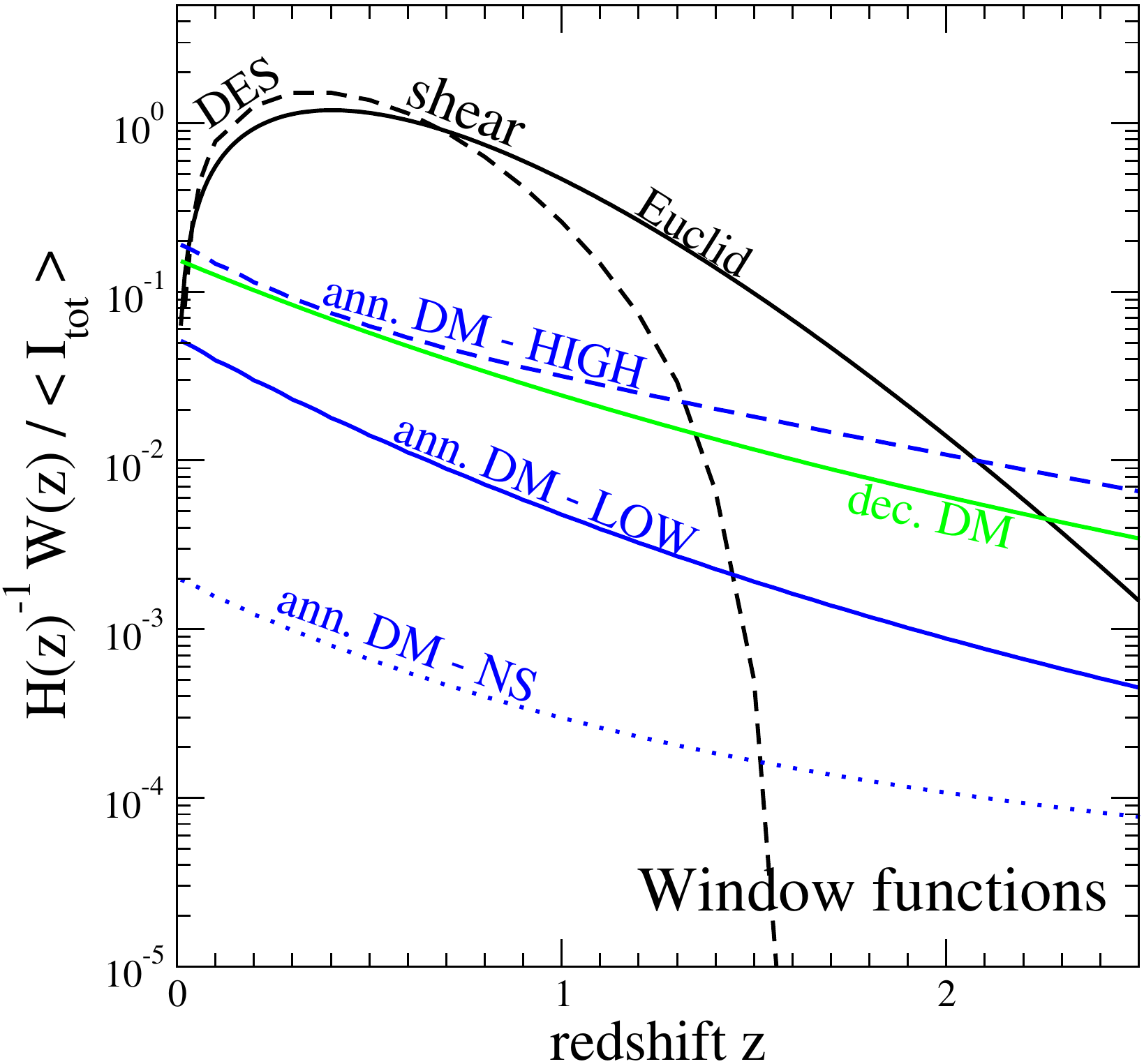}
\hspace{10mm}
\includegraphics[width=0.45\textwidth]{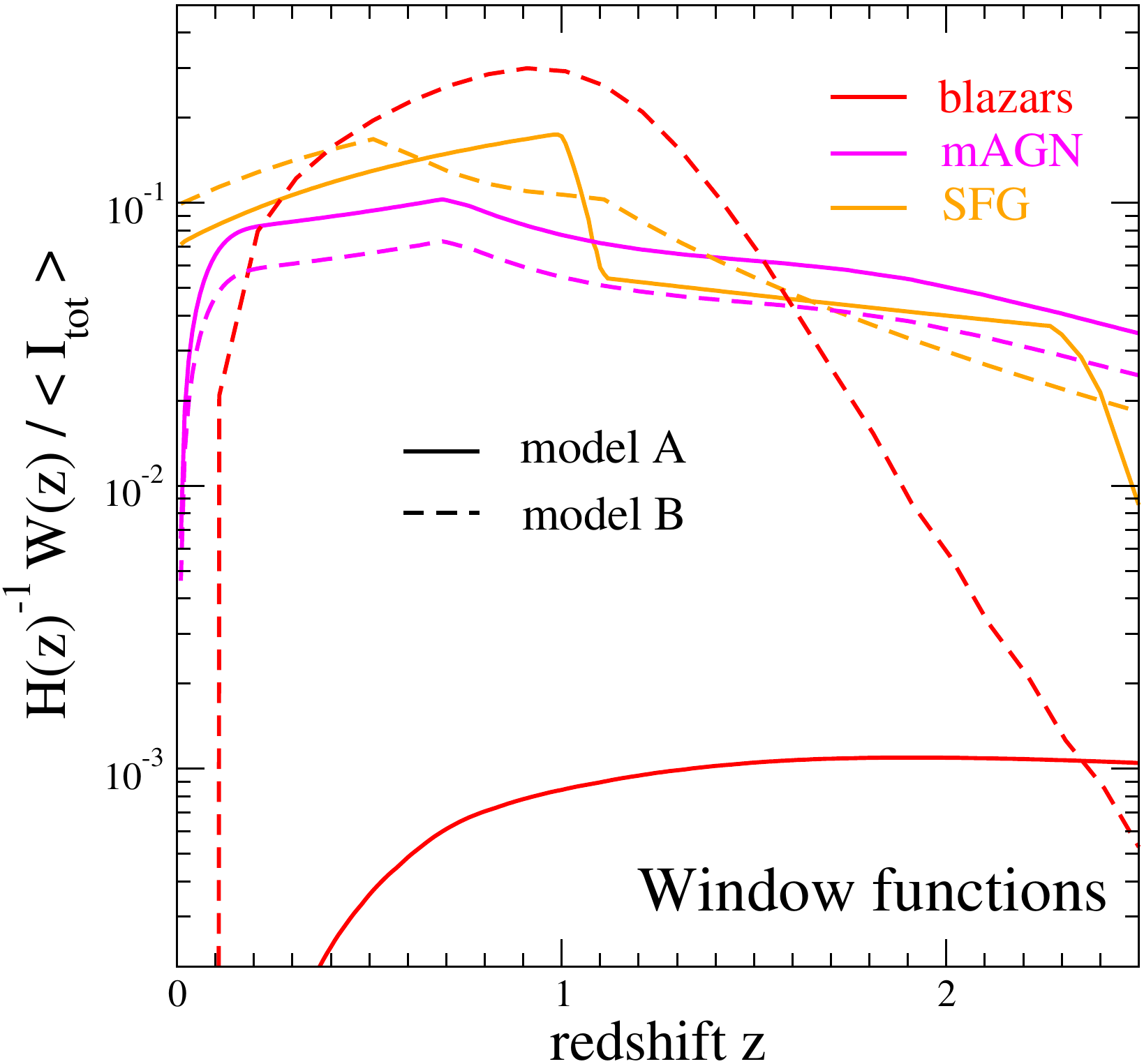}
\caption{Redshift dependence of the window function for the various signals described in the text. The window functions for \g\ rays are integrated above $E>1$ GeV and normalised to the {\it total} EGB intensity measured by the \Fermi-LAT telescope~\cite{Abdo:2010nz,Ackermann:2014usa} above 1 GeV, in order to ease comparison. The shear signal is normalised so that the integral of the window function over $\chi$ equals 1. The solid (dashed) black line in left panel indicates the window function for cosmic shear from \Euclid\ (\DES). The solid (dashed) blue line corresponds to annihilating DM in the \low\ (\high) scenario, whilst the dotted blue line represents the \ns\ model. The solid green line stands for decaying DM. The WIMP mass is set to 100 GeV (200 GeV) for annihilating (decaying) DM. The annihilation cross section is $\sv=3 \times 10^{-26}\,\mathrm{cm^3s^{-1}}$, whereas the decay rate is $\Gamma_d = 0.33 \times 10^{-27}\,\mathrm{s^{-1}}$. An annihilation/decay channel into $b\bar b$ quarks is assumed. In the right panel, red, yellow and magenta curves represent the contributions from blazars, SFGs and mAGNs. model A is reported with solid lines, while model B with dashed. See text for the description of the models.}
\label{fig:window}
\end{figure}

The blue curves in the left panel of Fig.~\ref{fig:window} show the window functions of Eq.~\eqref{eqn:window_annihilating_DM} for annihilating DM in the \high, \low, and \ns\ scenarios. The corresponding contributions to the average \g-ray intensity is shown in the left panel of Fig. \ref{fig:intensity}. Results are presented for a benchmark DM particle with $\mdm=100$ GeV, `thermal' $\sv=3 \times 10^{-26}\,\mathrm{cm^3s^{-1}}$ and an annihilation channel into $b \bar{b}$ quarks. Note that the three clustering scenarios (\ns, \low\ and \high\ for dotted, solid and dashed curves, respectively) share approximately the same redshift dependence, but they correspond to different intensities for the clumping factor and, consequently, for the DM-induced \g-ray flux. Note that a comparison with previous works in the literature can be non-trivial, as different groups employ different prescriptions for the DM clustering and, in particular, for the boost factor (see, e.g., Refs.~\cite{Fornasa:2012gu,Ando:2013ff}). Fig.~\ref{fig:intensity} can be useful as a normalisation test when confronting the result presented in the rest of the paper with other works.

\subsubsection{Decaying dark matter}
\label{sec:decaying_DM}
\begin{figure}
\centering
\hspace{-4mm}
\includegraphics[width=0.45\textwidth]{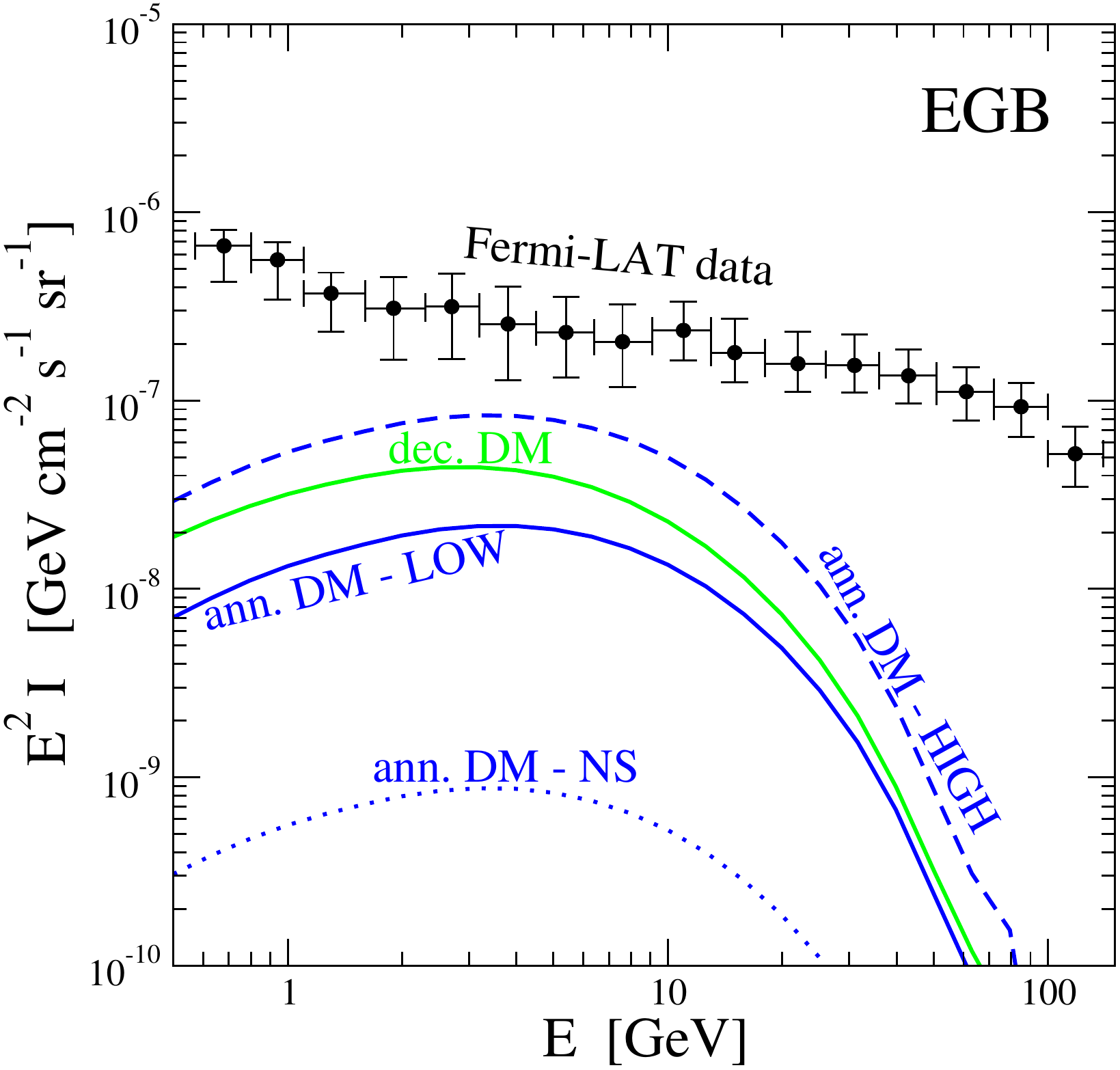}
\hspace{10mm}
\includegraphics[width=0.45\textwidth]{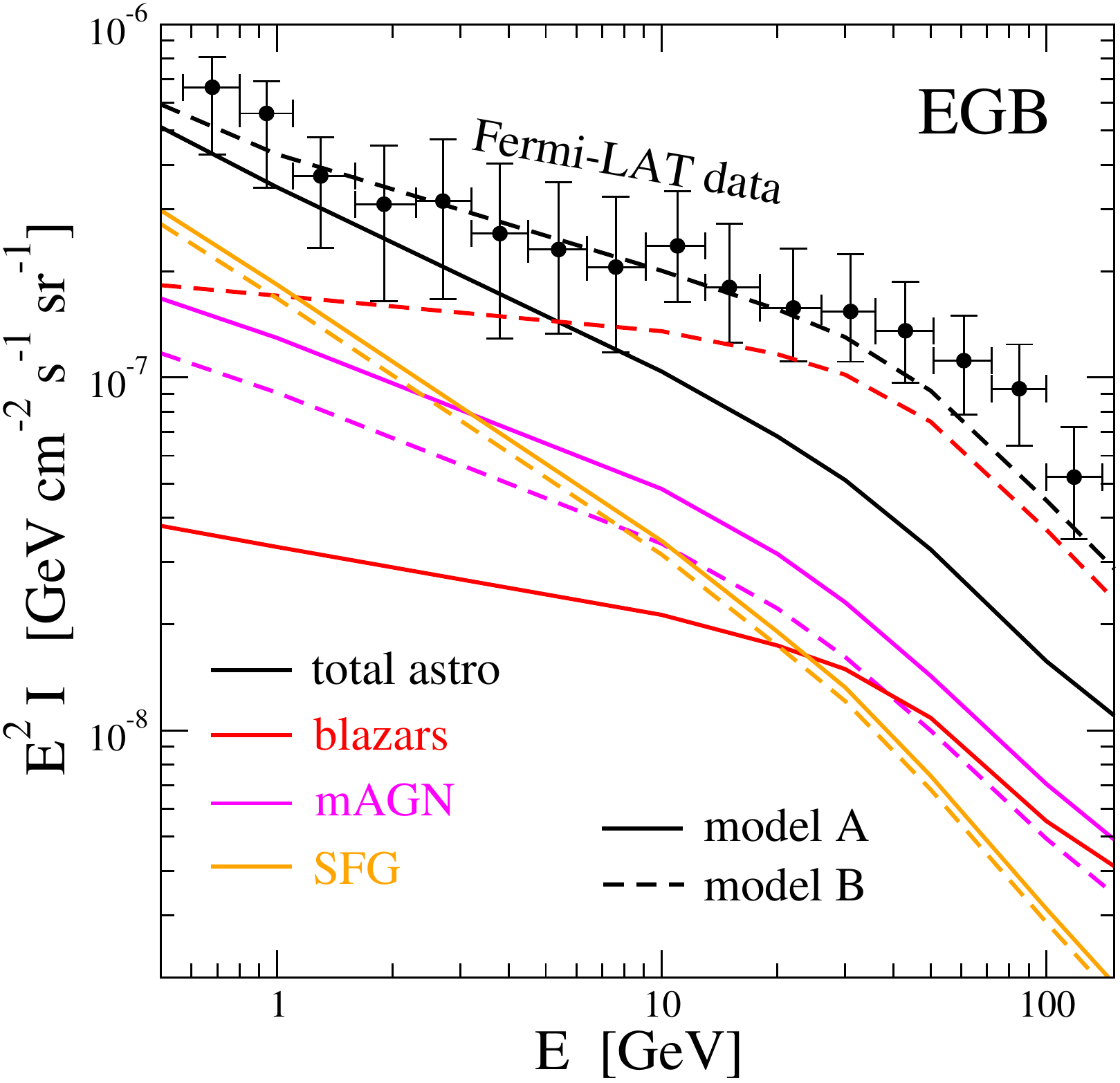}
\caption{Contributions to the EGB intensity from annihilating DM and decaying DM (left panel), and from blazars, mAGNs, and SFGs (right panel). Same models and colour coding of Fig.~\ref{fig:window}. The black lines in the right panel show the sum of the three astrophysical components, whilst the \Fermi-LAT measurement of the EGB intensity from Ref.~\cite{Ackermann:2014usa} is depicted with black data points (adding in quadrature systematic and statistical uncertainties).}
\label{fig:intensity}
\end{figure}

If DM produces \g\ rays by particle decay ($X=\gamma_{{\rm DMd}}$), the source field traces the DM density linearly, i.e.\ $g_{\gamma_{{\rm DMd}}}(\chi,\hat{\mathbf n}) = \rho_{\rm DM}(\chi,\hat{\mathbf n})$. Thence, the window function is \cite{Fornengo:2013rga}
\begin{equation}
W^{\gamma_{{\rm DMd}}}(\chi) = \frac{\odm \rho_c}{4\pi} 
\frac{\Gamma_{\rm d}}{\mdm} \int_{\Delta E_\gamma} \de E_\gamma \, 
\frac{\de N_d}{\de E_\gamma} \left[E_\gamma(\chi) \right] 
e^{-\tau\left[\chi,E_\gamma(\chi)\right]}.
\label{eqn:window_decaying_DM}
\end{equation}
The photon yield for DM decays, $\de N_d / \de E_\gamma$, is assumed to be the same as for annihilating DM, but the energy available is now $\sqrt{s}=\mdm$ instead of $2\mdm$. In other words, $\de N_d / \de E_\gamma(E_\gamma) = \de N_a / \de E_\gamma(2E_\gamma)$ and the kinematic end-point is at $\mdm/2$. The window function and the average emission for decaying DM are shown as green curves in the left panels of Figs~\ref{fig:window} and \ref{fig:intensity}. We choose a benchmark particle DM model with $\mdm=200$ GeV, decay rate $\Gamma_d = 0.33 \times 10^{-27}\,\mathrm{s^{-1}}$ and decays into $b\bar b$ quarks. Note the redshift scaling similar to the case of annihilating DM.

\subsection{Astrophysical sources}
\label{sec:intastro}
For astrophysical sources, we choose the source luminosity $\mathcal L$ as the parameter characterising the different source populations. For the angular scales of interest in this analysis ($\ell < 1000$), the emission from AGNs and galaxies can be safely considered as point-like. For a power-law energy spectrum with spectral index $\alpha$, the window function takes the following form \cite{Fornengo:2013rga}
\begin{equation}
W^{\gamma_i}(\chi) = \frac{A_i(\chi) \langle g_{\gamma_i}(\chi)\rangle}{4\pi E_0^2}
\int_{\Delta E_\gamma} \! \! \de E_\gamma \, 
\left( \frac{E_\gamma}{E_0} \right)^{-\alpha_i}
e^{-\tau\left[\chi,E_\gamma(\chi)\right]},
\label{eq:Wastro}
\end{equation}
where $i$ stands for blazars, mAGNs or SFGs. The normalisation factor $A_i$ depends on the exact definition of luminosity and $\langle g_{\gamma_i}(\chi) \rangle$ indicates the mean luminosity produced by unresolved objects located at a distance $\chi$. It can be written as
\begin{equation}
\langle g_{\gamma_i}(\chi) \rangle = 
\int_{\mathcal{L}_{\rm min}}^{\mathcal{L}_{\rm max}(F_{\rm max},z)} \de \mathcal{L} 
\, \mathcal{L} \, \rho_{\gamma_i}(\mathcal{L},z),
\label{eqn:average_g_blazars}
\end{equation}
where $\rho_{\gamma_i}$ is the \g-ray luminosity function for source class $i$. The upper bound, $\mathcal{L}_{\rm max}(F_{\rm sens},z)$, is the luminosity above which an object can be resolved, assuming a sensitivity $F_{\rm sens}=2 \times 10^{-9}\,\,\mathrm{cm^{-2}s^{-1}}$ above 100 MeV typical of \Fermi-LAT after 5 years of data taking.\footnote{When we consider a larger period of data taking or the case of \Fermissimo, we assume that the sensitivity scales as the inverse of the square of the exposure of the instrument (see Table~\ref{tab:gammaspec}).} Conversely, the minimum luminosity $\mathcal{L}_{\rm min}$ depends on the properties of the source class under investigation. For the interested reader, the three populations of astrophysical \g-ray emitters (i.e.\ blazars, mAGNs and SFGs) are discussed in the following Secs~\ref{sec:blazars}, \ref{sec:MAGNs} and \ref{sec:SFG}. For each of them we present our choice for $\alpha_i$ and for the \g-ray luminosity function.

\subsubsection{Blazars}
\label{sec:blazars}
In the case of blazars, we introduce two different models, labelled `A' and `B':
\begin{itemize}
\item[A]
Following Ref.~\cite{Abazajian:2010pc}, we consider flat-spectrum radio quasars and BL Lacertae cumulatively as one unique class of blazars ($X=\gamma_{\rm BLA}$)\footnote{Considering one single population of blazars is well motivated when the shape of the energy spectrum is allowed to depend on the luminosity of the source. In that case, the flexibility of model well reproduces the differences between the two subclasses of blazars \cite{Abazajian:2010pc,Ajello:2011zi,Ajello:2013lka}. Our approach assumes one common spectral index $\alpha_{\rm BLA}$ and, therefore, is a more simplified description.}. The common spectral index $\alpha_{\rm BLA}$ is fixed to 2.2 and the \g-ray luminosity $\mathcal{L}$ is computed at 100 MeV, which implies $A_{\rm BLA}(\chi)=[1+z(\chi)]^{-\alpha_{\rm BLA}}$. The lower limit of the integral in Eq.~\eqref{eqn:average_g_blazars} is set to $\mathcal{L}_{\rm min}=10^{42}\,\mathrm{erg\, s}^{-1}$, corresponding to the lowest luminosity detected (locally) by \Fermi-LAT. This assumes that no fainter blazar exists at any redshift. The \g-ray luminosity function of blazars is computed following the model described in Ref.~\cite{Inoue:2008pk}, taking advantage of the well-established correlation between \g-ray and X-ray luminosities. The X-ray luminosity function is taken from Ref.~\cite{Ueda:2003yx}. The free parameters in the \g-ray luminosity function are fixed to the values provided in Ref.~\cite{Harding:2012gk}, for which unresolved blazars account for about 6\% of the \Fermi-LAT EGB intensity from Ref.~\cite{Abdo:2010nz} above 1 GeV, but to 100\% of the EGB auto-correlation APS~\cite{Ackermann:2012uf}.

\item[B]
In this case, we consider only BL Lacertae from Ref.~\cite{Ajello:2013lka}, since flat-spectrum radio quasars would provide a subdominant contribution if described as in Ref.~\cite{Ajello:2011zi}. The \g-ray luminosity function is derived from a parametric fit to the redshift and luminosity distributions of resolved blazars in the First \Fermi-LAT Source Catalogue \cite{2010ApJ...720..435A}. The lower limit of the integral in Eq.~\eqref{eqn:average_g_blazars} is set to $\mathcal{L}_{\rm min}=7\times10^{42}\,\mathrm{erg\, s}^{-1}$ and $\alpha_{\rm BLA}=2.1$ from the average spectral index in Ref.~\cite{Ajello:2013lka}.
\end{itemize}

The window function and the average emission of unresolved blazars is shown by the red lines in the right panels of Figs~\ref{fig:window} and \ref{fig:intensity}. We note that the window function decreases rapidly below $z=1$. This is due to the fact that we focus only on unresolved sources and \Fermi-LAT has already detected a large number of blazars at low redshift.

\subsubsection{Misaligned AGNs}
\label{sec:MAGNs}
In the case of mAGNs ($X=\gamma_{\rm mAGN}$), we follow Ref.~\cite{DiMauro:2013xta}, which determines a correlation between the \g-ray luminosity and the core radio luminosity $L_{r,{\rm core}}$ at 5 GHz. By means of this correlation it is possible to infer the mAGN \g-ray luminosity function from the study of radio-loud mAGNs at radio frequencies \cite{Willott:2000dh}. We consider the best-fit $\mathcal{L}-L_{r,{\rm core}}$ relation from Ref.~\cite{DiMauro:2013xta} and we assume an average spectral index $\alpha_{\rm mAGN}$ of 2.37. $\mathcal{L}$ is defined between 0.1 and 100 GeV, which sets $A_{\rm mAGN}=(\alpha_{\rm mAGN}-2)/[1+z(\chi)]^2$. The magenta lines in Figs~\ref{fig:window} and \ref{fig:intensity} indicate the window function and the average emission for unresolved mAGNs.

\subsubsection{Star-forming galaxies}
\label{sec:SFG}
For SFGs ($X=\gamma_{\rm SFG}$) we take a spectral index $\alpha_{\rm SFG}=2.7$, and the \g-ray luminosity $\mathcal{L}$ is computed between 0.1 and 100 GeV. As done in Ref.~\cite{Ackermann:2012vca}, we assume that the \g-ray and infrared (IR) luminosities are correlated and we adopt the best-fit $\mathcal{L}-L_{\rm IR}$ relation from Ref.~\cite{Ackermann:2012vca}. Concerning the IR luminosity function, we consider two models: model A is taken from Ref.~\cite{Rodighiero:2009up} (see also Ref.~\cite{Ackermann:2012vca}), whilst model B is from Ref.~\cite{Gruppioni:2013jna} (adding up the contribution of spiral, starburst and ANG-hosting SFGs, from their Table 8).

The window function and average emission from unresolved SFGs are represented by yellow lines in the right panels of Figs~\ref{fig:window} and \ref{fig:intensity}. The discontinuity in the window function around $z=1$ in model A is inherited from the analytic fit to the IR data obtained in Ref.~\cite{Rodighiero:2009up} (see however the results of Ref.~\cite{Tamborra:2014xia}). Also, note that our predictions for the \g-ray flux associated to unresolved SFGs in model A is compatible with the results presented in Ref.~\cite{Ackermann:2012vca}. The three sub-classes of SFGs considered in model B peak at different redshifts and they contribute similarly to the EGB. It is possible to recognise the individual peaks in the SFG window functions in the right panel of Fig.~\ref{fig:window}.

\subsubsection{Further discussion on astrophysical emission}
In what follows, whenever we refer to model A (B), we are considering model A (B) for both the descriptions of blazars {\it and} of SFGs.

The \g-ray emission produced by the three extragalactic astrophysical populations in model A accounts for $\sim70\%$ of the EGB above 1 GeV (see Fig.~\ref{fig:intensity}). This leaves room for other emissions. Apart from the DM-induced emission described in Secs~\ref{sec:annihilating_DM} and \ref{sec:decaying_DM}, there may be a contribution associated with annihilations/decays in the DM halo of the Milky Way (see e.g.\  Ref.~\cite{Calore:2013yia}). This is not included in Fig.~\ref{fig:intensity} and, since it does not correlate with the weak gravitational lensing signal, we do not consider such a component in the present work. Other astrophysical components that can significantly contribute to the EGB intensity are galaxy clusters \cite{Keshet:2002sw,Zandanel:2013wja}, milli-second pulsars \cite{FaucherGiguere:2009df,Calore:2014oga} and cascades induced by the interactions of ultra-high-energy cosmic rays with the CMB \cite{Coppi:1996ze,Ahlers:2010fw}. However, we expect the inclusion of such additional terms to change the results presented in this paper only marginally.

For model B, the sum of the \g-ray emission of the astrophysical populations would slightly overshoot the measured EGB~\cite{Ackermann:2014usa}. As a simple fix, we multiply the luminosity function of each of the three astrophysical components by 0.71. The number is the best-fit overall normalisation so that the total astrophysical component is in agreement with the black data point in Fig.~\ref{fig:intensity}.

\subsection{Cosmic shear}
\label{sec:cosmic_shear}
The presence of intervening matter along the path of photons emitted by distant sources causes gravitational lensing distortions in the images of such high-redshift objects. In the weak lensing r\'egime, lensing effects can be evaluated on the unperturbed null-geodesic of the unlensed photon \citep{Bartelmann:1999yn}. Such distortions, directly related to the distribution of matter on large scales and to the geometry and dynamics of the Universe, can be decomposed into the so-called convergence $\kappa$ and shear $\gamma$ \citep{Kaiser:1996tp,Bartelmann:1999yn}. Convergence is a direct estimator of the fluctuations in the Newtonian potential, integrated along the line of sight. On scales $\ell\gtrsim10$, its APS coincides with that of cosmic shear and, thus, it can be estimated via the statistical analysis of correlations in the observed source ellipticities (see also Sec.~\ref{sec:cosmic_shear_3Dpower spectrum}). Thanks to Poisson's equation, which relates the gravitational potential to the distribution of matter, the weak lensing ($X=\kappa$) window function (for both shear and convergence) is
\begin{equation}
W^\kappa(\chi) = \frac{3}{2} \ho^2 \om \left[ 1+z(\chi) \right] \chi
\int_{\chi}^\infty \de \chi' \, \frac{\chi'-\chi}{\chi'} 
\frac{\de N_g}{\de\chi'}(\chi')
\label{eq:W_lens},
\end{equation}
where $\de N_g / \de \chi$ is the redshift distribution of background galaxies, and is normalised such that $\int d\chi\, W^\kappa(\chi) = 1$.

The window function of Eq.~\eqref{eq:W_lens} is shown as black lines in Fig.~\ref{fig:window} for the expected distribution of background galaxies of \DES\ (dashed curve) and \Euclid\ (solid curve). See Sec.~\ref{sec:surveys} for a comprehensive description of the surveys.

\section{Auto- and cross-correlation angular power spectra}
\label{sec:auto_and_cross_correlations}
In this section, we describe the formalism used to compute the auto-correlation APS of anisotropies in the \g-ray emission and their cross-correlation with cosmic shear. The APS quantifies the fluctuation amplitude in 2-dimensional sky maps. We define the fluctuation along the line-of-sight direction $\hat{\mathbf n}$ as 
$\delta \mathcal I^X(\hat{\mathbf n}) = \mathcal I^X(\hat{\mathbf n}) - 
\langle \mathcal I^X \rangle$ and we expand it in spherical harmonics as
\begin{equation}
\delta \mathcal I^X(\hat{\mathbf n}) = 
\sum_{\ell,m} a_{\ell,m} Y_{\ell,m}(\hat{\mathbf{n}}).
\end{equation} 
The $a_{\ell,m}$ coefficients can be written as
\begin{equation}
a_{\ell,m} =
\int \de \hat{\mathbf{n}} \, \delta \mathcal{I}^X(\hat{\mathbf{n}}) 
Y_{\ell,m}(\hat{\mathbf{n}}) = 
\int \de \hat{\mathbf{n}}\,\de \chi \, f_{X}(\chi,\hat{\mathbf{n}}) 
\, W^X(\chi) Y_{\ell,m}(\hat{\mathbf{n}}),
\label{eqn:a_lm}
\end{equation}
where $f_{X}(\chi,\hat{\mathbf{n}})$ is the fluctuation field, defined as $f_{X}(\chi,\hat{\mathbf{n}}) = g_X(\chi,\hat{\mathbf{n}}) / \langle g_X \rangle - 1$.

From the definition of the APS, $C_\ell^{XY} = (2\ell+1)^{-1} \sum_{m=-\ell}^\ell\langle a_{\ell, m}^{X} a_{\ell, m}^{Y*}\rangle$, we eventually obtain (see, e.g., Ref.\cite{Fornengo:2013rga}):
\begin{equation}
C_{\ell}^{XY} = 
\int\!\!\de \chi\,\frac{W^X(\chi)W^Y(\chi)}{\chi^2} 
P^{XY}\!\!\left(k=\frac{\ell}{\chi},\chi\right).
\label{eqn:APS}
\end{equation}
Here, $P^{XY}(k,\chi)$ stands the 3-dimensional power spectrum (3D PS) of fluctuations in observables $X$ and $Y$ ($X=Y$ for auto-correlation) and it is the Fourier transform of the 2-point correlation function of the fluctuation fields, $\langle \tilde{f}_X(\chi,\mathbf{k}) \tilde{f}_Y(\chi,\mathbf{k^\prime}) \rangle = (2\pi)^3 \delta^3 (\mathbf{k} + \mathbf{k^\prime}) P^{XY}(k,\chi)$. For the computation of $P^{XY}$, we adopt the so-called halo-model approach. Following Ref.~\cite{Scherrer:1991kk}, we assume that each fluctuation field is a sum of discrete seeds identified by their position and one characteristic quantity. For astrophysical emission we consider their luminosity $\mathcal{L}$, while, for the emission produced by DM haloes, we take the halo mass $M$. The details of the computation of the 3D PS have been recently reviewed in Ref.~\cite{Fornengo:2013rga}: $P^{XY}$ can be written as the sum of two contributions, namely a 1-halo $P^{XY}_{1h}$ and a 2-halo $P^{XY}_{2h}$ term. The former accounts for the correlation inside the same object, whilst the latter describes the correlations between two distinct sources.

Eq.~\eqref{eqn:APS} is obtained under Limber's approximation \cite{Limber:1953,Kaiser:1987qv}, which links the physical wavenumber $k$ to the angular multipole $\ell$. Such an approximation is known to hold for $\ell \gg 1$. We note that this condition is normally satisfied in this paper, since we consider angular scales for which $\ell\gtrsim50$.

The window functions entering Eq.~\eqref{eqn:APS} have already been described in Sec.~\ref{sec:intensity}. In the following, we present a detailed description of the various ingredients required for the computation of $P_{1h}$ and $P_{2h}$ (from Secs~\ref{sec:APS-gamma} to \ref{sec:discussion} included). The reader can skip to Sec.~\ref{sec:tomospec} for the introduction of the tomographic-spectral technique.

\subsection{Angular power spectrum of \g-ray emission}
\label{sec:APS-gamma}
Since we consider four classes of \g-ray emitters (three astrophysical populations plus DM), the auto-correlation APS of the total \g-ray emission is given by the sum of the auto-correlation APS of each class of emitters and their relative cross-correlations. The APS of the total \g-ray emission reads as follows (we do not specify if DM is annihilating or decaying):
\begin{multline}
C^\gamma_\ell = C^{\gamma_\mathrm{DM}}_\ell + C^{\gamma_{\rm BLA}}_\ell + 
C^{\gamma_{\rm mAGN}}_\ell + C^{\gamma_{\rm SFG}}_\ell + 2C^{\gamma_\mathrm{DM}\gamma_{\rm BLA}}_\ell 
+ 2C^{\gamma_\mathrm{DM}\gamma_{\rm mAGN}}_\ell + 2C^{\gamma_\mathrm{DM}\gamma_{\rm SFG}}_\ell \\+
2C^{\gamma_{\rm BLA}\gamma_{\rm mAGN}}_\ell + 2C^{\gamma_{\rm BLA}\gamma_{\rm SFG}}_\ell + 
2C^{\gamma_{\rm mAGN}\gamma_{\rm SFG}}_\ell.
\label{eq:clgamma}
\end{multline}

\subsubsection{Dark matter}
In the case of decaying DM, the density field is given by the DM density distribution, namely $g_{\gamma_{{\rm DMd}}}(\mathbf x) = \rho_{\rm DM}(\mathbf x)$. This means $f_{\gamma_{{\rm DMd}}}(\mathbf x) = \delta(\mathbf x)$, where $\delta(\mathbf x)$ is the DM density contrast. Therefore, the 3D PS quantifies the correlation between two density fluctuations $\delta$, and its 1- and 2-halo terms are
\begin{align}
P^{\delta\delta}_{1h}(k,z)& = \int_{M_{\rm min}}^{M_{\rm max}} dM \, \frac{\de n}{\de M} 
\tilde v^2(k|M),
\label{eq:psdelta1} \\
P^{\delta\delta}_{2h}(k,z)& = \left[ \int_{M_{\rm min}}^{M_{\rm max}} dM \, 
\frac{\de n}{\de M} b_h(M) \tilde v(k|M) \right]^2P_{\rm lin}(k,z),
\label{eq:psdelta2}
\end{align}
where $\tilde v(k|M)$ is the Fourier transform of $\rho_{\rm DM}(\mathbf x|M)/\bar \rho_{\rm DM}$, $P_{\rm lin}$ is the linear matter power spectrum and $b_h$ is the bias between haloes and the DM distribution. The bias is taken from Ref.~\cite{Cooray:2002dia}.

Conversely, in the case of annihilating DM, the source field scales with the square of the DM density, viz.\ $g_{\gamma_{{\rm DMa}}}(\mathbf x)=\rho_{\rm DM}^2(\mathbf x)$. Then, the 3D PS correlates fluctuations in the squared density field $\delta^2$, i.e.
\begin{align}
P^{\delta^2\delta^2}_{1h}(k,z)& = \int_{M_{\rm min}}^{M_{\rm max}} dM \, \frac{\de n}{\de M} 
\left[\frac{\tilde u(k|M)}{\Delta^2} \right]^2,
\label{eq:power spectrumh1} \\
P^{\delta^2\delta^2}_{2h}(k,z)& = \left[ \int_{M_{\rm min}}^{M_{\rm max}} dM \, 
\frac{\de n}{\de M} b_h(M) \frac{\tilde u(k|M)}{\Delta^2} \right]^2 \, 
P_{\rm lin}(k,z),
\label{eq:power spectrumh2}
\end{align}
where $\tilde u(k|M)$ is the Fourier transform of $\rho_{\rm DM}^2(\mathbf x|M)/\bar \rho_{\rm DM}^2$.

\begin{figure}
\centering
\hspace{-4mm}
\includegraphics[width=0.45\textwidth]{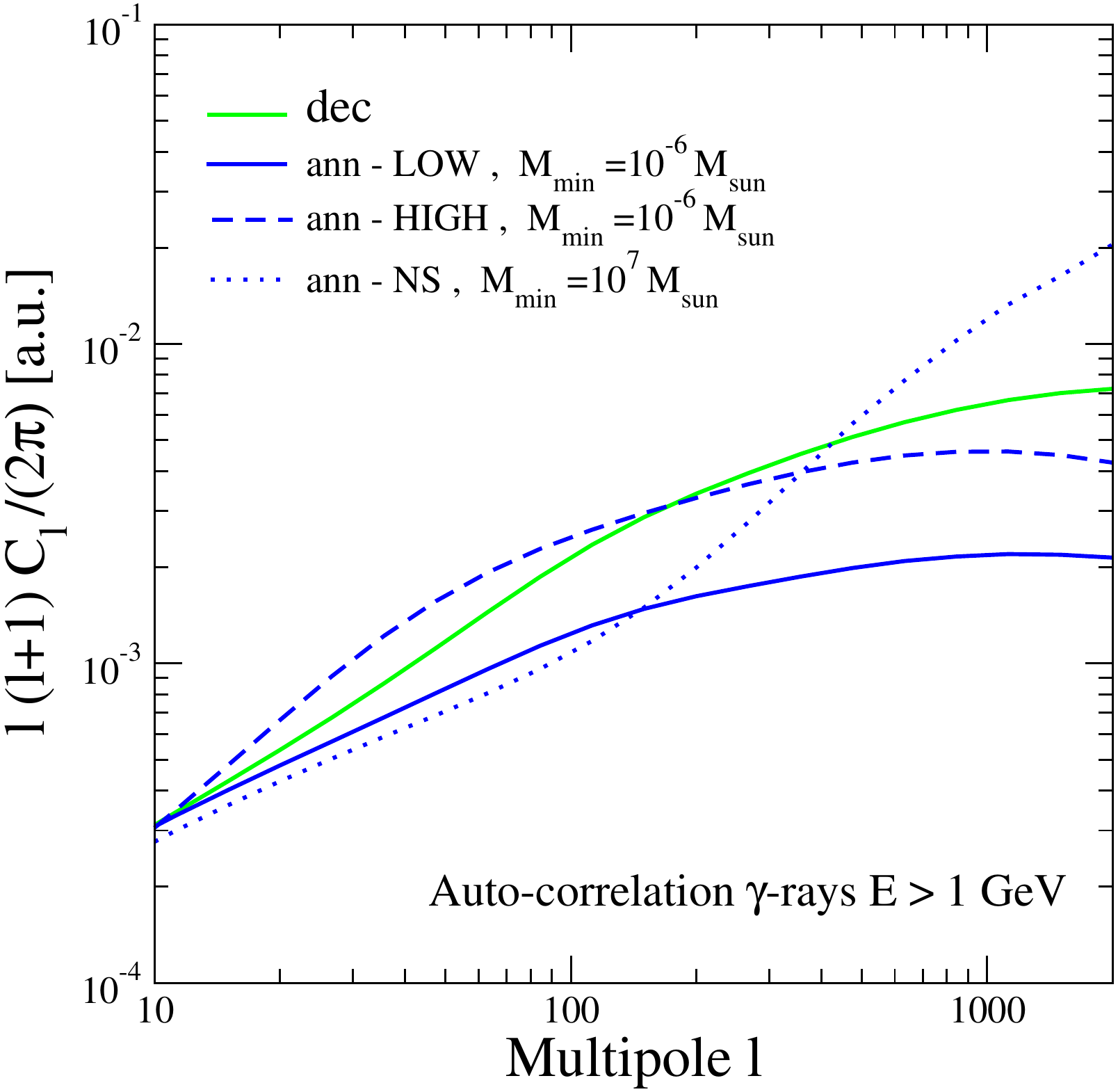}
\hspace{10mm}
\includegraphics[width=0.45\textwidth]{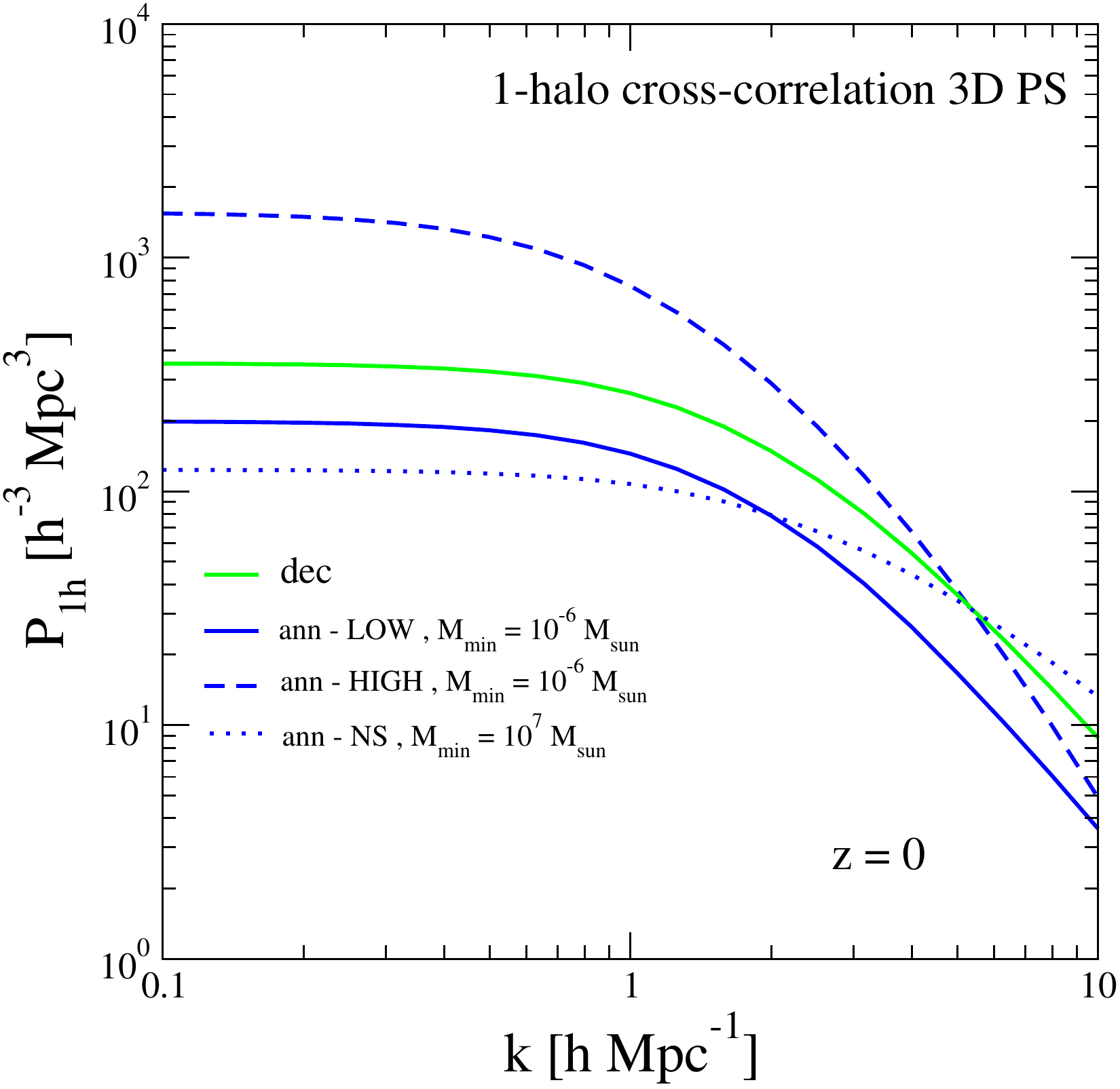}
\caption{Left: Arbitrarily normalised auto-correlation APS of the \g-ray emission for annihilating (blue) and decaying (green) DM. The properties of the DM particle are chosen as in Fig.~\ref{fig:window}. The solid (dashed) blue line refers to the \low\ (\high) scenario, whilst the dotted one is for \ns. Right: 1-halo term of the 3D PS for the cross-correlation of cosmic shear with DM-induced \g-ray emission at $z=0$. We use the same scenarios and line styles as in the left panel.}
\label{fig:cl_DM}
\end{figure}

The linear matter power spectrum is well constrained by cosmological observations and $N$-body simulations for masses above $10^{10}\,M_\odot$. In the case of a decaying DM candidate, the bulk of the signal comes from large haloes. Therefore, the predicted APS is rather robust and the auto-correlation APS (for decaying DM) is plotted as the solid green line in the left panel of Fig.~\ref{fig:cl_DM}. In the case of annihilating DM, the APS is affected by uncertainties related to the computation of the flux multiplier at small masses, i.e.\ the amount of subhaloes and the value of $M_{\rm min}$ (see Sec.~\ref{sec:annihilating_DM} and Ref.~\cite{Ando:2013ff} for further discussion). These uncertainties affect both the normalisation and shape of the APS. The former is proportional to the intensity of the emission, so its effect is expected to be similar to what we see in Fig.~\ref{fig:intensity} for the EGB intensity.

In Fig.~\ref{fig:cl_DM} (left panel) we plot the auto-correlation APS for annihilating DM in the case of the \ns\ (dotted, blue line), \low\ (solid, blue line) and \high\ (dashed, blue line) scenarios. To compare the shape of the different APS, we use a common arbitrary normalisation at low multipoles. The 1-halo term dominates at large multipoles (i.e. low angular scales).
In particular, its intensity at multipoles below few hundreds increases if the bulk of the DM-induced emission comes from haloes with large masses. 
This implies that, since the contribution of subhalos in the \high\ case boosts mainly the emission from large haloes, the APS grows more rapidly than the \low\ case for $\ell\lesssim 100$.
In general, in the presence of DM subhaloes, DM haloes with medium to large masses dominate the signal.
This also implies that, for the \low\ and \high\ scenarios, the 3D PS drops at small scales. In the \ns\ scenario, on the other hand, (relatively) low masses give a significant contribution and the associated 1-halo term extends to smaller scales. 
Consequently, the APS raises at large multipoles. This can be seen in the left panel of Fig.~\ref{fig:cl_DM} in the large-$\ell$ r\'egime, where the 1-halo term dominates over the 2-halo term. The (normalized) APS of the \ns\ scenario---where there are no subhaloes---is larger than the \low\ and \high\ cases for $\ell>300$. This trend is compatible with what found in Ref.~\cite{Fornasa:2012gu}.

We emphasise that, contrarily to some other works in the literature, we choose to normalise the 3D PS by $\langle g_X \rangle$ (i.e. proportional to $\Delta^2$, in the case of annihilating DM), which is re-absorbed in the window function. By doing so, the $P_{2h}$ reduces (approximately) to the linear matter power spectrum, except for the multiplicative bias term.

\subsubsection{Astrophysical sources}
For \g\ rays induced by astrophysical sources, we work under the hypothesis that their distribution traces the underlying matter field. Then, equations similar to Eqs~\eqref{eq:psdelta1} to \eqref{eq:power spectrumh2} can still be used to quantify the 3D 1-halo and 2-halo PS, with some caveats. First, the quantity that characterises the sources is not longer the halo mass, but their luminosity $\mathcal{L}$. Secondly, we have to consider the \g-ray luminosity function instead of the halo mass function $\de n/\de M$. Finally, instead of the halo bias $b_h$, each class of sources is characterised by its own bias factor $b_{\gamma_i}(\mathcal{L},z)$. Furthermore, under the hypothesis of point-like sources, the 1-halo term looses any dependence on $k$ and becomes Poisson-like. Hence, the 3D PS can be written as
\begin{align}
P^{\gamma_i\gamma_i}_{1h}(k,z)&= \int_{\mathcal{L}_{\rm min}(z)}^{\mathcal{L}_{\rm max}(z)}\!\!
\de \mathcal{L}\,\rho_{\gamma_i}(\mathcal{L},z)
\left( \frac{\mathcal{L}}{\langle g_{\gamma_i} \rangle} \right)^2,
\\
P^{\gamma_i\gamma_i}_{2h}(k,z)&= \left[\int_{\mathcal{L}_{\rm min}(z)}^{\mathcal{L}_{\rm max}(z)}\!\!
\de\mathcal{L} \, \rho_{\gamma_i}(\mathcal{L},z) \, b_{\gamma_i}(\mathcal{L},z)\frac{\mathcal{L}}{\langle g_{\gamma_i} \rangle} \right]^2 P_{\rm lin}(k,z).
\label{eq:power spectrumB}
\end{align}
If the \g-ray emission of a population of sources is dominated by a few very bright objects (but still below the detection threshold of the telescope), the 1-halo term will be larger than the 2-halo term and the auto-correlation APS will therefore be Poissonian, with no dependence on the source clustering. On the other hand, the auto-correlation APS can have some sensitivity to the distribution of the sources if they are faint but numerous. Indeed, in this case, the 2-halo term will dominate over the Poissonian 1-halo term. Following our description of astrophysical sources in Sec.~\ref{sec:intastro}, the auto-correlation APS of blazars and mAGNs results to be Poissonian over the multipole range considered here ($\ell < 1000$), whilst for SFGs, some dependence on the 2-halo term is present below $\ell\simeq500$.

The cross-correlation terms in Eq.~\eqref{eq:clgamma} includes correlations between the \g-ray emission of two different astrophysical populations or between an astrophysical population and DM. 

\begin{figure}
\centering
\hspace{-4mm}
\includegraphics[width=0.45\textwidth]{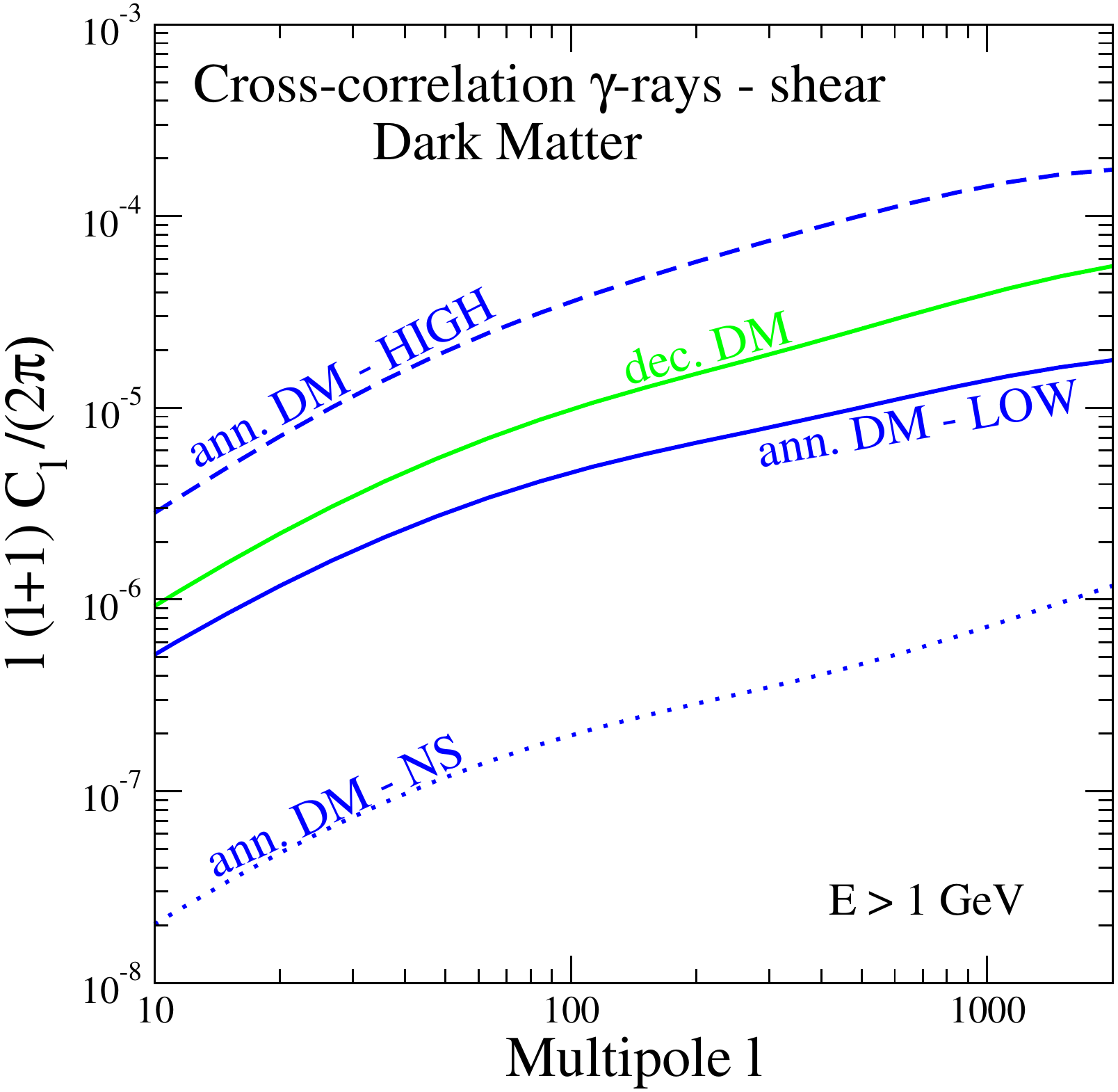}
\hspace{10mm}
\includegraphics[width=0.45\textwidth]{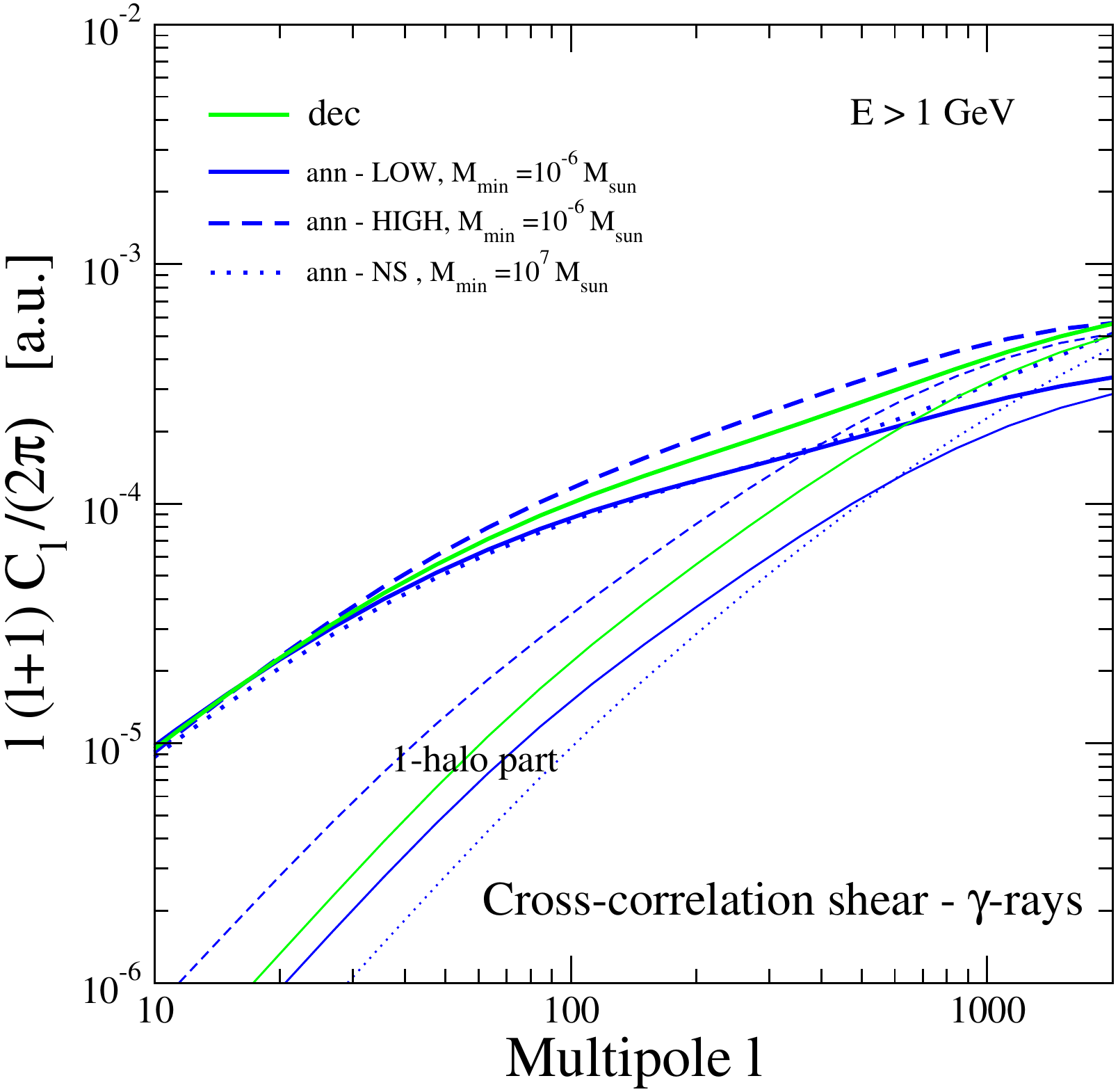}
\caption{Cross-correlation APS of Eq.~\eqref{eq:clkX} for the DM-induced \g-ray emission at $E>1$ GeV and the weak lensing signal of \Euclid. The DM microscopic properties are chosen as in Fig.~\ref{fig:window}. The solid (dashed) blue line refers to the \low\ (\high) scenario, whilst the dotted one is for \ns. In the right panel, the total APS is showed with thicker lines, while the 1-halo terms is denoted with thinner lines. Curve are arbitrarily normalised to better compare their angular shapes.}
\label{fig:cl_DM2}
\end{figure}

The 3D PS for the correlations between a class of astrophysical emitters and decaying DM reads
\begin{align}
P^{\gamma_i\delta}_{1h}(k,z) & = 
\int_{\mathcal{L}_{\rm min}(z)}^{\mathcal{L}_{\rm max}(z)}\!\!\de\mathcal{L} \, 
\rho_{\gamma_i}(\mathcal{L},z) \frac{\mathcal{L}}{\langle g_{\gamma_i} \rangle} \, 
\tilde v[k|M(\mathcal{L})]
\label{ps:1hdS} \\
P^{\gamma_i\delta}_{2h}(k,z) &= 
\left[\int_{\mathcal{L}_{\rm min}(z)}^{\mathcal{L}_{\rm max}(z)}\!\!\de\mathcal{L} \, 
\rho_{\gamma_i}(\mathcal{L},z)\, b_{\gamma_i}(\mathcal{L},z) 
\frac{\mathcal{L}}{\langle g_{\gamma_i} \rangle} \right]
\left[ \int_{M_{\rm min}}^{M_{\rm max}}\!\!\de M \,\frac{\de n}{\de M} b_h(M)
\tilde v(k|M) \right] \, P_{\rm lin}(k,z).
\label{ps:2hdS}
\end{align}
On the other hand, for annihilating DM:
\begin{align}
P^{\gamma_i\delta^2}_{1h}(k,z) &= 
\int_{\mathcal{L}_{\rm min}(z)}^{\mathcal{L}_{\rm max}(z)}\!\!\de\mathcal{L} \, 
\rho_{\gamma_i}(\mathcal{L},z) \frac{\mathcal{L}}{\langle g_{\gamma_i} \rangle}
\frac{\tilde u[k|M(\mathcal{L})]}{\Delta^2} 
\label{ps:1hd2S} \\
P^{\gamma_i\delta^2}_{2h}(k,z) &= 
\left[\int_{\mathcal{L}_{\rm min}(z)}^{\mathcal{L}_{\rm max}(z)}\!\!\de\mathcal{L} \, 
\rho_{\gamma_i}(\mathcal{L},z)b_{\gamma_i}(\mathcal{L},z)
\frac{\mathcal{L}}{\langle g_{\gamma_i} \rangle} \right] 
\left[ \int_{M_{\rm min}}^{M_{\rm max}}\!\!\de M \, \frac{\de n}{\de M} b_h(M) 
\frac{\tilde u(k|M)}{\Delta^2} \right]P_{\rm lin}(k,z).
\nonumber
\end{align}
Lastly, the 2-halo term of the cross-correlation 3D PS between two 
astrophysical populations is
\begin{equation}
P^{\gamma_i\gamma_j}_{2h}(k,z)= 
\left[ \int_{\mathcal{L}_{\rm min}(z)}^{\mathcal{L}_{\rm max}(z)}\!\!\de\mathcal{L} \, 
\rho_{\gamma_i}(\mathcal{L},z)b_{\gamma_i}(\mathcal{L},z)
\frac{\mathcal{L}}{\langle g_{\gamma_i} \rangle} \right]
\left[\int_{\tilde{\mathcal{L}}_{\rm min}(z)}^{\mathcal{L}_{\rm max}(z)}\!\!\de\mathcal{L} \, 
\rho_{\gamma_j}(\mathcal{L},z) b_{\gamma_j}(\mathcal{L},z)
\frac{\mathcal{L}}{\langle g_{\gamma_j} \rangle} \right] P_{\rm lin}(k,z)\;.
\end{equation}
Here, the term $P^{\delta\delta}_{1h}=0$ since we consider astrophysical sources as point-like and two distinct objects cannot be in the same point.

Note that in Eqs~\eqref{ps:1hdS} and \eqref{ps:1hd2S}, a relation between the source luminosity $\mathcal{L}$ and the mass $M$ of host is required. Such a relation is not well established and different prescriptions can lead to very different results for the 1-halo terms. We will discuss it in more details in the following section.

To summarise, in order to estimate the contribution of astrophysical sources to the APS, four ingredient are needed: $i)$ the \g-ray luminosity function of the specific source class; $ii)$ its energy spectrum; $iii)$ the relation between the \g-ray luminosity of the source and the mass of the host DM halo, $M(\mathcal{L})$ and $iv)$ the bias $b_{\gamma_i}$. The last quantity can be computed in terms of the halo bias $b_{\gamma_i}(\mathcal{L},z) = b_h[M(\mathcal{L}),z]$, depending, once more, on $M(\mathcal{L})$. We note that, since at low $z$ the halo bias $b_h$ is $\mathcal{O}(1)$, the 2-halo term is often only very mildly dependent on $M(\mathcal{L})$.

\subsection{Cross-correlation of cosmic shear and \g-ray emission}
\label{sec:cosmic_shear_3Dpower spectrum}
In a perturbed Universe around a (flat) Friedmann-Lema\^{\i}tre-Robertson-Walker background, the line element reads
\begin{equation}
\de s^2 = -(1+2\Phi) \,\de t^2 + a^2(t) \left[ (1-2\Phi) 
\left( \de\chi^2 + \chi^2 \, \de \Omega^2\ \right) \right],
\end{equation}
where $\Phi$ is the Newtonian potentials and $\Omega$ is the solid angle. In such a perturbed Universe, a light ray coming towards us from a distant source is deflected by an angle proportional to the transverse gradient of the Newtonian potential $\Phi$ (see e.g.\ Refs~\cite{Bartelmann:1999yn,Bartelmann:2010fz}). By means of Poisson's equation, fluctuations in the potential correspond to fluctuations in the matter distribution, which plays directly the r\^ole of the fluctuation field $f_{\kappa}(\hat{\mathbf{n}},\chi)$ in Eq.~\eqref{eqn:a_lm}.

The cross-correlation APS between the cosmic shear and a \g-ray emitter $X$ follows from Eq.~\eqref{eqn:APS} (see also Paper I and Ref.~\cite{Fornengo:2013rga}):
\begin{equation}
C_{\ell}^{X\kappa} = 
\frac{1}{\langle \mathcal{I}^{EGB}\rangle} 
\int\!\!\de \chi\,\frac{W^X(\chi) W^{\kappa}(\chi)}{\chi^2}  P^{X\kappa}\left(k=\frac{\ell}{\chi},\chi\right).
\label{eq:clkX}
\end{equation}
In Eq.~\eqref{eq:clkX}, we have included a normalisation $\langle \mathcal{I}^{EGB}\rangle$, given by the measured EGB intensity, namely $\langle \mathcal{I}^{EGB} \rangle=4\times10^{-7} {\rm cm^{-2} s^{-1} sr^{-1}}$ for $E>1$ GeV. This allows us to work with a dimensionless APS. We also remind the reader that the shear window function was defined so that the average shear intensity $\langle\mathcal I^\kappa\rangle\equiv \int d\chi\, W^\kappa(\chi)$ is 1. Notice, also, that we \textit{do not} divide by the average \g-ray intensity of each class of sources (as often done in the literature), because in this case the $C_{\ell}$'s would no longer be additive.

The total cross-correlation APS is the sum of contributions of the four classes of sources considered, i.e.
\begin{equation}
C^{\gamma \kappa }_\ell = C^{\gamma_{\rm DM}\kappa}_\ell + 
C^{\gamma_{\rm BLA}\kappa }_\ell + C^{\gamma_{\rm mAGN}\kappa }_\ell + C^{\gamma_{\rm SFG}\kappa }_\ell.
\end{equation}

\subsubsection{Dark matter}
In the case of decaying DM, both the DM-induced \g-ray emission and cosmic shear depend linearly on the DM density. Therefore, the 3D cross-correlation PS is equivalent to that of the auto-correlation for decaying DM and is given by Eqs~\eqref{eq:psdelta1} and \eqref{eq:psdelta2}. The 1-halo term of the 3D PS $P^{\delta\delta}_{1h}$ is represented by the solid green line in Fig.~\ref{fig:cl_DM} (right panel).

For the case of annihilating DM, the 1-halo and 2-halo terms of the 3D cross-correlation PS are
\begin{align}
P^{\delta\delta^2}_{1h}(k,z) & = \int_{M_{\rm min}}^{M_{\rm max}}\!\!\de M \, 
\frac{\de n}{\de M} \tilde v(k|M)\frac{\tilde u(k|M)}{\Delta^2},
\label{eq:power spectrumsh1} \\
P^{\delta\delta^2}_{2h}(k,z) & = \left[ \int_{M_{\rm min}}^{M_{\rm max}}\!\!\de M \,
\frac{\de n}{\de M} b_h(M) \tilde v(k|M) \right]
\left[ \int_{M_{\rm min}}^{M_{\rm max}}\!\!\de M \,\frac{\de n}{\de M} b_h(M) 
\frac{\tilde u(k|M)}{\Delta^2} \right]P_{\rm lin}(k,z).
\label{eq:power spectrumsh2}
\end{align}

The 2-halo term $P^{\delta\delta^2}_{2h}$ is fairly insensitive to the uncertainties associated to the clustering of haloes at small masses. Their effect mainly alter the bias factor, but no appreciable changes in the shape of the power spectrum are induced. On the other hand, those uncertainties have a heavier impact on the 1-halo term. The right panel of Fig.~\ref{fig:cl_DM} shows the 1-halo term for the 3D PS $P^{\delta\delta^2}_{1h}$. The blue solid (dashed) curve refers to the \low\ (\high) scenario, whilst the blue dotted line stands for the \ns\ case. As discussed above, large subhalo boosts increase the contribution of the most massive haloes and the 1-halo power spectrum is then approximately an order of magnitude larger for the \high\ scenario than for the \low\ case, at large scales, i.e.\ low values of $k$. The picture is opposite at small scales, where scenarios with a larger contribution from less massive haloes (especially the \ns\ model) have a larger 1-halo term.

Fig.~\ref{fig:cl_DM2} shows the cross-correlation APS between cosmic shear and DM. Blue (green) lines stand for annihilating (decaying) DM. The same line style as in Fig.~\ref{fig:cl_DM} is used to indicate the three different models for the clustering of DM haloes. Despite the differences amongst the three scenarios discussed in Fig.~\ref{fig:cl_DM} (left panel), the shape of the cross-correlation APS is now quite robust and it does not depend much on the amount of subhaloes or the value of $M_{\rm min}$. This is because those uncertainties mainly affect the 1-halo term and it can be easily seen in the right panel of Fig.~\ref{fig:cl_DM2}, in which we adopt an arbitrary normalisation to focus on the APS shape (as done in Fig.~\ref{fig:cl_DM}). The upper sets of lines describe the total cross-correlation APS, while the lower lines only consider the 1-halo term. The latter starts to dominate only at very large multipoles, $\ell \gtrsim 500$. Even if the shape of the APS does not depend dramatically on the clustering model, its normalisation does (as it is clear from the left panel), because it is proportional to the intensity of the emission (see also Fig.~\ref{fig:intensity}).

\subsubsection{Blazars}
\label{sec:M-L_blazars}
Moving to the discussion of the cross-correlation 3D PS between cosmic shear and astrophysical \g-ray emitters, the computation of the cross-correlation PS is given by Eqs.~(\ref{ps:1hdS}) and (\ref{ps:2hdS}). They rely on the knowledge of how the luminosity of the source, $\mathcal{L}$, relates to the mass of the host DM halo.

For the case of blazars, the $M(\mathcal{L})$ relation is modelled from the results of Ref.~\cite{Hutsi:2013hwa}. A simple power-law scaling $M = A(z) \mathcal{L}_X^{\Gamma(z)}$ between the X-ray luminosity, $\mathcal{L}_X$, and the mass of the host DM halo, $M$, is found to reproduce well the abundance of X-ray selected AGNs at different redshifts (see their Fig.~6 and Table~1). Then, the X-ray luminosity is linked to that of \g\ rays through the correlation already discussed and employed in Sec.~\ref{sec:blazars}. This will represent our fiducial model for the $M(\mathcal{L})$ relation for blazars. The corresponding $P_{1h}(k,z=0.2)$ is plotted in Fig.~\ref{fig:3D_power spectrum_cross_correlation_astro1} as a solid line (left panel).\footnote{For the sake of conciseness, we shown the impact of $M(\mathcal{L})$ on $P_{1h}$ only for model A (in both cases of blazars and SFGs). The impact on model B is very similar to what shown for model A.}

Conversely, Ref.~\cite{Ando:2006mt} consider a different approach, linking the \g-ray luminosity of a blazar to the mass of the super-massive black hole (SMBH) powering the AGN. In turn, that is related to the DM halo mass, leading to $M = 10^{11.3} M_\odot (\mathcal{L} / 10^{44.7}\,\mathrm{erg\,s^{-1}}^{-1})^{1.7} $ (what the authors of Ref.~\cite{Ando:2006mt} call `model A') or to $M=10^{13}M_\odot (\mathcal{L} / 10^{44}\,\mathrm{erg\,s^{-1}})^{1.7}$ (their `model B'). Both dashed and lower dotted lines in Fig.~\ref{fig:3D_power spectrum_cross_correlation_astro1} (left panel) produce a $P_{1h}$ that is smaller than the one obtained with the $M(\mathcal{L})$ relation from Ref.~\cite{Hutsi:2013hwa}. The prediction for their model A (i.e.\ the smaller of the two) is approximately 4 orders of magnitude below our fiducial model, for $k<10$ Mpc$/h$. We consider this as our lower limit for the 1-halo 3D PS for blazars. Our upper limit is a modification of model B from Ref.~\cite{Ando:2006mt}, with a larger normalisation, namely $M = 10^{13} M_\odot (\mathcal{L} / 10^{43}\,\mathrm{erg\,s^{-1}})^{1.7}$ (see the upper dotted line in the left panel of Fig.~\ref{fig:3D_power spectrum_cross_correlation_astro1}). This corresponds to a PS approximately one order of magnitude larger than our fiducial model. The uncertainty generated by the $M(\mathcal{L})$ relation is large and it propagates to the 1-halo term of the 3D cross-correlation PS. However, as we shall see later, such uncertainties will not affect our conclusions.

\subsubsection{Misaligned AGNs}
\label{sec:cross_correlation_mAGNs}
The relation $M(\mathcal{L})$ for mAGNs can be inferred through a chain of correlations. We first make use of the relation between the \g-ray luminosity and the core radio luminosity $L_{r,{\rm core}}$, as done already in Sec.~\ref{sec:MAGNs}. Then, $L_{r,{\rm core}}$ is related to the mass of the accreting SMBH, $M_\bullet$, as in Ref.~\cite{Franceschini:1998ub}. The $L_{r,{\rm core}}-M_\bullet$ relation is affected by considerable scatter, as found e.g.\ in Ref.~\cite{Bettoni:2002rk}. We find that allowing the normalisation of $M_\bullet(L_{r,{\rm core}})$ to vary in the range between 0.1 and 2.5 generates an uncertainty band that encompasses well the data (see, e.g. Fig. 8 of Ref.~\cite{Bettoni:2002rk}).

\begin{figure}
\centering
\centering
\hspace{-4mm}
\includegraphics[width=0.45\textwidth]{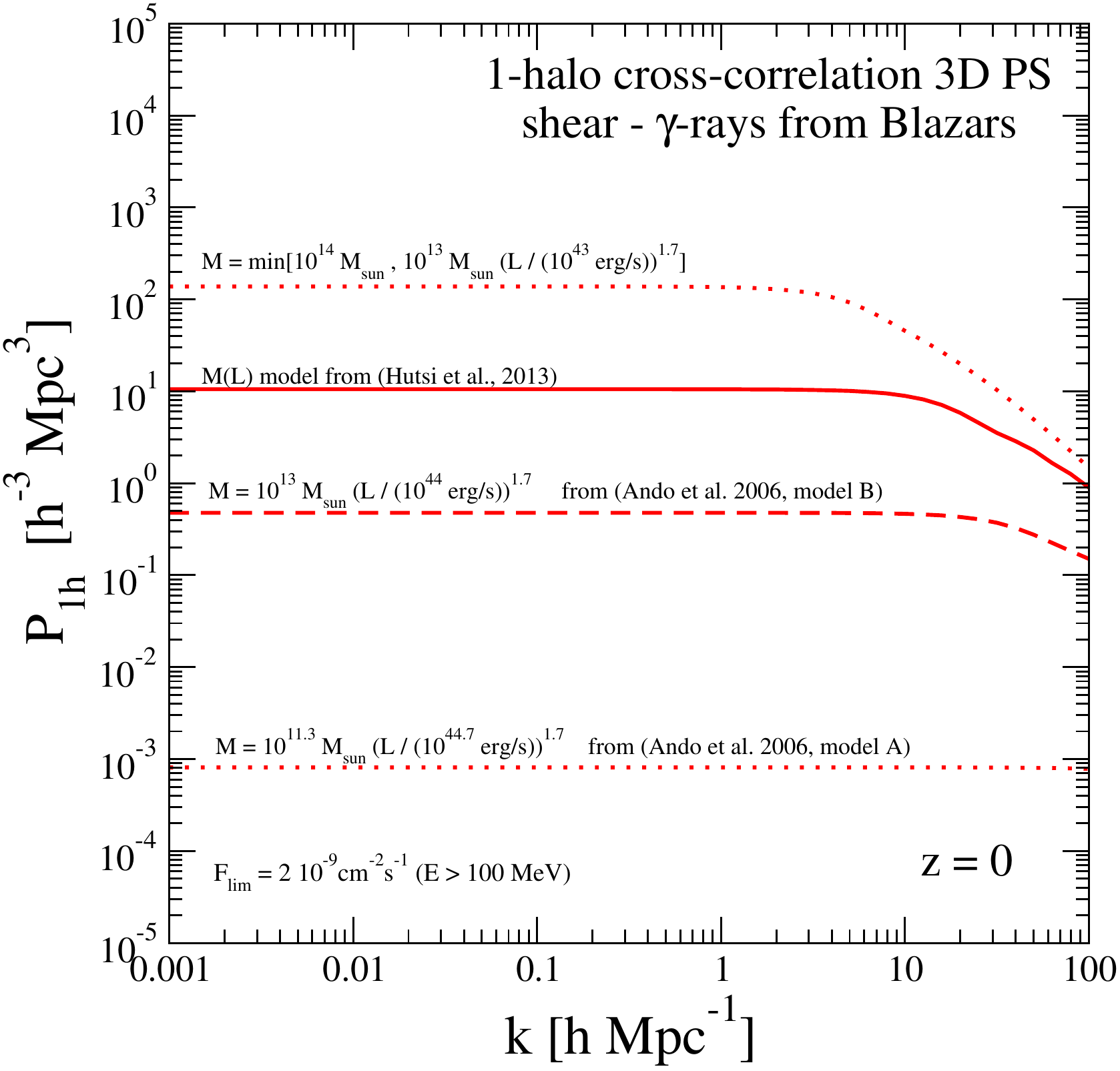}
\hspace{10mm}
\includegraphics[width=0.45\textwidth]{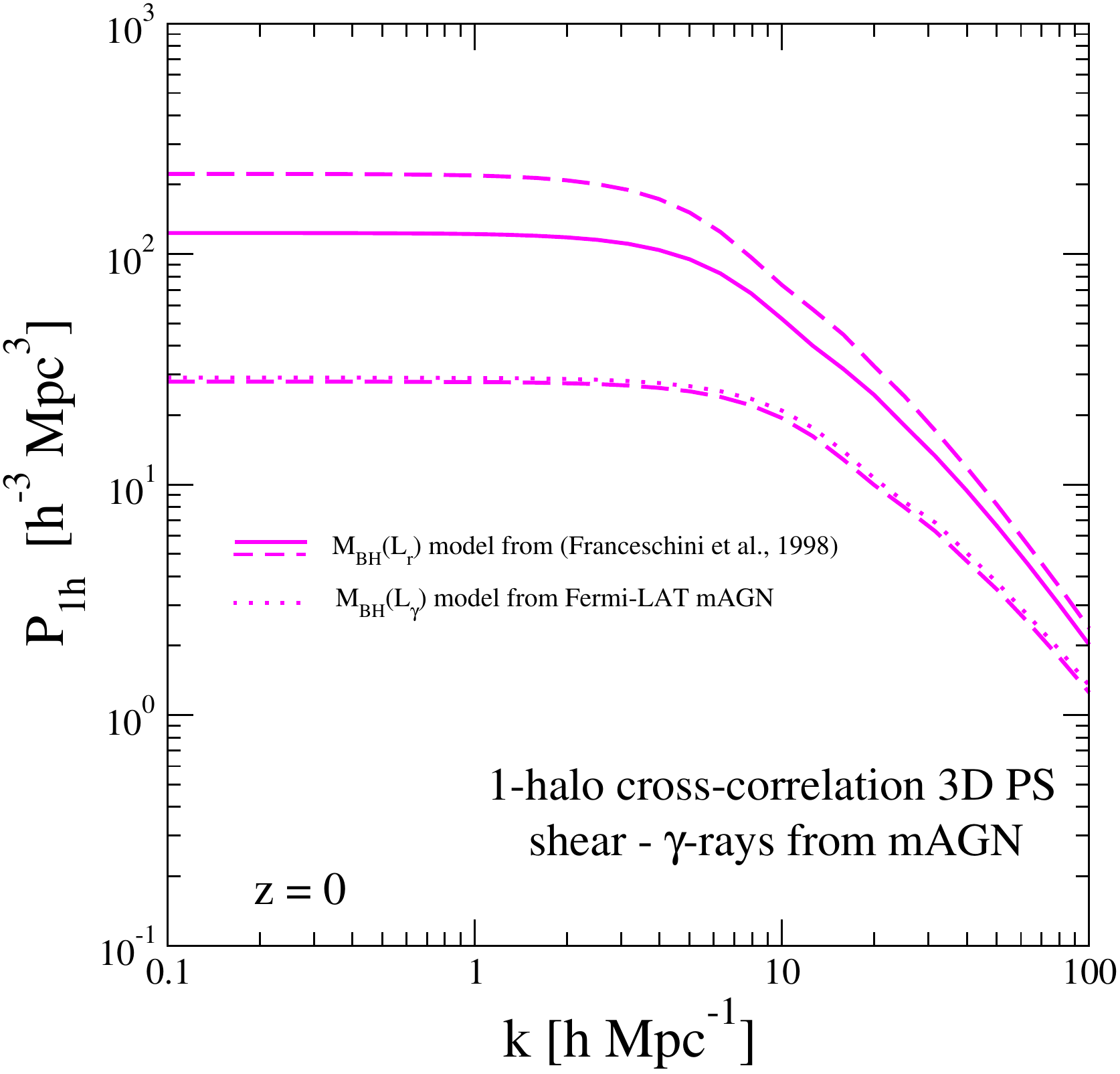}
\caption{1-halo term of the 3D cross-correlation PS between cosmic shear and \g\ rays from unresolved blazars (left) and mAGNs (right). The different lines correspond to different choices of the $M(\mathcal{L})$ relation. See text for details.}
\label{fig:3D_power spectrum_cross_correlation_astro1}
\end{figure}

Finally, we relate the mass of the SMBH to the mass of the host DM halo by following Ref.~\cite{Hutsi:2013hwa} (in turn based on Ref.~\cite{Bandara:2009sd}). However, we found that this procedure, when applied to the sample of mAGNs detected by \Fermi-LAT \cite{DiMauro:2013xta}, predicts SMBHs that are too massive. This is probably because the method is based on a galaxy population different from mAGNs. Thus, we renormalise the $M-M_\bullet$ relation of Ref.~\cite{Hutsi:2013hwa} in order to reproduce the characteristics of M 87, with a SMBH with a mass of $6.6 \times 10^9 M_\odot$ \cite{Gebhardt:2011yw} and a DM-halo of $2.2 \times 10^{13} M_\odot$ \cite{Gebhardt:2009cr}. The 1-halo 3D PS obtained by means of this $M(\mathcal{L})$ relation is plotted in the right panel of Fig.~\ref{fig:3D_power spectrum_cross_correlation_astro1} as a solid, magenta line (see also Appendix \ref{sec:appendix_MAGNs}) and it will represent our fiducial model. The two dashed lines quantify the uncertainty from the scatter in the $L_{r,{\rm core}}-M_\bullet$ relation mentioned above.

We also derive the relation between the \g-ray luminosity and the mass of the SMBH empirically, by using the measurements available in the literature on the mass of SMBHs for the mAGNs detected by \Fermi-LAT (see Appendix \ref{sec:appendix_MAGNs} and Fig.~\ref{fig:luminosity_mass_MAGN}). The result can be seen in the dotted line in Fig.~\ref{fig:3D_power spectrum_cross_correlation_astro1} (right panel) and is within the uncertainty band of the model described above.

\subsubsection{Star-forming galaxies}
\label{sec:cross_correlation_SFGs}
\begin{figure}
\centering
\hspace{-4mm}
\includegraphics[width=0.45\textwidth]{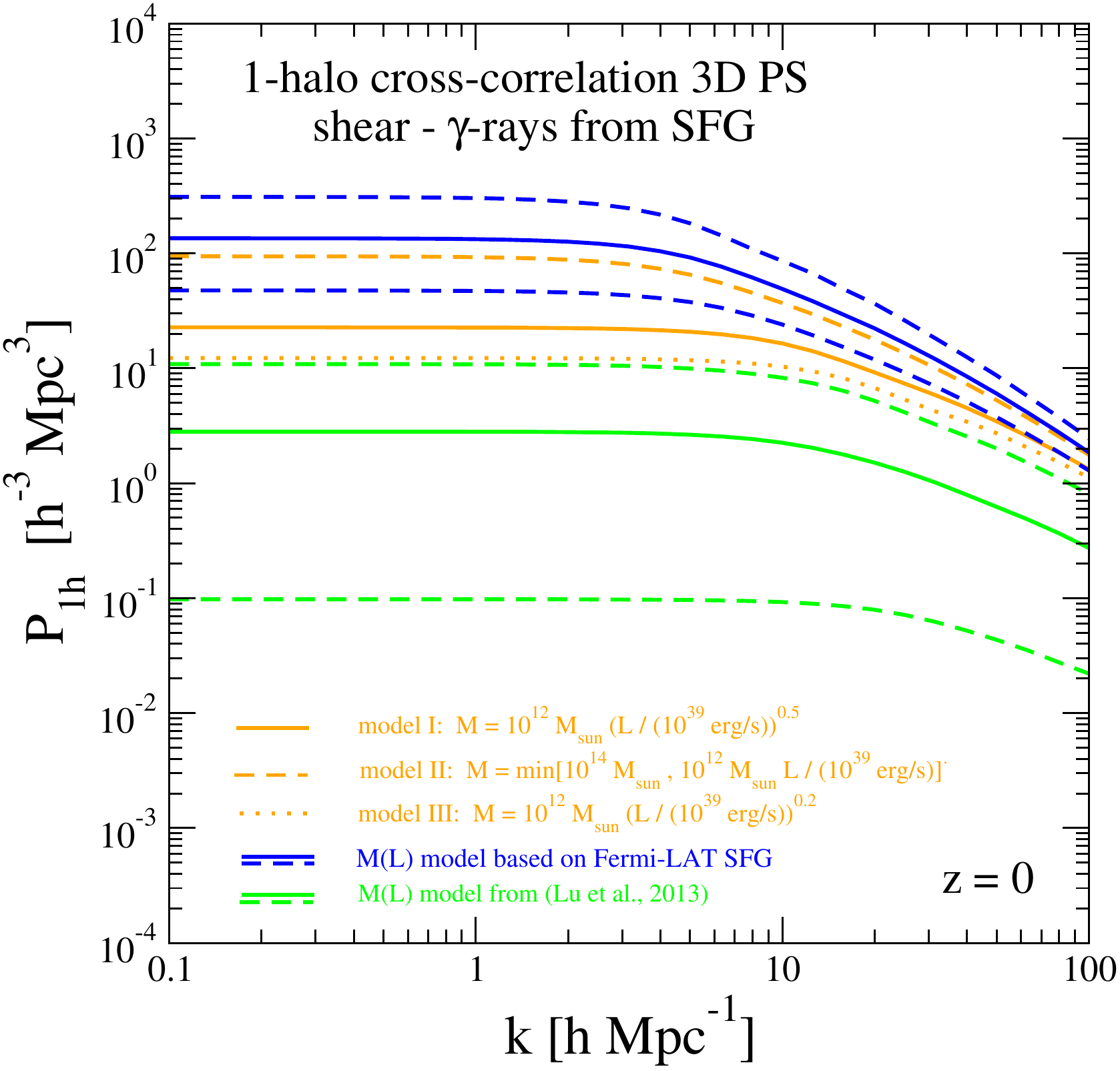}
\hspace{10mm}
\includegraphics[width=0.45\textwidth]{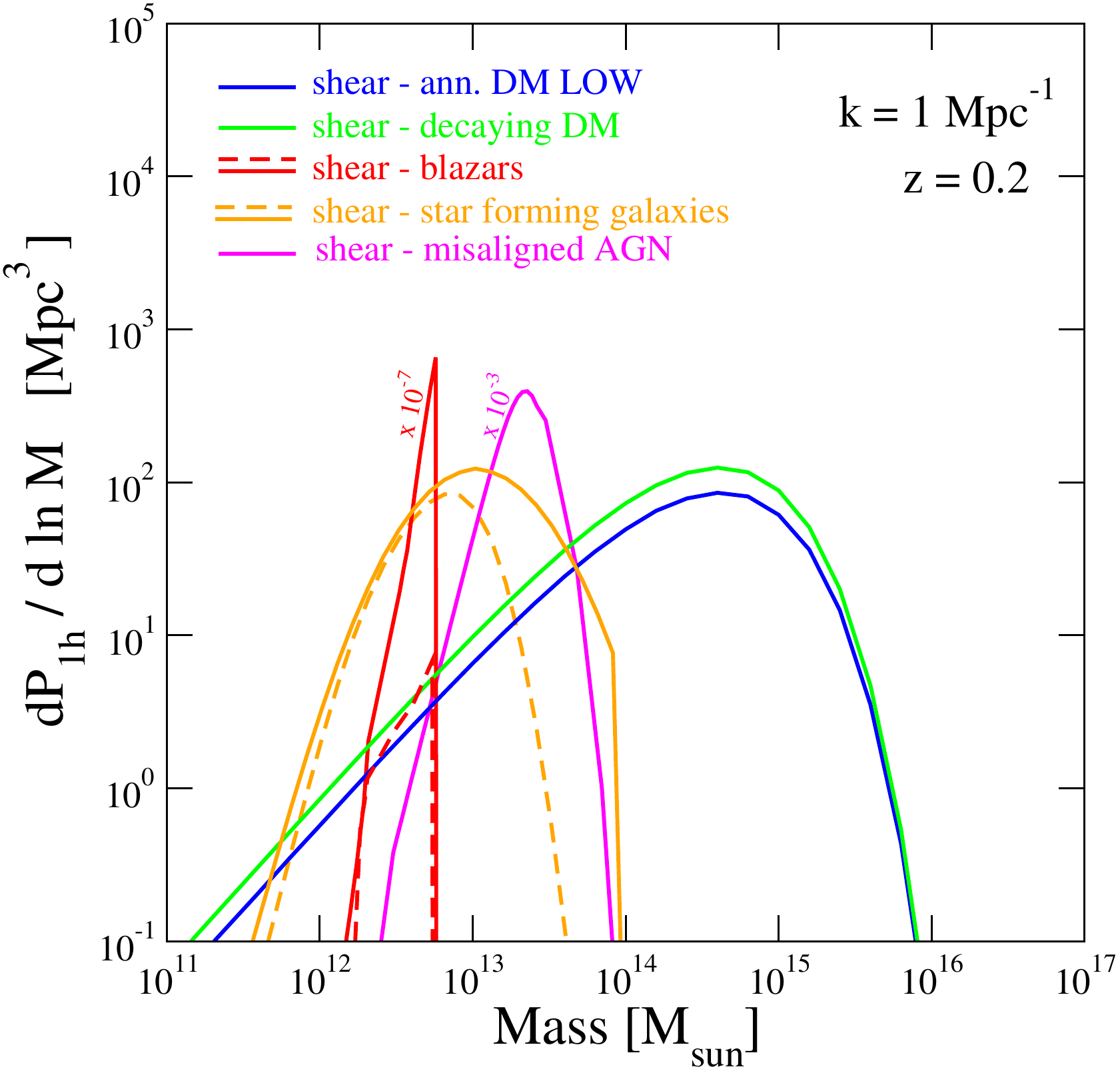}
\caption{Left: 1-halo term of the 3D cross-correlation PS between cosmic shear and \g\ rays from unresolved SFGs. The different lines correspond to different choices of the $M(\mathcal{L})$ relation. See text for details. Right: The quantity $\de P_{1h}/\de \ln M$ as a function of the mass of the DM halo, $M$ (at $z=0.2$ and $k=1\,{\rm Mpc^{-1}}$). $\de P_{1h}/\de \ln M$ is the integrand of the 1-halo term $P_{1h}$ of the 3D cross-correlation PS and it is plotted here for the different \g-ray emitters considered in this work. The blue line corresponds to annihilating DM (for a \low\ scenario), whilst the green one is for decaying DM. The red (magenta) line stands for blazars (mAGNs) and it has been multiplied by $10^{-7}$ ($10^{-3}$) to ease the comparison. The yellow one is for SFGs.}
\label{fig:3D_power spectrum_cross_correlation_astro2}
\end{figure}

In Ref.~\cite{Ackermann:2012vca}, the 8 SFGs detected by \Fermi-LAT (together with the upper limit from the non-detection of 64 bright IR galaxies) were employed to determine a relation between the \g-ray luminosity and the star-formation rate (SFR) (see Figs 3 and 4 of Ref.~\cite{Ackermann:2012vca}). A correlation between $\mathcal{L}$ and SFR is expected in many models describing the \g-ray emission of SFGs \cite{Fields:2010bw,Lacki:2012si}. Then, the mass of the DM halo can be derived from the SFR through the Kennicutt-Schmidt law, which links the SFR and the mass of the SFG gas contenct, and then assuming a certain DM-to-gas ratio. The resulting $M(\mathcal{L})$ relation is normalised to the properties of the Milky Way and it is used in Paper I. It shows that $M$ can be well approximated by $\simeq 10^{12}M_\odot\sqrt{\mathcal{L}/10^{39}\,\mathrm{erg\,s^{-1}}}$. This will represent our fiducial case and the corresponding 1-halo 3D PS is shown as a solid, yellow line in Fig.~\ref{fig:3D_power spectrum_cross_correlation_astro2} (left panel). The dashed and dotted yellow curves are taken from Paper I to provide reasonable uncertainties to this model.

An alternative relation between SFR and the mass of the host DM halo can be obtained from the empirical model for star formation proposed in Ref.~\cite{Lu:2013aoa} (see their Fig.~9). Together with the $\mathcal{L}$-SFR relation from Ref.~\cite{Ackermann:2012vca}, this provide a new $M(\mathcal{L})$ relation, and the corresponding 1-halo 3D PS is plotted in Fig~\ref{fig:3D_power spectrum_cross_correlation_astro2} (left panel) by means of green lines. Model III from Ref.~\cite{Lu:2013aoa} is adopted and the dashed green lines indicate the uncertainty induced when the parameters of the model are left free to vary (within the 95\% C.L. region).

As done in the previous section, we can also proceed empirically and derive the $M(\mathcal{L})$ relation by means of a power-law fit considering the SFGs detected by \Fermi-LAT for which we have estimates of the mass of their DM halo (see Appendix \ref{sec:appendix_SFGs} and Fig.~\ref{fig:Mass_luminosity_SFGs}). We obtain $M=10^{11.9 \pm 0.5} M_\odot(\mathcal{L} / 10^{38}\,\mathrm{erg\,s^{-1}}^{-1})^{0.71}$. Such a relation is valid only at $z=0$ because the SFGs employed in the fit are local. Thus, we assume the same dependence on $z$ as in the models from Ref.~\cite{Lu:2013aoa}. The corresponding 1-halo 3D PS is shown in the left panel of Fig.~\ref{fig:3D_power spectrum_cross_correlation_astro2} as blue lines.

\subsubsection{Discussion}
\label{sec:discussion}
\begin{figure}
\centering
\hspace{-4mm}
\includegraphics[width=0.45\textwidth]{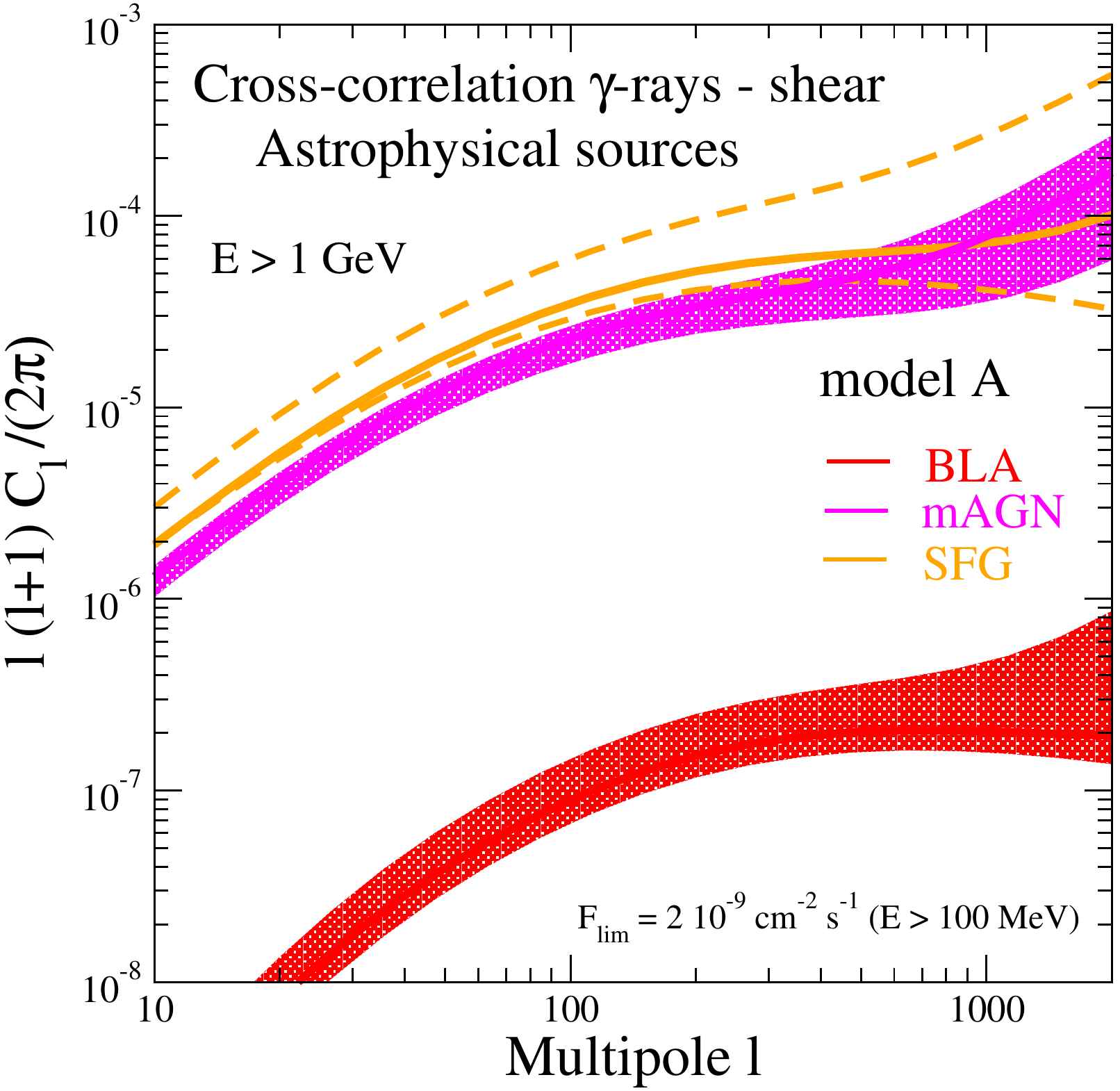}
\hspace{10mm}
\includegraphics[width=0.45\textwidth]{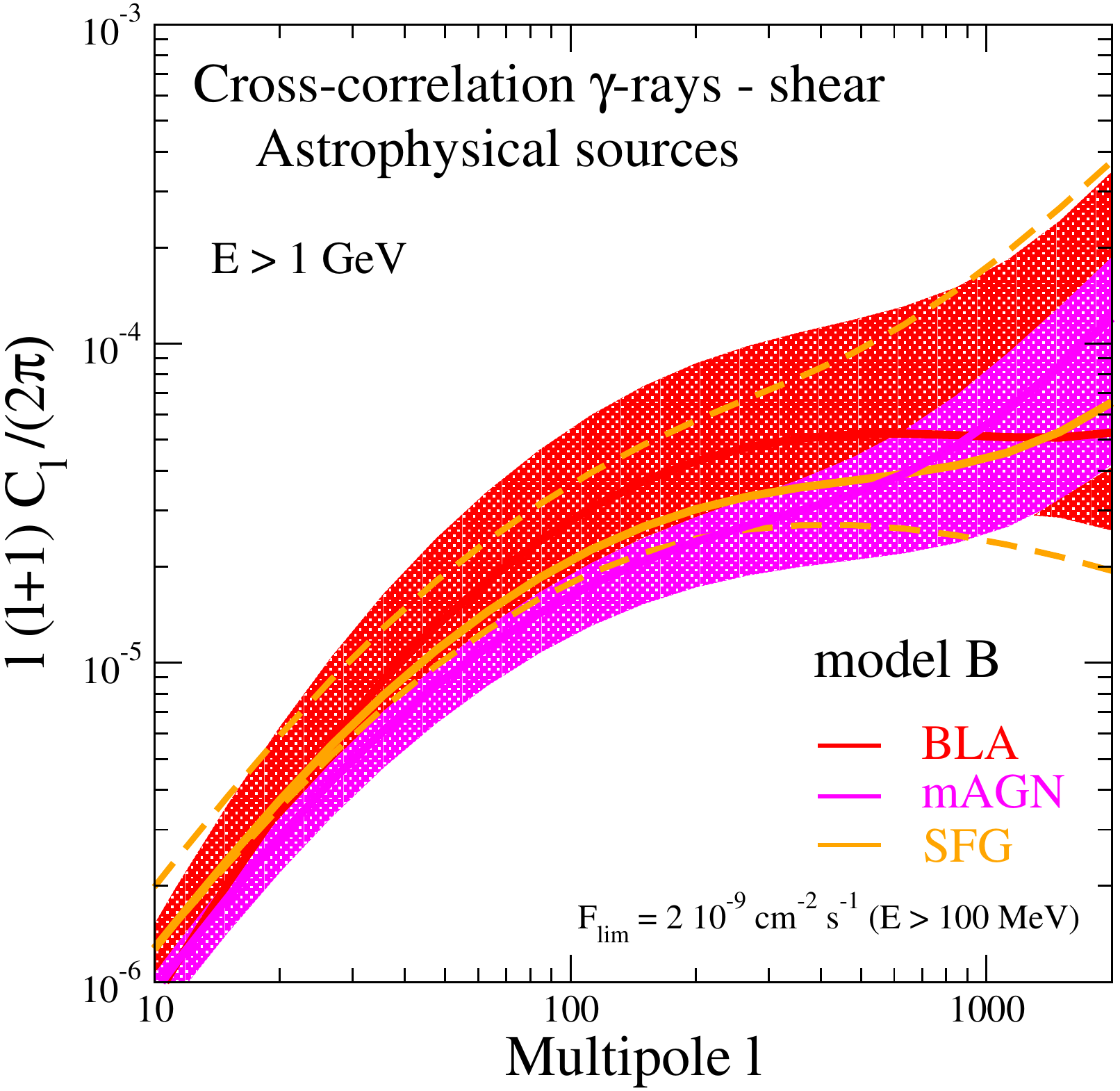}
\caption{Cross-correlation APS of Eq.~\eqref{eq:clkX} between cosmic shear and \g\ rays from blazars (red), mAGNs (magenta) and SFGs (yellow) in model A (left) and model B (right). The upper and lower limits of the bands come from different choices of $M(\mathcal{L})$ from Fig.~\ref{fig:3D_power spectrum_cross_correlation_astro1} and the left panel of Fig.~\ref{fig:3D_power spectrum_cross_correlation_astro2}a (see text for details) with solid lines being our fiducial models (the SFG band is not filled just for the sake of clearness).}
\label{fig:cl_comp}
\end{figure}

In Fig.~\ref{fig:cl_comp}, we show the impact of the assumed $M(\mathcal{L})$ relation on the predicted cross-correlation APS for the astrophysical \g-ray emitters (red for blazars, magenta for mAGNs and yellow for SFGs). The solid lines correspond to the fiducial cases described above, whilst for the upper (lower) dashed lines we considered the largest (smallest) $M(\mathcal{L})$ amongst the alternatives presented in the previous sections (see Figs.~\ref{fig:3D_power spectrum_cross_correlation_astro1} and the left panel of \ref{fig:3D_power spectrum_cross_correlation_astro2}). It is important to notice that, for {\it all} the classes of sources considered here, the lower edge of the uncertainty band corresponds to a case where the 1-halo term is subdominant in the multipole range of interest ($\ell \lesssim 1000$). The cross-correlation APS starts to be sensitive to the 1-halo term only at around $\ell=500$, i.e.\ where the sensitivity of \Fermi-LAT is already quite low (see later). On the other hand, the effect of the $M(\mathcal{L})$ relation on the 2-halo term is only through the bias. It can affect low multipoles through a change of order $\mathcal{O}(1)$ in the overall normalisation (as visible in Fig.~\ref{fig:cl_comp}). For simplicity this effect at low multipoles will be neglected in the following. We note that, overall, the variability of the cross-correlation APS due to the uncertainty on $M(\mathcal{L})$ is significantly smaller than the case of the 1-halo 3D PS. This guarantees that our prediction will not be spoiled by the huge uncertainties associated to $M(\mathcal{L})$. 

The cross-correlation APS between cosmic shear and astrophysical sources $C^{\gamma_i\kappa}_\ell$ will be computed assuming the fiducial models for $M(\mathcal{L})$ described above. However, in order to take into account the uncertainty associated to the $M(\mathcal{L})$ relations, we introduce the coefficients $\mathcal{A}_{\rm BLA}$, $\mathcal{A}_{\rm mAGN}$ and $\mathcal{A}_{\rm SFG}$. They will act as normalisation factors multiplying the 1-halo term of the APS, shifting its intensity with respect to the fiducial case (corresponding to $\mathcal{A}_i=1$). The range of variation for these normalisations is determined in order to encompass the uncertainty on $M(\mathcal{L})$ discussed above and it is derived from Figs~\ref{fig:3D_power spectrum_cross_correlation_astro1} and \ref{fig:3D_power spectrum_cross_correlation_astro2}. Specifically, for $\mathcal A_{\rm BLA}$ and $\mathcal A_\mathrm{\rm SFG}$ we assume an interval between 0.1 and 10, and for $\mathcal A_\mathrm{\rm mAGN}$ between 0.33 and 3.

By comparing the right panel of Fig.~\ref{fig:cl_DM} with Fig~\ref{fig:3D_power spectrum_cross_correlation_astro1} and the left panel of Fig.~\ref{fig:3D_power spectrum_cross_correlation_astro2}, we see that, contrarily to the case of astrophysical sources, the 1-halo term of the cross-correlation APS with DM-induced \g-ray emission is large at multipoles of few hundreds. This is related to the fact that, for DM, the cross-correlation with cosmic shear is dominated by haloes with $M\gtrsim10^{14}\,M_\odot$, which generate a large gravitational lensing signal. However, these are much larger than the haloes hosting unresolved astrophysical emitters. The same effect can be seen in the right panel of Fig.~\ref{fig:3D_power spectrum_cross_correlation_astro2}, where we plot $\de P_{1h}(M)/ \de \ln M$, namely, the integrand of the 1-halo term of the 3D PS of the cross-correlation between cosmic shear and \g-ray emission (at $z=0$ and for $k=1 \mbox{Mpc}^{-1}$). The solid, blue line corresponds to the case of annihilating DM for the \low\ scenario, whilst the green line is for decaying DM. Red, magenta and yellow curves stand for blazars, mAGNs and SFGs. Note that, as discussed above, the blue and green lines peak at larger masses and are characterised by a larger area below the curves.

The right panel of Fig.~\ref{fig:3D_power spectrum_cross_correlation_astro2} also helps us to understand the scaling in Figs~\ref{fig:3D_power spectrum_cross_correlation_astro1} and \ref{fig:3D_power spectrum_cross_correlation_astro2}: a $M(\mathcal{L})$ relation that implies a larger $M$ for a given $\mathcal{L}$ (i.e.\ a large $\mathcal{A}_i$) will shift the peak of the curves in the right panel of Fig.~\ref{fig:3D_power spectrum_cross_correlation_astro2} to larger masses, increasing the overall $P_{1h}$.

For the sake of clarity, in Table~\ref{tab:summary_sources} we list the classes of \g-ray emitters considered in this paper and we summarise their main characteristics.

\begin{table*}
\centering
\begin{tabular}{|c|p{0.8\textwidth}|}
\hline
Acronym & \multicolumn{1}{c|}{Brief description} \\
\hline
DMd & Decaying DM: halo mass function from Ref. \cite{Sheth:1999mn}; NFW density profile; $M_{\rm min}=10^{-6} M_\odot$; $c(M)$ relation from Ref.~\cite{MunozCuartas:2010ig} above $10^{10} M_\odot$ and from Ref.~\cite{Bullock:1999he} below; bias from Ref. \cite{Cooray:2002dia}. \\
\hline
DMa \ns\ & Annihilating DM, no-substructure scenario: halo mass function from Ref. \cite{Sheth:1999mn}; NFW density profile; $M_{\rm min}=10^{7} M_\odot$; $c(M)$ relation from Ref.~\cite{MunozCuartas:2010ig} above $10^{10} M_\odot$ and from Ref.~\cite{Bullock:1999he} below; no subhaloes; bias from Ref. \cite{Cooray:2002dia}. \\
\hline
DMa \low\ & Annihilating DM, \low\ substructure scenario: halo mass function from Ref. \cite{Sheth:1999mn}; NFW density profile; $M_{\rm min}=10^{-6} M_\odot$; $c(M)$ relation from Ref.~\cite{MunozCuartas:2010ig} above $10^{10} M_\odot$ and from Ref.~\cite{Bullock:1999he} below; subhaloes included as in Ref. \cite{SanchezConde:2011ap}; bias from Ref. \cite{Cooray:2002dia}. \\
\hline
DMa \high\ & Annihilating DM, \high\ substructure scenario: halo mass function from Ref. \cite{Sheth:1999mn}; NFW density profile; $M_{\rm min}=10^{-6} M_\odot$; $c(M)$ relation from Ref.~\cite{MunozCuartas:2010ig} above $10^{10} M_\odot$ and from Ref.~\cite{Bullock:1999he} below; subhaloes included as in Ref. \cite{Gao:2011rf}; bias from Ref. \cite{Cooray:2002dia}. \\
\hline
BLA & Blazars: power-law energy spectrum with $\alpha_{\rm BLA}=2.2$ ($\alpha_{\rm BLA}=2.1$); \g-ray luminosity function from Ref.~\cite{Harding:2012gk} (Ref.~\cite{Ajello:2013lka}); $\mathcal{L}_{\rm min}=10^{42} \mbox{erg\, s}^{-1}$ ($\mathcal{L}_{\rm min}=7\times10^{42} \mbox{erg\, s}^{-1}$) for model A (B). \\
\hline
mAGN & Misaligned Active Galactic Nuclei: power-law energy spectrum with $\alpha_{\rm mAGN}=2.37$; \g-ray luminosity function from Ref. \cite{DiMauro:2013xta}. \\
\hline
SFG & Star forming galaxies: power-law energy spectrum with $\alpha_{\rm SFG}=2.7$; \g-ray luminosity function from Ref. \cite{Ackermann:2012vca} rescaling the IR luminosity function of Ref.~\cite{Rodighiero:2009up} (Ref.~\cite{Gruppioni:2013jna}) for model A (B). \\
\hline
\end{tabular}
\caption{Summary of the names and characteristics of the classes \g-ray emitters considered.}
\label{tab:summary_sources}
\end{table*}

\section{The tomographic-spectral approach}
\label{sec:tomospec}
In Paper I, we proposed for the first time the study of the cross-correlation between cosmic shear and \g-ray anisotropies. We demonstrated that the signal is within reach, when combining  weak lensing surveys such as \DES\ and \Euclid\ with \g-ray data from the \Fermi-LAT satellite. In the present analysis, we capitalise on that idea and determine the sensitivity of the cross-correlation to DM detection. In the case of a positive signal, we will also estimate the precision with which it is possible to infer the properties of the DM particle, e.g. its mass and \g-ray production rate.

To this aim, we want to extract as much information as possible from the experiments. The redshift measurements performed by weak lensing surveys allow us to break down the cosmic shear signal into redshift bins, a procedure usually referred to as `redshift tomography' \citep{Hu:1999ek}. Similarly, we also consider binning in energy and call this a `tomographic-spectral' approach. We assume $N_z$ redshift bins, computing $N_z$ window functions $W^{\kappa_i}$ (with $i=1\ldots N_z$) integrating Eq.~\eqref{eq:W_lens} over the $i$-th redshift bin. In the same way, we consider $N_E$ energy intervals over which we integrate the \g-ray window functions. We define the cross-correlation tomographic-spectral matrix $\mathbf{C}_\ell^{\gamma\kappa}$ as
\begin{equation}
\mathbf C_\ell^{\gamma\kappa} = 
\left( \begin{array}{ccc}
C_\ell^{\gamma_1\kappa_1} & \ldots & C_\ell^{\gamma_1\kappa_{N_z}} \\
\vdots & \ddots & \vdots\\
C_\ell^{\gamma_{N_E}\kappa_1} & \ldots & C_\ell^{\gamma_{N_E}\kappa_{N_z}}
\end{array} \right),
\end{equation}
where each element $C_{\ell}^{\gamma_a\kappa_i}$ is the cross-correlation APS obtained from Eq.~\eqref{eq:clkX} with window function $W^{\kappa_i}$ and $W^{\gamma_a}$, i.e.\ within the $i$-th redshift bin and the $a$-th energy bin. Similarly, the auto-correlation matrices $\mathbf C_\ell^{\gamma\gamma}$ and $\mathbf C_\ell^{\kappa\kappa}$ can be constructed for the \g-ray emission and the cosmic shear signal, respectively. Note that $\mathbf C_\ell^{\gamma\gamma}$ and $\mathbf C_\ell^{\kappa\kappa}$ are square matrices, whereas $\mathbf C_\ell^{\gamma\kappa}$ is a $N_E \times N_z$ object.

\subsection{Surveys specifications}
\label{sec:surveys}
In order to estimate the auto- and cross-correlation APS and to assess properly the sensitivity to a DM signal, we need to define the characteristics of the detectors. As in Paper I, we focus on \DES\ \citep{Abbott:2005bi} and \Euclid\ \cite{Laureijs:2011gra,Amendola:2012ys}, considered as representatives of current and future cosmic shear surveys. The former is a ground-based experiment provided with an extremely sensitive 570-Megapixel digital camera mounted on the Blanco 4 m telescope at Cerro Tololo Inter-American Observatory high in the Chilean Andes. It started taking data in September 2012 and it will continue for 5 years, surveying $5,000$ square degrees over the Southern sky. On the other hand, \Euclid\ is a European Space Agency space-based, medium-class astronomy and astrophysics mission, whose launch is planned for 2020. \Euclid\ will observe $15,000$ deg$^2$ of the darkest sky that is free of contamination by light from our Galaxy and the Solar System. Its weak lensing survey will measure several billion photometric redshifts of galaxies as far as $z \gtrsim 2$. 

The expected source redshift distributions for \DES\ and \Euclid\ can be written as \cite{Smail:1994sx}:
\begin{equation}
\frac{\de N_g}{\de z}(z)\propto z^2
\exp\left[{-\left(z/z_0\right)^{1.5}}\right],
\label{eq:n_z-Euclid}
\end{equation}
where $z_0=z_m/\sqrt{2}$ and $z_m$ is the median redshift of the survey, respectively 0.8 for \DES\ and 0.9 for \Euclid. We report in Table~\ref{tab:WL-specs} the details of our modelling of the surveys and of the redshift binning. We consider 5 (10) equally-populated bins for \DES\ (\Euclid) and we also include a photometric scatter $\sigma_z\propto(1+z)$.\footnote{Note that, according to recent estimates for \Euclid, 0.05 is a more realistic value of $\sigma_z/(1+z)$. However, we checked that this does not influence significantly our results.}

\begin{table*}
\centering
\begin{tabular}{|c|c|c|c|}
\hline
Parameter & Description & \DES\ & \Euclid\ \\
\hline
$f_{\rm sky}$ & Surveyed sky fraction & 0.12 & 0.36 \\
$\bar{N}_g$ [arcmin$^{-2}$] & Galaxy density & 13.3 & 30 \\
$\de N_g/\de z(z)$ & Redshift distribution & Ref.~\citep{Smail:1994sx} & Ref.~\citep{Smail:1994sx} \\
$N_z$ & Number of bins & 5 & 10 \\
$\sigma_z/(1+z)$ & Redshift uncertainty & 0.05 & 0.03 \\
$\sigma_\epsilon$ & Intrinsic ellipticity & 0.3 & 0.3 \\
\hline
\end{tabular}
\caption{Summary of the specifications for \DES\ and \Euclid\ used in this analysis.}
\label{tab:WL-specs}
\end{table*}

\Fermi-LAT is the principal scientific instrument on the Fermi Gamma Ray Space Telescope spacecraft, launched into a near-Earth orbit in June 2008. The design life of the mission is 5 years and the goal for mission operations is 10 years. \Fermi-LAT is an imaging high-energy \g-ray telescope covering the energy range from about 20 MeV to more than 300 GeV. Its field of view covers about $20\%$ of the sky at any time, and in survey mode it scans the whole sky every three hours. An exposure of 10 years (after analysis cuts) represents our benchmark configuration, that we shall refer to as \Fermitenyr. As a second benchmark, we consider an hypothetical upgrade (dubbed \Fermissimo), representative of possible future space-based large-area full-sky \g-ray detectors, with improved capabilities with respect to \Fermi-LAT. Proposals currently under study include Gamma400 \cite{Gamma400}, HERD \cite{HERD} and DAMPE \cite{DAMPE}. We do not specifically assume any of these designs, but we take inspiration from them in defining the characteristics of \Fermissimo.

Table~\ref{tab:gammaspec} summarises the specifics of two experimental configurations adopted in our analysis. We consider 6 energy bins for \Fermitenyr\ (specifically, 1--2 GeV, 2--5 GeV, 5--10 GeV, 10--50 GeV, 50--100 GeV and 100--300 GeV) and we add one additional bin at lower energies (between 0.3 and 1 GeV) and one at higher energies (between 300 GeV and 1 TeV) for \Fermissimo. The energy resolution of the detector is significantly smaller than the bins widths, namely $\Delta E/E \lesssim 20\%$. However, considering bins that are too small implies a reduced statistics, especially at high energies, with no consequent gain in information. In Table~\ref{tab:gammaspec}, we also present the properties of the assumed \g-ray campaigns. Note that we quote the average angular beam size $\langle\sigma_b\rangle$ but, in our analysis, we adopt the values taken from Ref.~\cite{Atwood:2009ez}, allowing $\sigma_b$ to change with the energy; for example, the angular resolution of \Fermi-LAT reaches 0.1 deg at high energies.

In our analysis, we will investigate the prospects of the cross-correlation technique for two different situations: in the first one we combine the data from \Fermitenyr\ and DES and, therefore, this scenario can be achieved in few years from now. The second possibility consider the data of \Fermissimo\ and of \Euclid, and it requires a longer timescale.

\begin{table*}
\centering
\begin{tabular}{|c|c|c|c|}
\hline
Parameter & Description & \Fermitenyr & \Fermissimo \\
\hline
$f_{\rm sky}$ & Surveyed sky fraction & 1 & 1 \\
$E_{\min}-E_{\max}$ [GeV]& Energy range  & $1-300$ & $0.3-1000$ \\
$N_E$ & Number of bins & 6 & 8 \\
$\varepsilon$ [cm$^2$ s] & Exposure & $3.2\times10^{12}$ & $4.2\times10^{12}$ \\
$\langle\sigma_b\rangle$ [deg] & Average beam size & 0.18 & 0.027 \\
\hline
\end{tabular}
\caption{Summary of the different configurations assumed for the \g-ray experiments. We mention here that, in reality, the analysis will not be full sky (e.g., the galactic plane is typically masked out), but the considered fraction of surveyed sky considered still remains significantly larger than for galaxy surveys in Table~\ref{tab:WL-specs}.}
\label{tab:gammaspec}
\end{table*}

\begin{figure}
\centering
\includegraphics[width=0.6\textwidth]{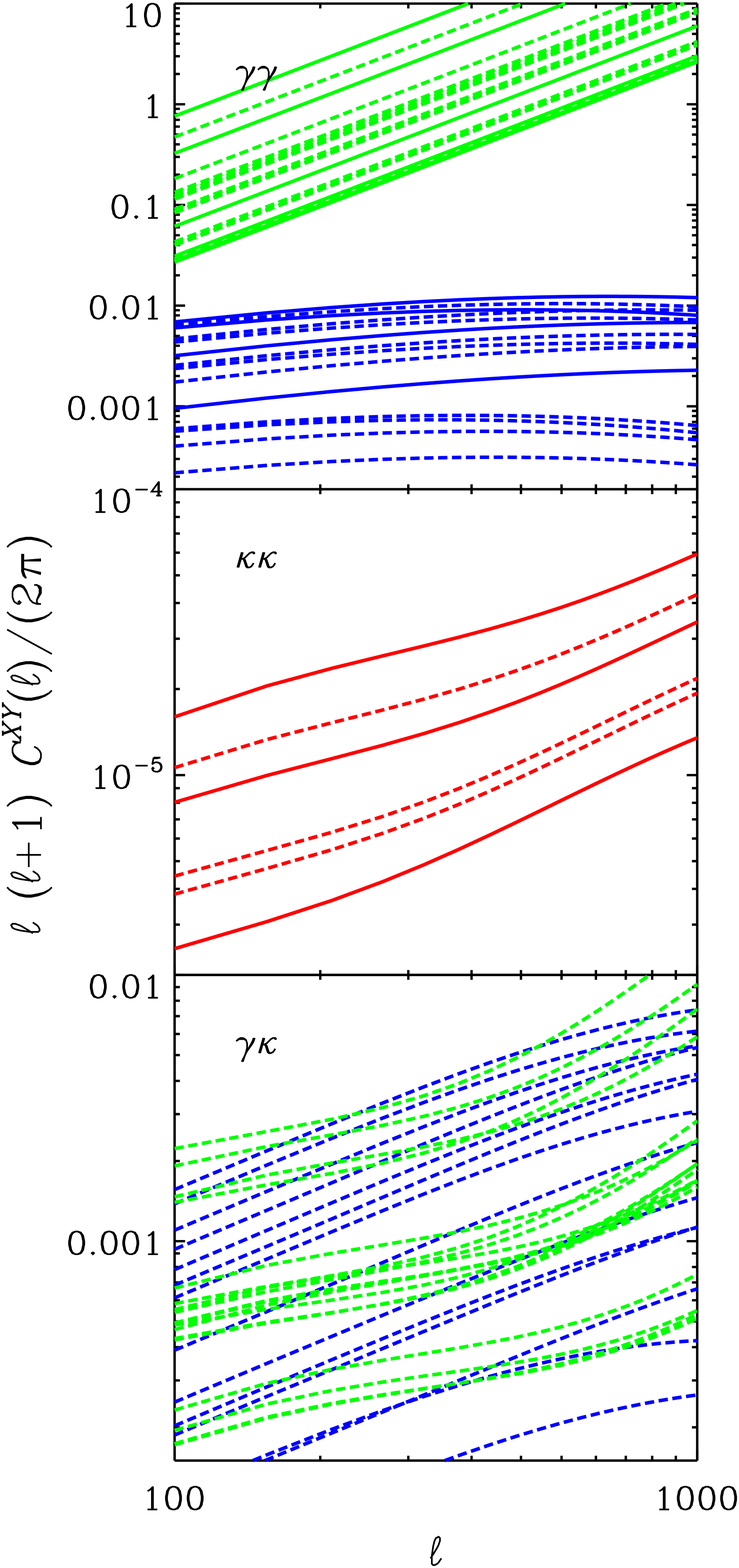}
\caption{Illustrative example of tomographic-spectral matrix elements $\mathbf C_\ell^{\gamma\gamma}$ (top panel), $\mathbf C_\ell^{\kappa\kappa}$ (middle panel) and $\mathbf C_\ell^{\gamma\kappa}$ (bottom panel), for annihilating DM (blue), astrophysical background (green) and cosmic shear (red). Each curve corresponds to one element in the matrices. Solid lines refer to diagonal entries and dashed lines to off-diagonal ones. The plots refer to the \high\ substructure scheme, to model A for astrophysical sources and to the combination \DES\ + \Fermitenyr.}
\label{fig:spectra}
\end{figure}

For illustrative purposes, in Fig.~\ref{fig:spectra} we show all the entries of the tomographic-spectral matrices $\mathbf{C}_\ell^{\gamma\kappa}$ for the \DES+\Fermitenyr\ configuration. The top panel shows the \g-ray auto-correlation APS $C_\ell^{\gamma_a\gamma_b}$, the middle panel the auto-correlation APS of cosmic shear $C_\ell^{\kappa_i\kappa_j}$ and the cross-correlation APS $C_\ell^{\gamma_a\kappa_j}$ is plotted in the bottom panel. For the auto-correlation APS (top and middle panels) solid curves refer to the diagonal elements (the auto-correlation APS are obtained considering the same energy or redshift bin), whilst dashed curves are for off-diagonal terms. Blue lines refer to the case the \g-ray emission is induced by annihilating DM (in the \high\ scenario), while green lines stand for the \g-ray emission produced by the sum of all the astrophysical sources. Note that, although astrophysical sources largely dominate the auto-correlation signal (top panel), the cross-correlation spectra for astrophysical sources and DM are characterised by similar intensities (bottom panel). This confirms that the cross-correlation with cosmic shear is a powerful way to reduce the impact of the astrophysical background (as proved also in Fig. 5 of Paper I) and it suggests that the cross-correlation is more promising than auto-correlation to detect a possible DM signal.

\section{Analysis technique}
\label{sec:analysis}
We are now ready to discuss the potentiality of the cross-correlation technique
to uncover a DM signal. We shall proceed along two lines:
\begin{enumerate}
\item For each DM mass
$\mdm$ we determine the minimal $\sv$ (or $\gd$) for which the 
cross-correlation signal is able to provide a {\it detection} of the DM particle with a 
confidence level (CL) equal or larger than $5\sigma$;
\item We investigate the 
precision that can be achieved by the various experimental setups in the {\it reconstruction} of the particle DM parameters (i.e.\ its mass and annihilation cross section or decay rate) for a set of representative benchmark cases.
\end{enumerate}
As a byproduct of the first point, we also determine the expected $2\sigma$ CL upper bounds in the DM parameter space which can be obtained with the cross-correlation technique, would the DM signal be significantly suppressed and therefore not detectable.

The technique we employ to forecast the prospects for detection is the 
Bayesian Fisher matrix method, which is discussed in the 
following section.

\subsection{Fisher matrix fundamentals}
\label{sec:fisher}
The Fisher matrix approach for parameter estimation assumes the presence of
a likelihood function $L(\boldsymbol\vartheta)$ that quantifies the
agreement between a certain set of experimental data and the set of parameters of
the model, $\boldsymbol\vartheta=\{\vartheta_\alpha\}$. It is also assumed that
that the behaviour of the likelihood near its maximum characterises the whole 
likelihood function sufficiently well to be used to estimate errors on the 
model parameters \citep{Jeffreys:1961,Vogeley:1996xu,Tegmark:1996bz}. 

Under the hypothesis of a Gaussian likelihood, the Fisher matrix 
is defined as the inverse of the parameter covariance matrix. Thence, it is
possible to infer the statistical accuracy with which the data encoded in the 
likelihood can measure the model parameters. If the data is
taken to be the expected measurements performed by future experiments, the Fisher 
matrix method can be used, as we do here, to determine its prospects for 
detection and the corresponding level of accuracy. The $1\sigma$ marginal error on 
parameter $\vartheta_\alpha$ reads
\begin{equation}
\sigma(\vartheta_\alpha) = \sqrt{ \left( \mathbf F^{-1} \right)_{\alpha\alpha}},
\label{eq:marginal}
\end{equation}
where $\mathbf F^{-1}$ is the inverse of the Fisher matrix, and no summation 
over equal indices is applied here.

Our experimental data will come from the measurement of the cross-correlation 
APS $C^{XY}_\ell$ between the observables $X$ and $Y$, as discussed in the 
previous section. More specifically, we shall use the cross-correlation
tomographic-spectral matrix $\mathbf C^{\gamma\kappa}_\ell$ defined in 
Sec.~\ref{sec:tomospec}. The parameters in our model 
are $\boldsymbol\vartheta=\{ \mdm, \, \Gamma_d, \, \mathcal{A}_{\rm BLA}, \, \mathcal{A}_{\rm SFG}, \,\mathcal{A}_{\rm mAGN} \}$ for decaying DM and $\{ \mdm, \, \sv, \, 
\mathcal{A}_{\rm BLA}, \, \mathcal{A}_{\rm SFG}, \, \mathcal{A}_{\rm mAGN} \}$ for  annihilating DM. Unless stated otherwise, we fix the annihilation/decay energy spectrum to be that produced by the hadronisation of $b$ quarks. We also fix the description of astrophysical sources to the model discussed in 
Sec.~\ref{sec:intastro}, leaving only the quantities $\mathcal{A}_{\rm BLA}$, 
$\mathcal{A}_{\rm SFG}$ and $\mathcal{A}_{\rm mAGN}$ as free parameters. We remind the reader that they quantify the deviation from our fiducial case in the normalisation of the 1-halo term of the 3D cross-correlation PS of blazars, SFGs and mAGNs, respectively. 

Following Ref.~\citep{Hu:2003pt}, the generic element of the covariance matrix 
$\mathbf\Gamma^{\gamma\kappa}_{\ell\ell'}$ can be expressed as
\begin{equation}
\left[ \Gamma^{\gamma\kappa}_{\ell\ell'} \right]^{ai,bj} = 
\frac{ \widehat{C}^{\gamma_a\kappa_j}_\ell \widehat{C}^{\gamma_b\kappa_i}_\ell + \widehat{C}^{\gamma_a\gamma_b}_\ell \widehat{C}^{\kappa_i\kappa_j}_\ell}
{(2\ell+1) \Delta\ell f_{\rm sky}} \delta_K^{\ell\ell'},
\label{eq:covariance}
\end{equation}
where $\Delta\ell$ is the bin width (for angular multipole binned data), 
$f_\mathrm{sky}$ is the fraction of sky probed by the survey, and $\delta_K$ is the Kronecker symbol. Here, 
$\widehat{\mathbf C}^{XY}_\ell=\mathbf C^{XY}_\ell+\boldsymbol{\mathcal N}^{XY}_\ell$,
being $\boldsymbol{\mathcal N}^{XY}_\ell$ the experimental noise on the 
measurement of $\mathbf C^{XY}_\ell$. We assume that noises for \g\ rays and 
cosmic shear do not correlate, so that 
$\boldsymbol{\mathcal N}^{\gamma\kappa}_\ell=0$. This also implies 
$\widehat{\mathbf C}^{\gamma\kappa}_\ell=\mathbf C^{\gamma\kappa}_\ell$. Conversely, for the auto-correlation terms we have
\begin{align}
\mathcal{N}^{\gamma_a\gamma_b}_\ell &= \delta_K^{ab} 
\frac{4 \pi f_\mathrm{sky}}{\bar N_{\gamma_a}} \mathcal W_\ell^{-2},\label{eq:noise-gamma} \\
\mathcal{N}^{\kappa_i\kappa_j}_\ell &= \delta_K^{ij}
\frac{\sigma_\epsilon^2}{\bar N_{g_i}},\label{eq:noise-shear}
\end{align}
where the first term in Eq.~\eqref{eq:niose-gamma} describes the so-called photon noise and 
$\bar N_{\gamma_a}$ is the total number of \g\ rays expected in the $a$-th 
energy bin. The factor $\mathcal W_\ell=\exp(-\sigma_b^2\ell^2/2)$ is the 
window of a Gaussian point-spread function and $\bar N_{g_i}$ is the 
number of galaxies per steradian in the $i$-th redshift bin.

Now, it is worth to make a remark. We emphasise that our method consists in a cross-correlation between two distinct signals. For a start---as we stated above---this implies that $\boldsymbol{\mathcal N}^{\gamma\kappa}_\ell=0$, because cosmic-shear and \g-ray noises do not correlate. However, this has another major advantage, which is the fact that any {\it additive} systematic effect shall not correlate either. Thus, this method is more robust than a simple auto-correlation analysis.\footnote{Nonetheless, note that other, non-additive systematics may have an impact, such as shape bias in the case of cosmic shear.}

From an observational point of view, we can consider each single mode 
$\widehat{C}^{\gamma_a\kappa_j}_\ell$ in tomographic, spectral and multipole space as a parameter of the theory. Then, to recast the Fisher matrix in the space of the model parameters, $\boldsymbol\vartheta$, it is sufficient to multiply the inverse of the covariance matrix by the Jacobian of the change of variables, viz.\
\begin{equation}
\mathbf{F}_{\alpha\beta} = \sum_{\ell,\ell'=\ell_\mathrm{min}}^{\ell_\mathrm{max}}
\frac{\partial \mathbf{C}^{\gamma\kappa}_\ell}{\partial \vartheta_\alpha}
\left[ \mathbf{\Gamma}^{\gamma\kappa}_{\ell\ell'} \right]^{-1}
\frac{\partial \mathbf{C}^{\gamma\kappa}_{\ell'}}{\partial \vartheta_\beta},
\label{eq:fisher}
\end{equation}
where we sum over all the multipoles because $\Gamma^{\gamma\kappa}_{\ell\ell'}$ is diagonal in $\ell$ and $\ell'$.
%

\section{Results}
\label{sec:results}
As previously discussed, we characterise the potential of the cross-correlation technique in terms 
of two analyses. First, in order to determine the minimal $\sv$ (or $\gd$)
that correspond to a $5\sigma$ DM detection, we construct the Fisher matrix 
given in Eq.~\eqref{eq:fisher} by varying over the full parameter set, 
$\boldsymbol\vartheta$. Astrophysical parameters $\mathcal{A}_i$ will be 
marginalised over by integrating them out in the global likelihood, whilst the DM parameters $\mdm$ and $\sv$ or $\gd$ will be retained. For each value of the DM mass, we determine the value for $\sv$ or $\gd$ that is 5 times larger than its estimated error, as computed from Eq.~\eqref{eq:marginal}. This corresponds to the minimal $\sv$ or $\gd$ above which it will be possible to discriminate (with a $5\sigma$ CL, or larger) between an interpretation of the 
cross-correlation data in terms of astrophysics-only and an interpretation 
that requires a DM component. Notice that, when obtaining the 
$5\sigma$ CL detection, we do not assume a specific fiducial model for the DM 
particle parameters but we rather leave them free to vary. We 
call this first analysis the determination of the {\it detection reach} of the 
cross-correlation technique, and its results will be presented in 
Sec.~\ref{sec:forecasts}.

As a second step, we forecast the precision that can be achieved in the 
reconstruction of the particle DM parameters. To do this, we assume a few 
benchmark DM models by fixing $\mdm$, $\sv$ (or $\gd$). We derive the $1\sigma$ 
and $2\sigma$ marginal errors with which those quantities can be determined from the cross-correlation measurement, according to the marginal error estimate of 
Eq.~\eqref{eq:marginal}. As before, the astrophysical parameters 
$\mathcal{A}_i$ are marginalised over. We refer to this analysis as 
{\it parameter reconstruction}, and its results will be presented in 
Sec.~\ref{sec:parameter_forecasts}.

\subsection{Detection reach}
\label{sec:forecasts}
\subsubsection{Impact of clustering scenario and astrophysical modelling}
\begin{figure}
\centering
\includegraphics[width=0.6\columnwidth]{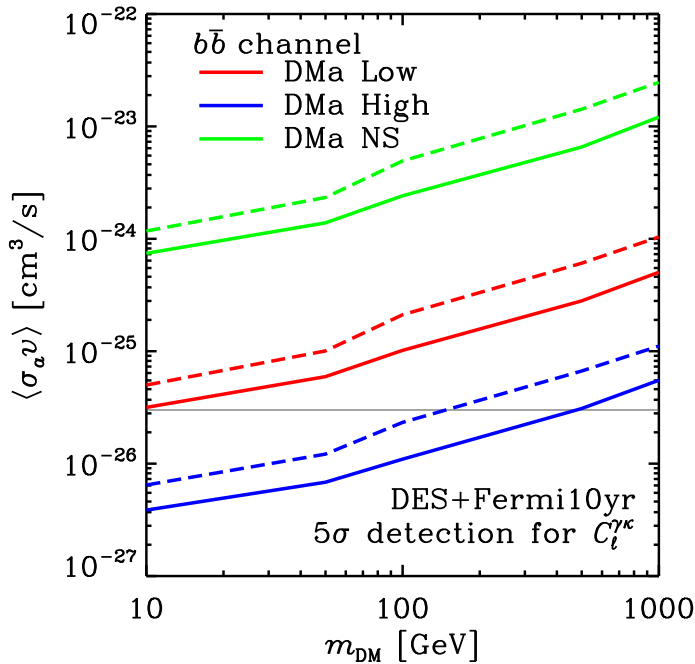}
\caption{Coloured lines denote the level above which the cross-correlation APS between cosmic shear and \g-ray emission provides a $5\sigma$ {\it detection} of an annihilating DM signal. A $b\bar b$ annihilation channel is assumed. Line colours refer to different clustering models: green for \ns, red for \low\ and blue for \high. The \g-ray emission from astrophysical sources is included in the computation of the cross-correlation signal and line styles differentiate between model A (solid) and model B (dashed). Results refer to the combination of \DES\ with \Fermitenyr. The horizonthal grey line denotes the `thermal' cross section, $3 \times 10^{-26}\,\mathrm{cm^3s^{-1}}$.}
\label{fig:Detection_Cl_1a_ALL_DESFermi5yr}
\end{figure}

We start by discussing the case of annihilating DM. In
Fig. \ref{fig:Detection_Cl_1a_ALL_DESFermi5yr} the region above the lines is the portion of the
$(\mdm,\sv)$ parameter space where the cross-correlation provides a $5\sigma$ (or larger)
detection of the DM particle. Here, we refer to the \DES+\Fermitenyr\ 
combination. All our results 
are obtained by marginalising over the $\mathcal{A}_i$ parameters which define our uncertainty on the size of the 1-halo of the APS of the \g-ray astrophysical sources. For these parameters we assume priors in the ranges outlined in Sec. \ref{sec:discussion}:
$\mathcal A_{\rm BLA}$ and $\mathcal A_\mathrm{SFG}$ between 0.1 and 10, and $\mathcal A_\mathrm{mAGN}$ between 0.33 and 3.
The three set of lines refer to the three different DM clustering scenarios: \high\ (lower, blue),
\low\ (middle, red) and \ns\ (upper, green), and are shown for the two different astrophysical scenarios called model A (solid) and model B (dashed) in Sec. \ref{sec:intastro}. 

As a first result, we notice that the difference between model A and model B for the astrophysical \g-ray emitters reflects into a factor of $\sim2$ vertical shift in the detection reach.
This observation provides an estimate of the uncertainty currently inherent in the modelling of astrophysical \g-ray sources. Assuming the \high\ model (blue line), our results show that a $5\sigma$ detection is possible 
for a DM particle with a thermal annihilation cross section of 
$3 \times 10^{-26}\, \,\mathrm{cm^3s^{-1}}$, if the $\mdm$ is smaller than 500 GeV (150 GeV in case of model B). In the more conservative case of the \low\ subhalo model, the detection reach scales up
by an order of magnitude (red line), whilst for the \ns\ case 
the green line exceeds $10^{-24}\ \mbox{cm}^{3}\mbox{s}^{-1}$ over most of the DM mass range. Note that the \ns\ scenario is very conservative and likely unphysical in assuming that no DM 
structures are present below $M_{\rm min} = 10^7 M_\odot$. However, it represents a {\it guaranteed} DM component that would be unrealistic, within 
the WIMP framework, to neglect.

\subsubsection{Reconstructed bounds in case of null-detection}
\begin{figure}
\centering
\includegraphics[width=0.49\columnwidth]{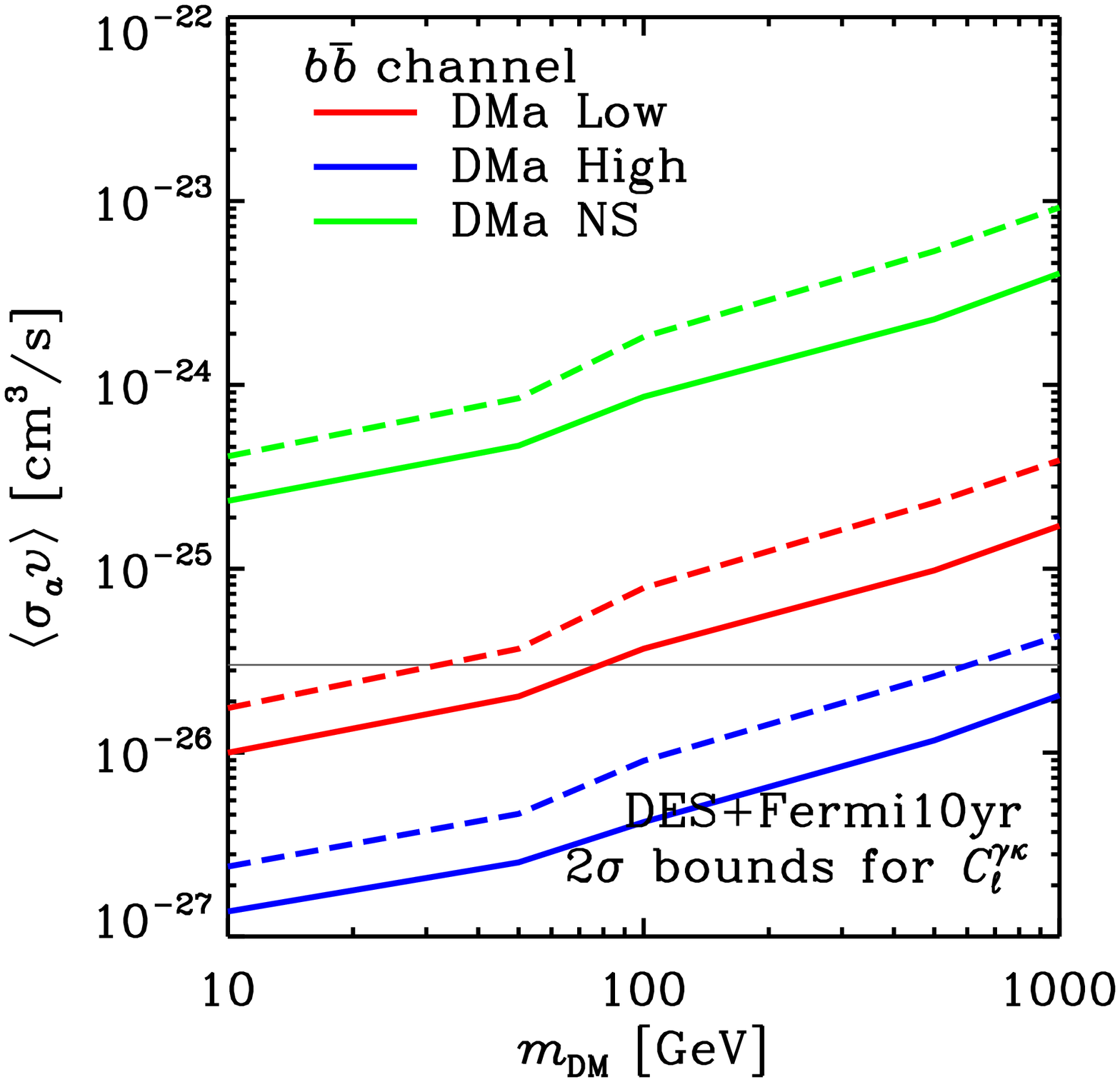}
\includegraphics[width=0.49\columnwidth]{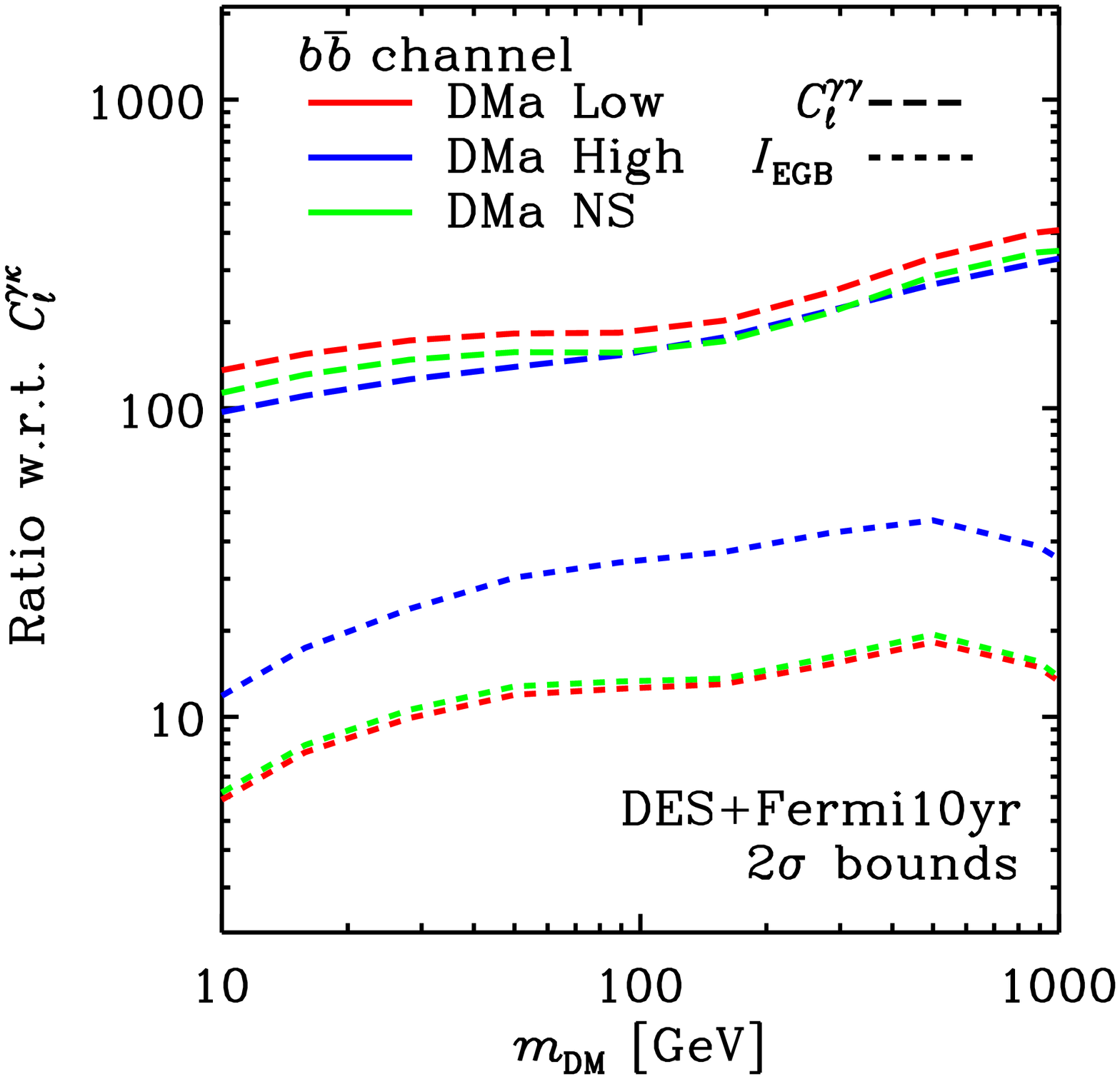}
\caption{Left: Solid lines show the 2$\sigma$ {\it upper limits} that can be derived 
from a non-detection of an annihilating DM signal from the analysis of the cross-correlation APS between cosmic shear and \g-ray emission, for the combination of \DES\ with \Fermitenyr. (Same notations as in Fig. \ref{fig:Detection_Cl_1a_ALL_DESFermi5yr}.) Right: Comparison of the 2$\sigma$ upper limit that can be derived from the cross-correlation study, with bounds obtainable from \g-ray autocorrelation and EGB total intensity. The plot shows the ratio of the bounds that can be derived from the measured \g-ray auto-correlation APS (dashed) and the measured EGB intensity (dotted) with the bounds obtained with the cross-correlation. Red, blue and green lines refer to the \low, \high\ and \ns\ substructure DM modelling, respectively. Model A of astrophysical emission is employed, but similar results are obtained for model B.}
\label{fig:Bounds_ALL_DESFermi5yr}
\end{figure}

Fig.~\ref{fig:Detection_Cl_1a_ALL_DESFermi5yr} illustrates the minimal cross
section for which a cross-correlation signal corresponds to a $5\sigma$ 
detection of the DM particle. If a signal is not detected, it is otherwise 
customary to derive upper bounds on the annihilation rate as a function of the 
DM mass. Fig.~\ref{fig:Bounds_ALL_DESFermi5yr} shows the expected  \DES+\Fermitenyr\ $2\sigma$ 
upper limits that can be obtained from the cross-correlation between 
\g-ray emission and cosmic shear. Within the Fisher matrix formalism, 
forecast can be computed only assuming a fiducial model. The upper limits on 
the annihilation rate are thus obtained from the Fisher matrix of 
Eq.~\eqref{eq:fisher} with a $\sv=0$ DM fiducial model. The left panel of Fig.~\ref{fig:Bounds_ALL_DESFermi5yr} shows the $2\sigma$ upper bounds by adopting the  same notations of Fig. \ref{fig:Detection_Cl_1a_ALL_DESFermi5yr}.

The right panel instead compares the ratio of the $2\sigma$ upper bounds for  cross-correlation with the bounds that can be derived from the measured \g-ray auto-correlation APS and the
measured EGB intensity. These limits are obtained again from a Bayesian analysis, but using public data, thus without the need of a fiducial model. We employed the EGB estimated by the Fermi-LAT Collaboration~\cite{Ackermann:2014usa} (adding up in quadrature statistical and systematic errors given in their Table 3) and the auto-correlation APS estimated in four energy bins in Ref. \cite{Ackermann:2012uf} (as provided in their Table II, averaged in the multipole range $155\leq\ell\leq 504$). For both probes, the model prediction has been computed using the same DM and astrophysical modeling as in the cross-correlation analysis. The plot depicts the ratio 
of the bounds obtained from the \g-ray auto-correlation APS (dashed lines) or EGB (dotted lines) to those provided by the cross-correlation APS discussed in this paper. Red, blue and green lines refer to the \low, \high\ and \ns\ substructure DM modelling, respectively. Model A of astrophysical emission is employed here, but similar results are obtained for model B.

This plot clearly shows that the cross-correlation technique always provides more stringent bounds: by a factor of 5-50 (depending on the particle DM mass and DM clustering model) more constraining than those from the intensity alone; by a factor 100-300 tighter than those derived from the \g-ray auto-correlation APS. As explained in Paper I and in Secs~\ref{sec:intensity} and \ref{sec:tomospec}, the information provided by the cross-correlation with the cosmic shear manages to isolate the DM signal in the EGB, even when the EGB intensity and its auto-correlation APS 
are dominated by astrophysical sources. In turn, this arises from the fact 
that the window functions of DM-induced emission and of astrophysical 
sources have quite different behaviours. In particular, DM preferably emits 
at low redshifts, whilst the astrophysical components peak at a larger 
redshift (see Fig.~\ref{fig:window}). This information is 
lost when measuring the EGB intensity alone (either as total emission, or 
through the auto-correlation of its anisotropies), but it can be recovered by 
cross-correlating it with cosmic shear, which provides valuable redshift information. In addition, the importance of the 1-halo term and the predicted shape of the APS are different between 
the case of DM-induced and astrophysical emission. This provides an extra handle that is lost when averaging over the sky.

\subsubsection{Impact of experimental setup}
\begin{figure}
\centering
\includegraphics[width=0.49\textwidth]{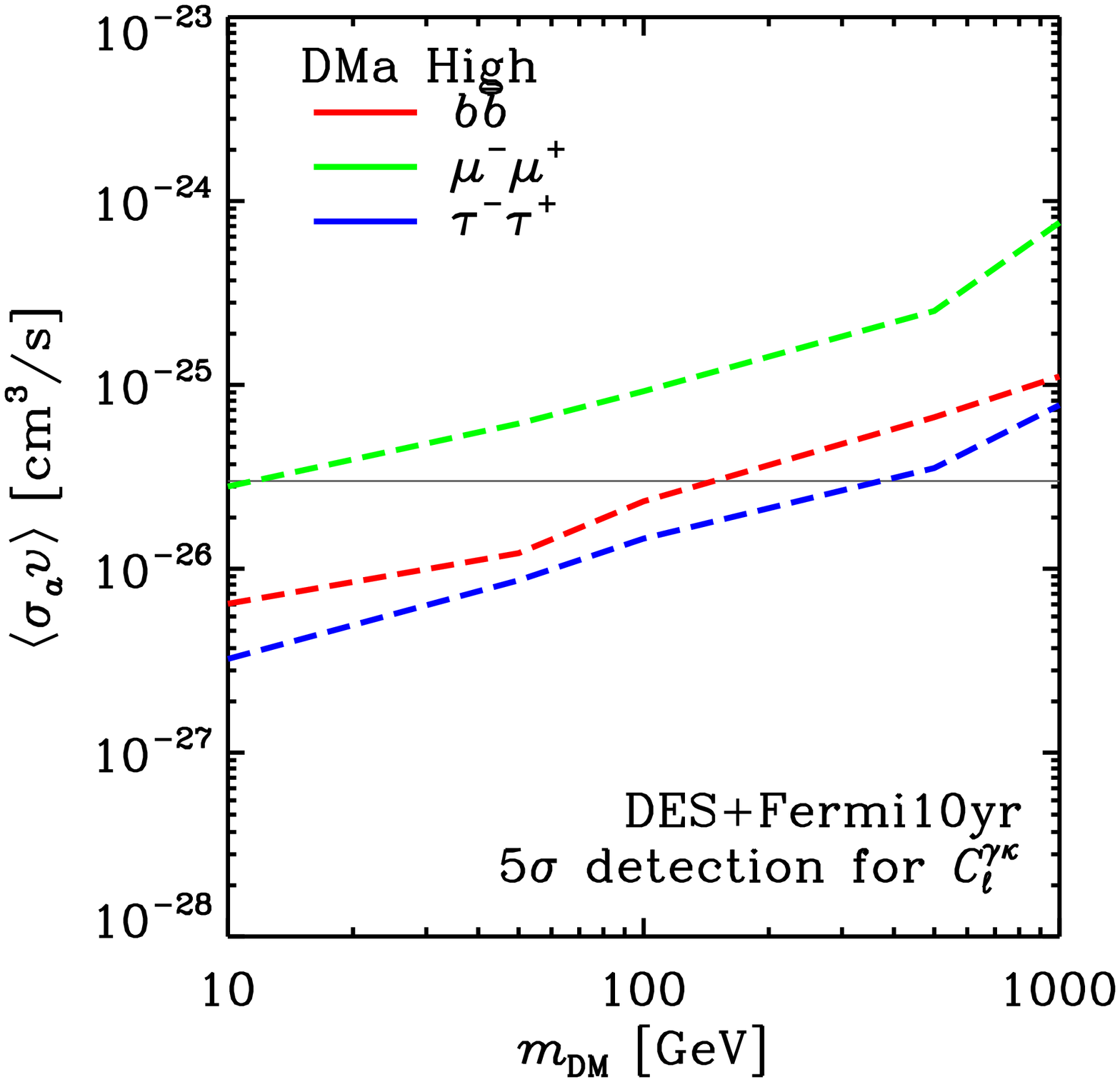}
\includegraphics[width=0.49\textwidth]{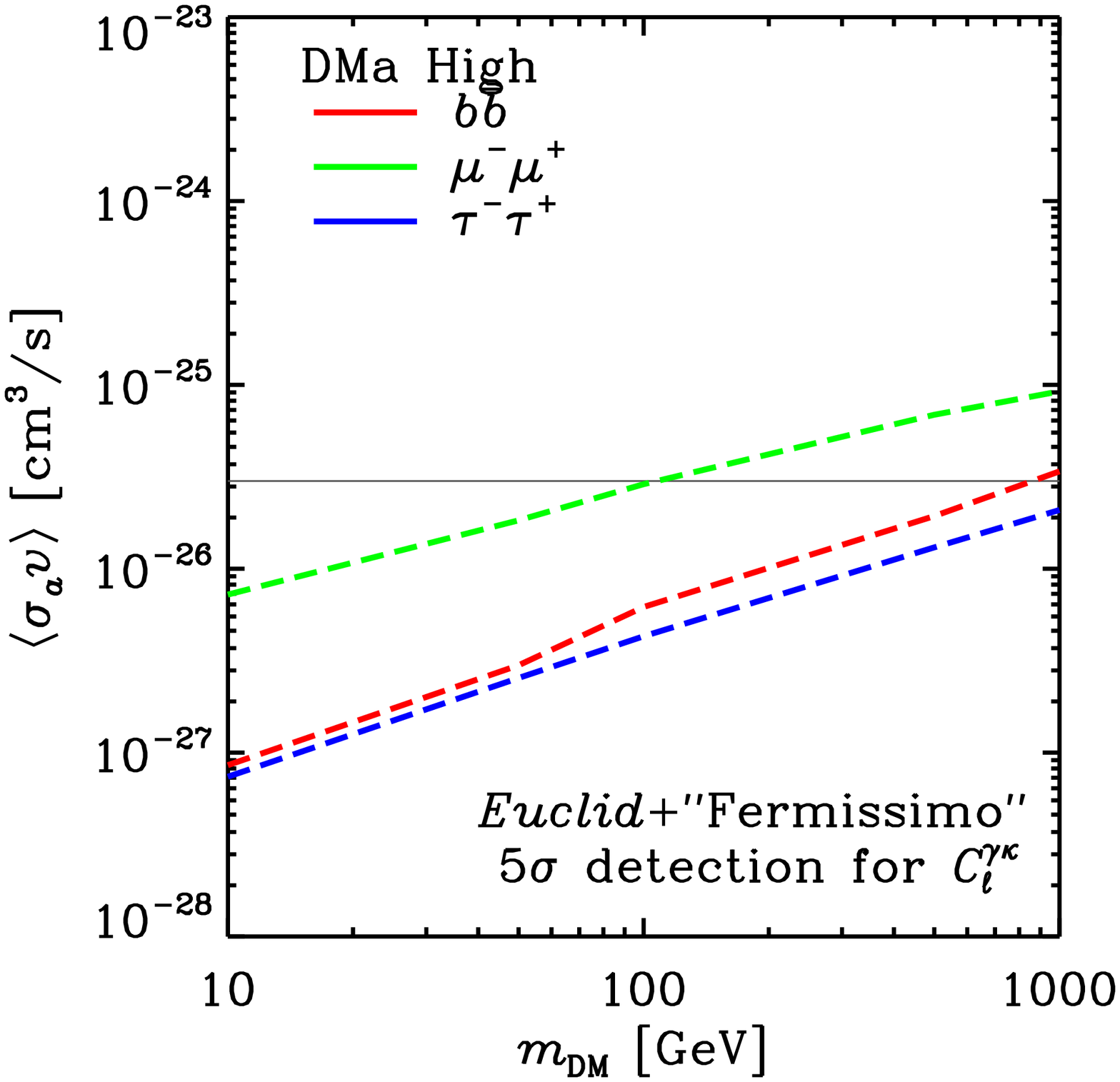}
\caption{Coloured lines denote the level above which the cross-correlation APS between cosmic shear and \g-ray emission provides a $5\sigma$ {\it detection} of an annihilating DM signal. In each panel, from top to bottom, lines stand for the $\mu^+\mu^-$ (green), $b\bar{b}$ (red) and $\tau^+\tau^-$ (blue) annihilation channels. DM clustering model is \high\ and astrophysical sources are modelled with model B. The thin grey line indicates the `thermal' cross section, $3 \times 10^{-26}\,\mathrm{cm^3s^{-1}}$. The two panels refer to the case of \DES+\Fermitenyr\ (left) and \Euclid+\Fermissimo\ (right).}
\label{fig:Detection_Cl_1bbis_HIGH_ALL}
\end{figure}
The $5\sigma$ detection reach for additional annihilation channels and detector configurations is shown in Fig. \ref{fig:Detection_Cl_1bbis_HIGH_ALL}: the left panel refers to \DES+\Fermitenyr, 
whilst the right panel shows what can be achieved by future 
detectors by reporting the results for the combination of \Euclid\ and \Fermissimo. The DM clustering model is set to \high\ and the astrophysical \g-ray sources are modelled according to model B. In both panels, from top to bottom the lines refer to the $\mu^+\mu^-$, $b\bar b$ and
$\tau^+\tau^-$ annihilation channels. We notice that detection reach is similar for hadronic  channels and the tau scenario, whilst it is about one order of magnitude weaker for the muon case. 
Let us also note that for masses larger than the TeV, inverse Compton scattering on the CMB due to the electrons produced by the muon decays (here neglected for simplicity as mentioned in Sec.~\ref{sec:annihilating_DM}) can increase the \g-ray emission for this channel and therefore improve the detection reach in the high-mass range.

\begin{figure}
\centering
\includegraphics[width=0.60\textwidth]{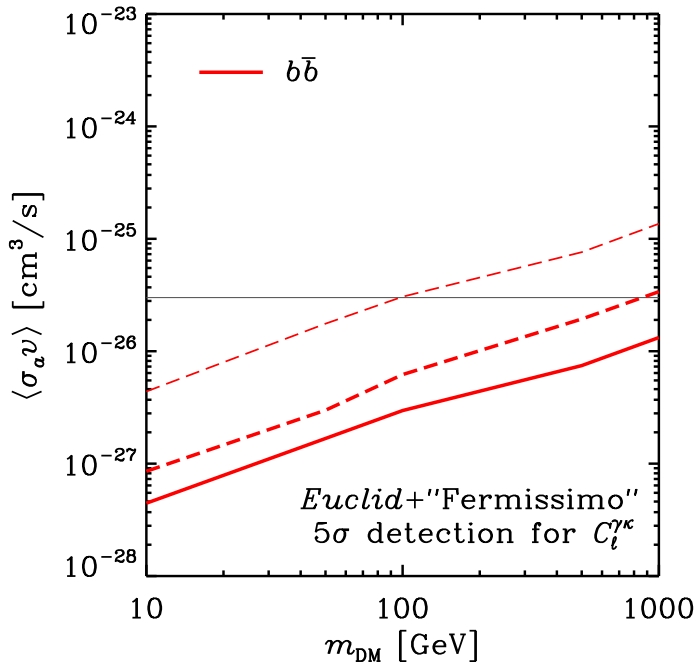}
\caption{ Level above which the cross-correlation APS between cosmic shear and \g-ray emission provides a $5\sigma$ {\it detection} of an annihilating DM signal in the $b\bar b$ channel.
The median line (dashed) refers to \Euclid+\Fermissimo, \high\ substructure scheme and model B for astrophysical sources. The upper line (thin dashed) shows the change when we replace the \high\ with the \low\ model. The lower (solid) line depicts the modification when model A is used instead of model B.}
\label{fig:12c}
\end{figure}

The improvement from \DES+\Fermitenyr\ to \Euclid+\Fermissimo\ is manifest: the weaker signal arising from the  $\mu^+\mu^-$ channel could arise in detection for thermal DM up to
masses of 100 GeV, while for the other channels there are detection prospects in wide portions of the parameter space, including thermal relics for masses up to the TeV scale. From the discussion of Fig. \ref{fig:Detection_Cl_1a_ALL_DESFermi5yr} we can also comment that in case of \low\ DM clustering, the detection curves worsen by about a factor of 6 (for a $b\bar b$ channel, this implies shifting the detectability of a thermal DM from a mass of 1 TeV down to 100 GeV), while in the case of model A they improve by a factor of 2 (raising the detectability of the thermal DM mass from 1 TeV to several TeV, again for a $b\bar b$ channel). This is explicitly shown in Fig.~\ref{fig:12c}, where the \Euclid+\Fermissimo\ $5\sigma$ detection reach for the $b\bar b$ channel already shown in the right panel of Fig.~\ref{fig:Detection_Cl_1bbis_HIGH_ALL} (red dashed line), is here confronted with the reach obtainable in the \low\ scenario (upper red thin dashed line) and for model A of astrophysical sources (lower red solid line).

\begin{figure}
\centering
\includegraphics[width=0.6\textwidth]{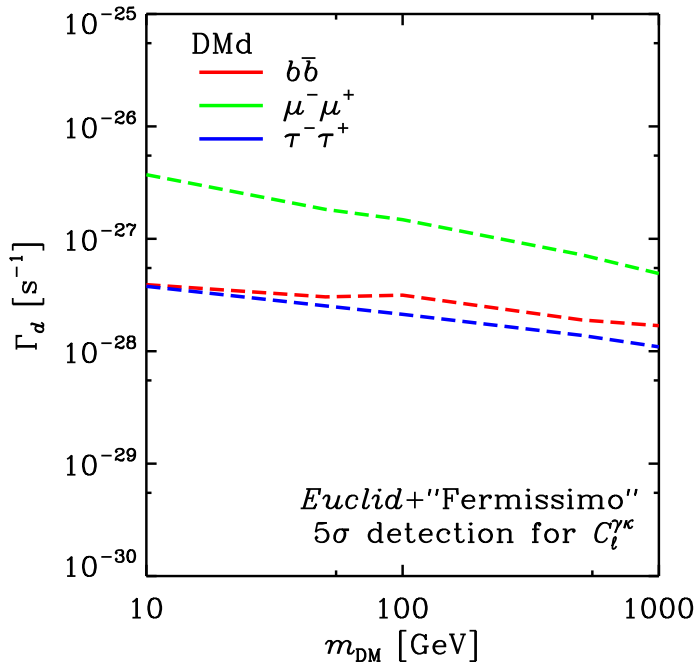}
\caption{Coloured lines denote the level above which the cross-correlation APS between cosmic shear and \g-ray emission provides a $5\sigma$ {\it detection} of  decaying DM signal with \Euclid+\Fermissimo. From top to bottom, the lines stand for the $\mu^+\mu^-$ (green), $b\bar{b}$ (red) and $\tau^+\tau^-$ (blue) decay channels. Astrophysical sources are modelled with model B.}
\label{fig:Detection_decay}
\end{figure}

The case for DM decay is shown in Fig. \ref{fig:Detection_decay}. From top to bottom, the lines show the $5\sigma$ detection reach for the $\mu^+\mu^-$ (green), $b\bar{b}$ (red) and $\tau^+\tau^-$ (blue) decay channels, for the \Euclid+\Fermissimo\ configuration. Astrophysical sources are modelled with model B.

\subsection{Parameter reconstruction}
\label{sec:parameter_forecasts}
We now turn to the discussion of the prospects of the different experimental 
setups for the reconstruction of the particle DM parameters. We
consider a set of representative benchmark cases: for annihilating DM we 
take $\mdm=10$, 100, 1000 GeV and a common value for
the annihilation cross section, viz.\ the thermal case $\sv = 3 \times 10^{-26} \,\mathrm{cm^3s^{-1}}$. For decaying DM, the masses considered are $\mdm=20$, 200, 2000 GeV (to give the same energy end-points in the \g-ray spectra as in the case of annihilation), and a 
representative value $\gd=0.33 \times 10^{-27}\,\mathrm{s^{-1}}$ for the decay rate, corresponding to a decay lifetime of $3 \times 10^{27}\,\mathrm{s}$.

\begin{figure}[h!]
\centering
\hspace{-5mm}
\includegraphics[width=0.6\columnwidth]{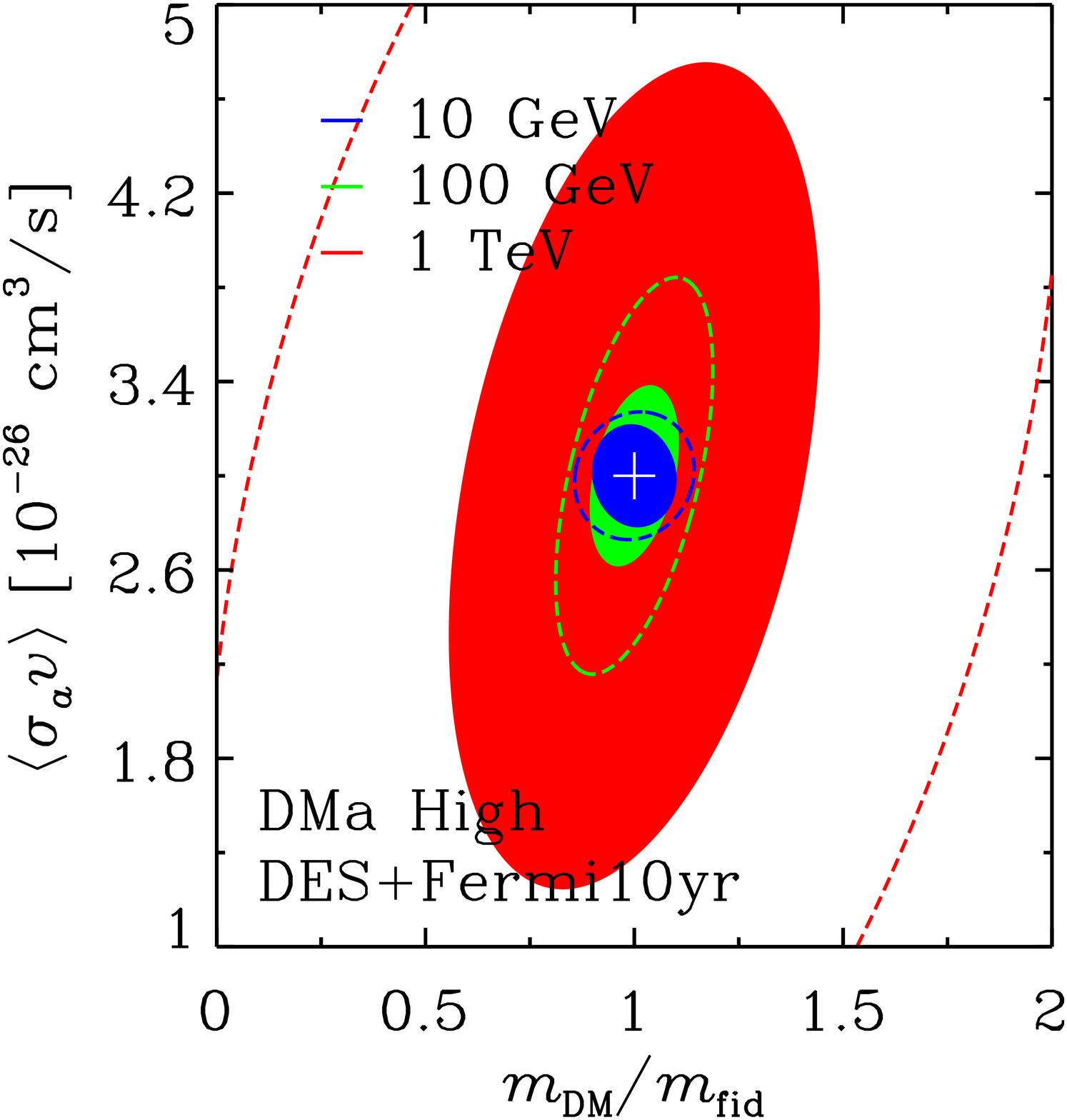}
\caption{Forecasts on the reconstruction of the DM mass and annihilation cross section, achieved by the combination of \DES\ and \Fermitenyr, for the $b\bar{b}$ annihilation channel and the \high\ clustering subhalo model. Results refer to a benchmark thermal cross section, $\sv = 3 \times 10^{-26} \,\mathrm{cm^3s^{-1}}$.  Marginal error contours are plotted in terms of the ratio $\mdm/m_{\rm fid}$ where $\mdm$ is the reconstructed mass and $m_{\rm fid}$ is the fiducial benchmark mass, i.e.\ 10 GeV (blue), 100 GeV (green) and 1 TeV (red). Filled areas (dashed ellipses) refer to model A (B) for the astrophysical \g-ray sources and show $1\sigma$ contours.}
\label{fig:Ellipses_HIGH_DESFermi5yr}
\end{figure}
We start by considering an annihilating DM candidate. Fig.~\ref{fig:Ellipses_HIGH_DESFermi5yr} shows the forecasts on the reconstruction of the DM properties achievable with the combination of \DES+\Fermitenyr\ for the $b\bar{b}$ annihilation channel and the \high\ clustering model. Filled areas (dashed ellipses) refer to model A (B) for the astrophysical \g-ray sources and denote joint $1\sigma$ contours in the $(\mdm/m_{\rm fid},\sv)$ plane, where $\mdm$ is the reconstructed mass and $m_{\rm fid}$ is the fiducial benchmark mass.\footnote{This is done for illustrative purposes only, as it renders easier for the reader to compare the relative constraining power as it varies as a function of the fiducial mass.} As in the previous section, we marginalise over the $\mathcal{A}_i$ factors. We notice that the tomographic-spectral approach has the potential of resolving a DM signal and of measuring DM properties over a wide range of masses. The two lighter benchmark cases, $\mdm=10$ and 100 GeV, can be accurately reconstructed with both mass and cross section determined with a precision better than 30\%. As can be seen in Eq.~\eqref{eqn:window_annihilating_DM}, this happens because lighter DM particles imply larger \g-ray fluxes, as a consequence of the dependence of the signal on the particle number density squared, namely $n^2 = \rho^2\,\mdm^{-2}$. For the 10 and 100 GeV mass benchmarks, the end-point of the DM-induced \g-ray spectrum is located within the energy range considered in our 
analysis (i.e.\ between 1 and 300 GeV) and it is thus easier to discriminate a 
DM-induced energy spectrum from the featureless power-laws characterising astrophysical sources. On the other hand, this is not the case for the benchmark at 1 TeV.

\begin{figure}
\centering
\hspace{-5mm}
\includegraphics[width=0.49\columnwidth]{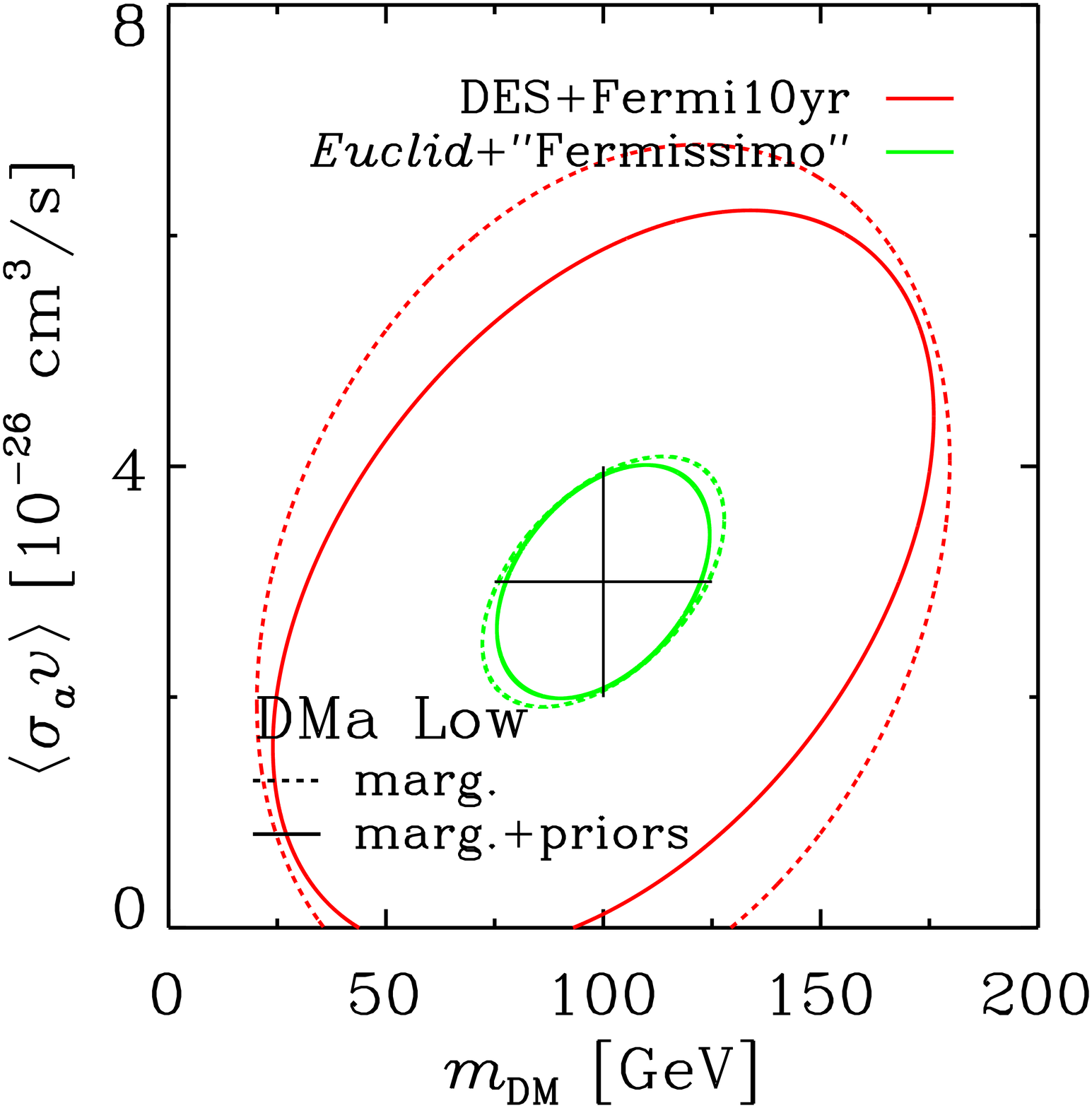}
\hspace{2.5mm}
\includegraphics[width=0.49\columnwidth]{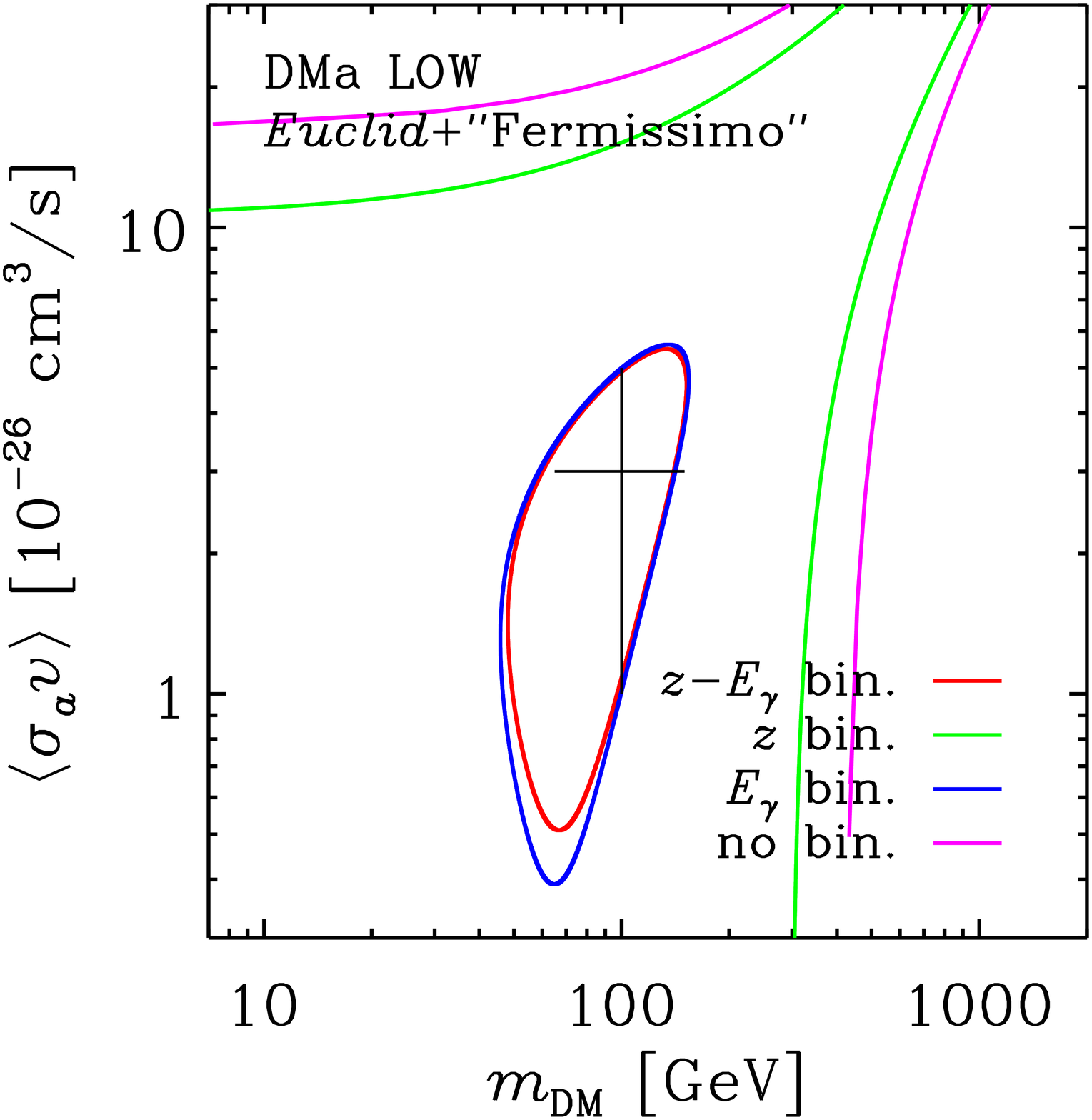}
\caption{Forecasts on the reconstruction of the DM mass and annihilation cross section, for a DM particle with a mass of 100 GeV, thermal annihilation cross section and $b\bar{b}$ annihilation channel. The \low\ clustering model is assumed {and astrophysical \g-ray emission is taken from model B}.  Contours show the $1\sigma$ CL reconstruction. Left: Red contours refer to \DES+\Fermitenyr\ , whilst green regions correspond to \Euclid+\Fermissimo. For each set of curves, solid lines correspond to marginalisation of the parameters $\mathcal{A}_i$ over the range of priors mentioned in the text, whilst dashed line refer to marginalisation without prior assumptions. Right: Magenta contours refer to the case in which neither redshift nor energy binning is considered. Green (blue) lines show the case where only the binning in redshift (energy) is considered, whilst for the red contours the full tomographic-spectral analysis is implemented. The combination \Euclid+\Fermissimo\ is assumed. In this plot, $\mathcal{A}_i$ are marginalised over without additional prior assumptions.}
\label{fig:Ellipses_LOW}
\end{figure}

The left panel of Fig.~\ref{fig:Ellipses_LOW} illustrates the improvement that can be achieved by 
considering future experiments. Results refer again to the benchmark scenario
with a mass of 100 GeV, thermal cross section and $b\bar b$ channel, but now a more conservative \low\ clustering scenario is assumed. Red contours are for a combination of \DES+\Fermitenyr\ 
(and can be compared to the analogous contours for the \high\ case in Fig.~\ref{fig:Ellipses_HIGH_DESFermi5yr}), whilst the green contour refers to \Euclid+\Fermissimo. The impact of considering \Euclid\ and a \g-ray detector with improved energy and angular resolution is dramatic: it would shrink the errors achievable with \DES+\Fermitenyr\ by more than a factor of 3, and  allow for a reconstruction of DM properties with a 20-30\% precision, even with the \low\ 
clustering model. The results outlined here refer to model B for astrophysical \g-ray sources; with model A the reconstruction capabilities would be further enhanced. The yields in the parameter reconstruction shown in Fig.~\ref{fig:Ellipses_LOW} (left panel) is due to a larger statistics, a finer 
redshift resolution that in turn enables us to perform a finer tomographic slicing for the surveyed 
redshift ranges (from 5 bins for \DES\ to 10 bins for \Euclid, see 
Tab.~\ref{tab:WL-specs}), and to improved \g-ray energy resolution (from 6 energy bins adopted in the \Fermitenyr\ analysis to 8 bins for \Fermissimo) and angular beam (from an average of 0.18
to 0.027). Once data will become available, these specifications will clearly be optimised, e.g.\ by adopting the measured energy dependence in the \g-ray detector angular resolution function, or 
by optimising the energy binning of the data.

Fig.~\ref{fig:Ellipses_LOW} (left panel) also allows us to see that the use of priors when marginalising over the parameters $\mathcal{A}_i$ only mildly affects the results. Furthermore, we verified that a different choice for the fiducial values of $\mathcal{A}_i$ (within the range of priors mentioned above) would have a very minor impact.

To have a deeper understanding on the interplay of the different elements entering the cross-correlation analysis, Fig. \ref{fig:Ellipses_LOW} (right panel) depicts how the adoption of tomographic and spectral information operated in the reconstruction of the DM particle-physics properties.
Results are presented for the same benchmark model of the left panel, in the \Euclid+\Fermissimo\ scenario. The magenta contour refers to the case in which no binning is considered in the 
analysis of the cross-correlation data, neither in redshift nor in energy. 
In this case, although bounds can be determined, the parameter reconstruction is limited by a strong degeneracy between mass and cross section. Indeed, in this case, the cross-correlation is simply 
constraining the strength of the DM component. This can be seen in
Eq.~\eqref{eqn:window_annihilating_DM}, where the signal depends on the quantity 
$S = \sv/\mdm^2\times\int_{E_{\rm min}}^{E_{\rm max}} \de N_a / \de E_\gamma$, where $E_{\rm max} = \mdm$ if $\mdm$ is smaller than the maximal detector energy $E_{\rm max}^{\rm det}$ (which in our analysis for \Euclid+\Fermissimo\ is 1 TeV, see Table \ref{tab:gammaspec}). This implies that the cross-correlation is constraining the quantity $S \sim \sv / \mdm$ (or $S \sim \sv/\mdm^{2}\times E_{\rm max}^{\rm det}$), giving rise to the degeneracy between mass and annihilation cross section.

Otherwise, if the binning in $z$ is performed (green contour), tomography shrinks the contour size. However, this information is still not enough to close the contour, since no new independent piece of information is provided on the mass or the cross section. On the other hand, if we only include the binning in energy (blue contour), the spectral information alone can allow the cross-correlation to determine closed contours in the reconstruction of both $\mdm$ and $\sv$. As we commented before, this comes from the fact that the DM-induced \g-ray spectrum is quite different from astrophysical spectra, which are typically simple power-laws. Finally, further including the binning in redshift enables the full exploitation of the complementarity between the spectral and tomographic information. The red contour are now closed and even tighter, corresponding to a good reconstruction of the DM properties.

We wish to comment that whenever a contour includes the case with $\sv=0$ (i.e.\ `open' ellipses), this implies that the technique we are using is only able to provide an upper limit on the annihilation cross section. This is because, in order to derive the contours, we employ the Fisher formalism, computing the derivatives in Eq.~\eqref{eq:fisher} at the fiducial $\mdm$ and $\sv$. A Gaussian likelihood is assumed, but this approximation clearly breaks down when $\sv$ goes to zero. Therefore when an ellipse includes $\sv=0$, only the upper part should be considered (as the corresponding upper limit), while the lower part of the ellipse has no statistical meaning. For example, the lower edge of the solid pink contour in Fig.~\ref{fig:Ellipses_LOW}, going rapidly to zero in the range $\mdm=300-500$ GeV, is just indicative.

Let us now move to additional annihilation channels.
\begin{figure}
\centering
\hspace{-5mm}
\includegraphics[width=0.49\columnwidth]{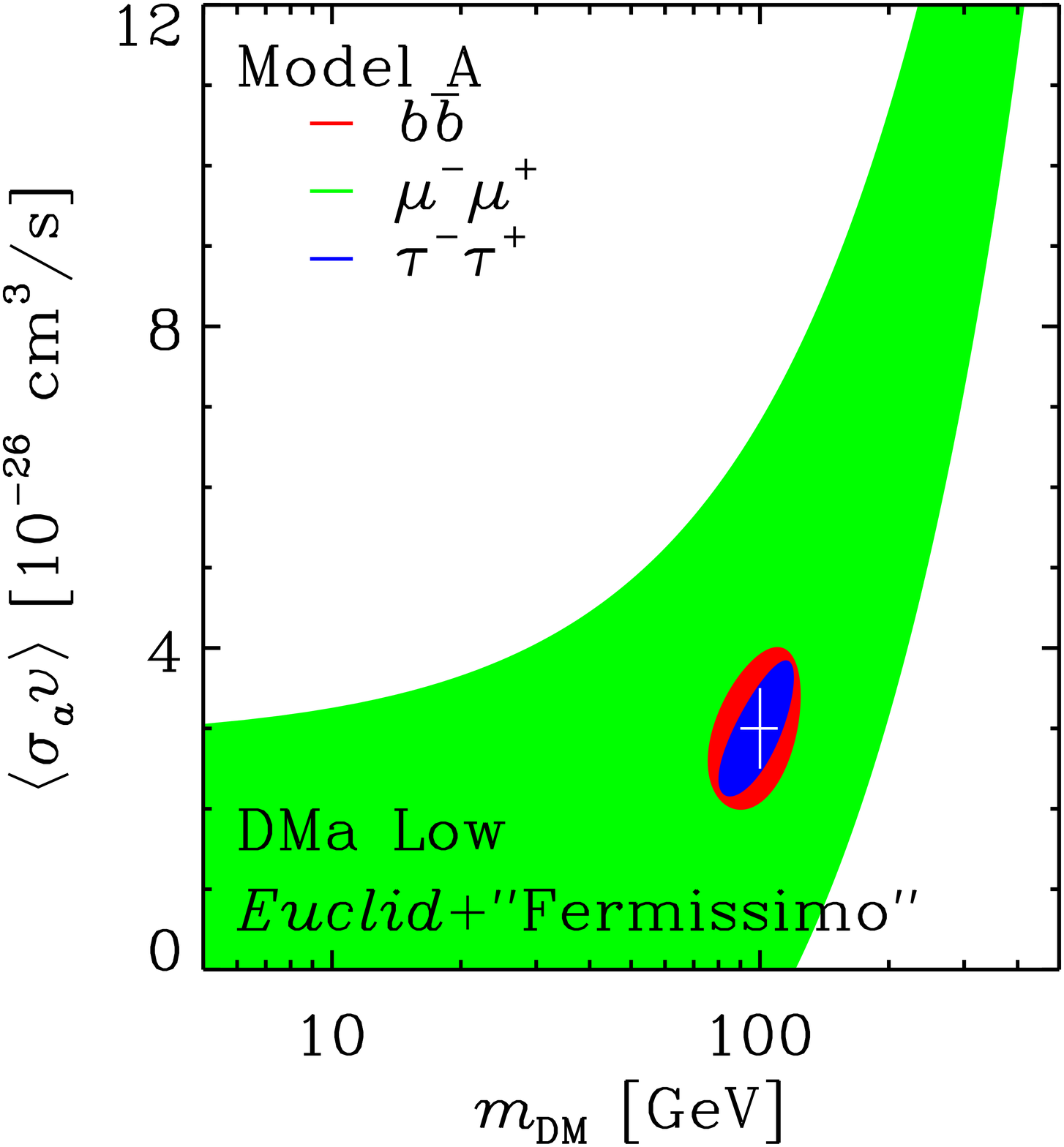}
\hspace{2.5mm}
\includegraphics[width=0.49\columnwidth]{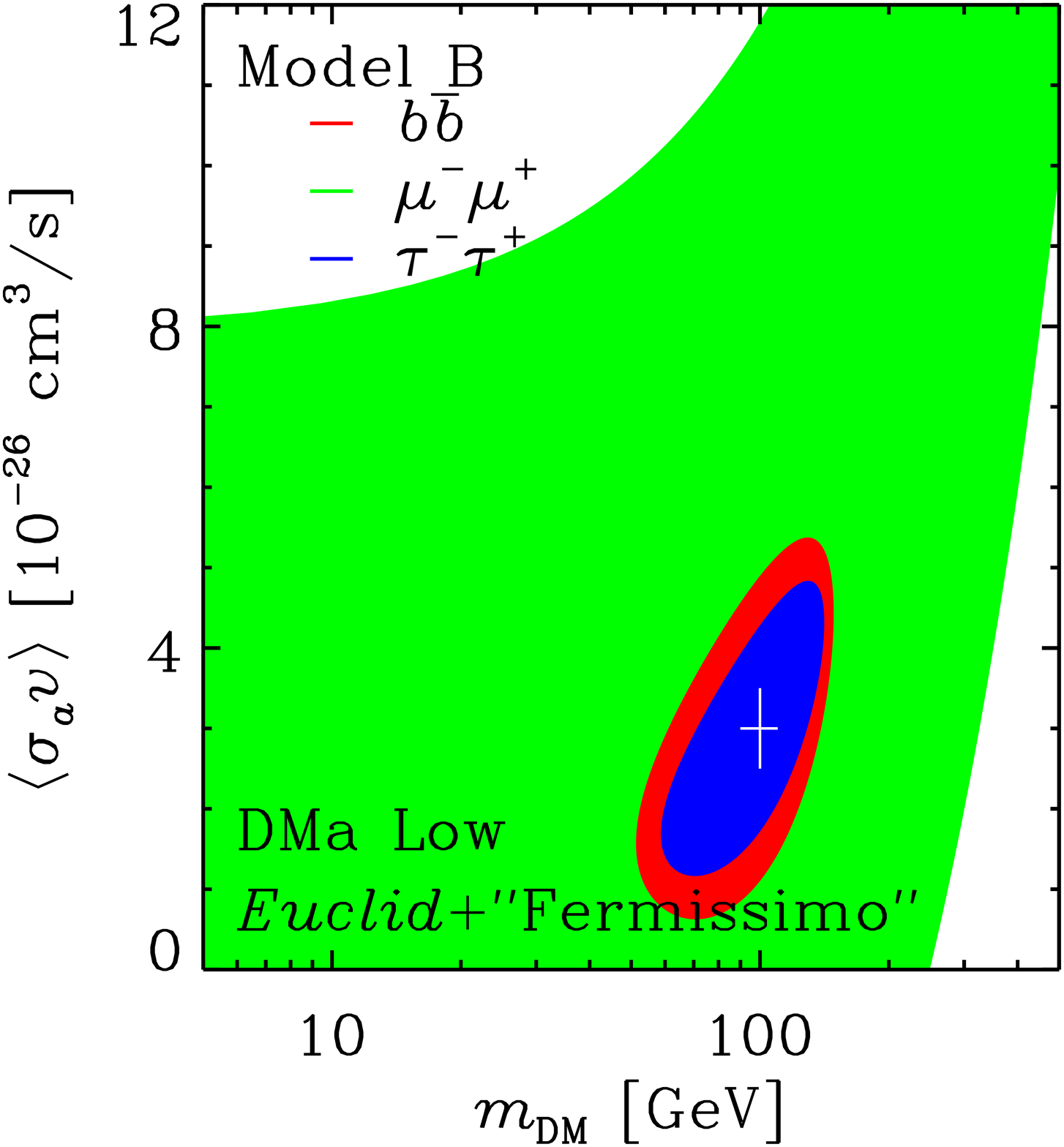}
\caption{Forecasts on the reconstruction of the DM mass and annihilation cross section, for a DM particle with a mass of 100 GeV and thermal annihilation cross section. The \low\ clustering model, model A (left panel) and B (right panel) for astrophysical \g-ray emission and the combination \Euclid+\Fermissimo\ are assumed. The red, green and blue contours refer to the case of annihilations into $b\bar b$, $\tau^+ \tau^-$ and $\mu^+\mu^-$, respectively. The filled areas show the $1\sigma$ CL reconstruction.}
\label{fig:Ellipses_2e_LOW_EuclidFermissimo}
\end{figure}
In Fig.~\ref{fig:Ellipses_2e_LOW_EuclidFermissimo}, we show our forecasts for a DM
particle with a mass of 100 GeV an a thermal cross section. The \low\ scenario
and \Euclid+\Fermissimo\ are assumed. Red areas 
represent the case of a $b \bar{b}$ annihilation channel (the same considered in
the previous figures, reproduced to ease comparison), whilst blue stands for 
$\tau^+\tau^-$ and green for $\mu^+\mu^-$ annihilation channels. The left and right panel show the impact of the astrophysical assumption, model A and B respectively. The size of the contours is mostly set by the efficiency of the channel in producing photons. A larger photon yield, as for the $b \bar{b}$ and $\tau^+\tau^-$ cases, increases the sensitivity of the cross-correlation technique (see also Fig.~\ref{fig:Detection_Cl_1bbis_HIGH_ALL}) and in turn the capability of reconstructing mass and annihilation rate. The case of muons as final states produces a smaller amount of photons and is hence the one with the worst forecast. Indeed, it is easy to see that contours are not closed in the region plotted of Fig.~\ref{fig:Ellipses_2e_LOW_EuclidFermissimo}. Another aspect to be taken into account is that different channels have different spectral shapes and some of them can mimic the astrophysical emission, thus making the reconstruction of the DM signal more difficult. This has however less impact than the emitted photon multiplicity.

We now consider the case of a decaying DM candidate.
\begin{figure}
\centering
\hspace{-5mm}
\includegraphics[width=0.49\columnwidth]{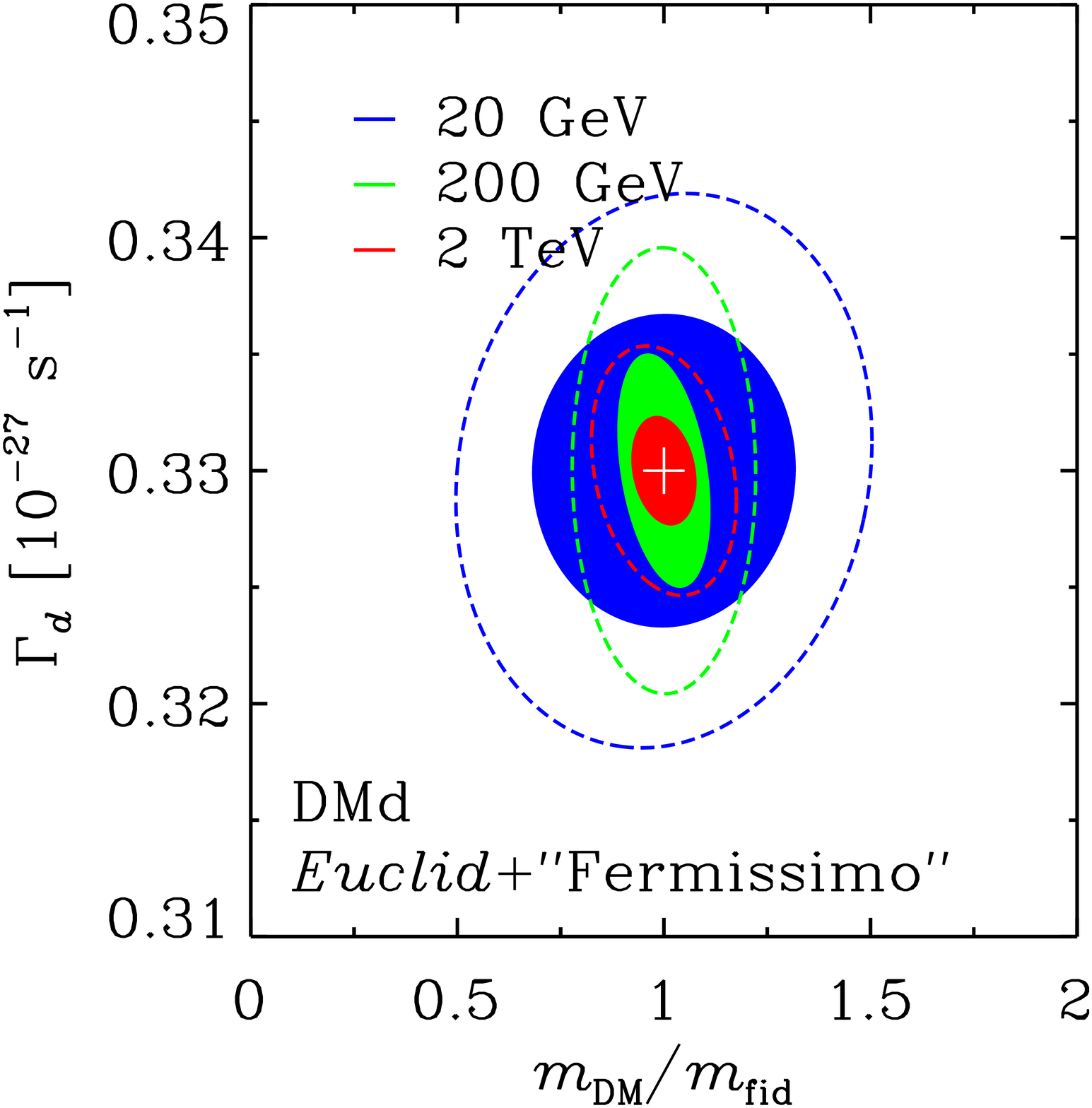}
\hspace{2.5mm}
\includegraphics[width=0.49\columnwidth]{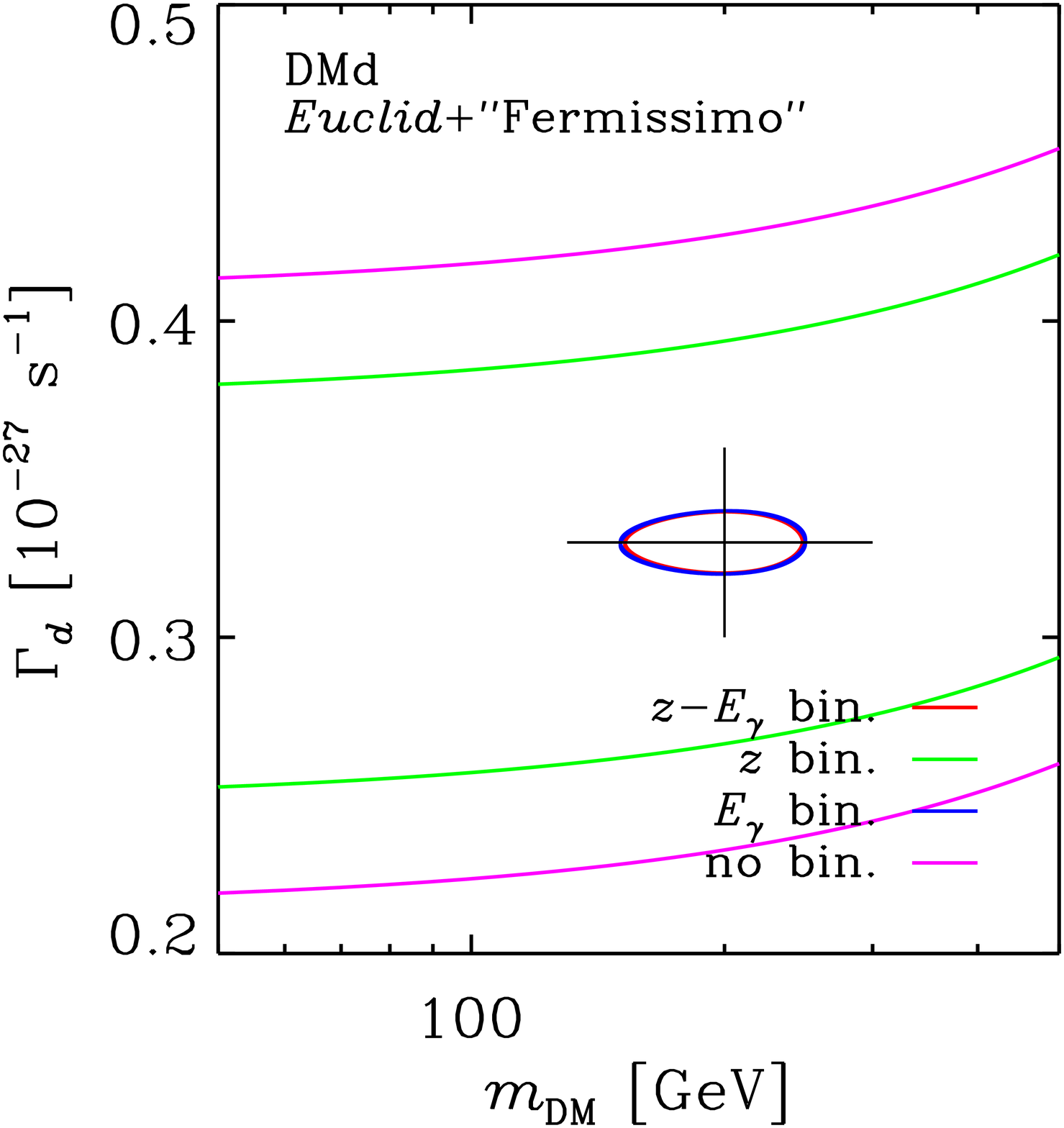}
\caption{Forecasts on the reconstruction of the DM mass and the decay rate, for a DM particle with a fiducial decay rate of $\gd = 0.33 \times 10^{-27}$\,s$^{-1}$ and decay channel into $b\bar b$. Results refer to the combination \Euclid+\Fermissimo. Left: Forecasts for three benchmark masses, 20 GeV (blue), 200 GeV (green) and 2 TeV (red). Filled areas (dashed contours) refer to model A (B) for the astrophysical \g-ray sources. Marginal error contours are plotted in terms of the ratio $\mdm/m_{\rm fid}$ where $\mdm$ is the reconstructed mass and $m_{\rm fid}$ is the fiducial benchmark mass.   Right: Forecasts for a DM particle with a mass of 200 GeV. The magenta contours refer to the case in which no binning is considered, neither in redshift nor in energy. Green (blue) curves show the case where only the binning in redshift (energy) is considered, whilst for the red contours the full tomographic-spectral analysis is implemented. The contours show the $1\sigma$ CL reconstruction. In this right plot, $\mathcal{A}_i$ are marginalised over without additional prior assumptions.}
\label{fig:Ellipses_dec_EuclidFermissimo}
\end{figure}
We compute the precision that can be achieved by the \g-ray and cosmic-shear cross-correlation in the reconstruction of its mass, $\mdm$, and decay rate, $\gd$, again for a few specific benchmark scenarios. In Fig.~\ref{fig:Ellipses_dec_EuclidFermissimo} (left panel), we illustrate the marginal error contours for a DM particle with a mass of 20, 200 GeV and 2 TeV (blue, green and red regions, respectively). The decay rate is fixed at a common value of $\gd = 0.33 \times 10^{-27}$\,s$^{-1}$, just for definiteness. The analysis refers to the combination of \Euclid+\Fermissimo. Analogously to the case of annihilating DM, the reconstruction power is extremely good. Both the mass and the decay rate can be potentially determined with a high level of accuracy.

The right panel of Fig.~\ref{fig:Ellipses_dec_EuclidFermissimo} otherwise depicts the effect of redshift and energy binning, for a benchmark case at $\mdm=200$ GeV. As for annihilating DM, the spectral information is crucial in the reconstruction of the DM mass. The same comments of Fig.~\ref{fig:Ellipses_LOW} (right panel) apply here,
with the notable difference that in the case of decay, 
Eq.~\eqref{eqn:window_decaying_DM} shows that the signal has a different 
dependence on the DM mass, namely $S = \gd/\mdm\times \int_{E_{\rm min}}^{E_{\rm max}} \de N_a / \de E_\gamma$. This implies that whenever $\mdm < E_{\rm max}^{\rm det}/2$ ($E_{\rm max}^{\rm det}=1$ TeV, in the case of Fig.~\ref{fig:Ellipses_dec_EuclidFermissimo}), the signal behaves as $S \sim \gd$, with no relevant dependence on the DM mass. This explains the flat behaviour of the contours in for small DM masses. Otherwise, when $\mdm > E_{\rm max}^{\rm det}/2$, the signal roughly scales as $S \sim \gd/\mdm\times  E_{\rm max}^{\rm det}$, which explains the orientation of the contours. This argument also explains why the reconstruction capabilities (shown in the left panel) are similar at different DM masses. For smaller masses, the sensitivity is lost due to the fact that most of the produced photons fall below the detector threshold, which we fix at 300 MeV for \Fermissimo.

To summarise the results on the capability of reconstructing DM properties 
with the cross-correlation expected from \Euclid+\Fermissimo, we report 
in Table \ref{tab:constraints-dec} the $1\sigma$ marginal errors for three 
benchmark models in the case of annihilating DM (two leftmost columns) and 
decaying DM (two rightmost columns). A $b \bar{b}$ spectrum is assumed, together 
with the \low\ clustering scenario. As usual, the astrophysical 
components are marginalised over. In the annihilating case, the benchmark refers
to the thermal cross-sections. Tighter errors refer to model A for the astrophysical \g-ray sources, wheras looser constraints (quoted in parenthesis) to model B. In this case, we see that both $\mdm$ and $\sv$ can be reconstructed with an uncertainty smaller than 20\% (35\%) for model A (B) up to DM masses of the order of 100 GeV, while the reconstruction capabilities degrade for DM masses at the TeV scale. In the decaying DM case, the relative uncertainties, on the contrary, reduce at large masses, as already noticed in the left panel of Fig.~\ref{fig:Ellipses_dec_EuclidFermissimo}.

\begin{table*}
\centering
\begin{tabular}{|c|c||c|c|}
\hline
$\mdm$ [GeV] & $\sv \, [10^{-26}$ cm$^3$s$^{-1}]$ & $\mdm$ [GeV] & $\Gamma_d\,[10^{-27} \, \mathrm s^{-1}] $\\
\hline
$10 \pm 0.52~(0.78) $ & $ 3 \pm 0.22~(0.32)$ & $20 \pm 4.2~(6.7)$ & $0.33 \pm 6.2~(9.1)\times10^{-3}$\\
$100 \pm 18~(34)$ & $ 3 \pm 0.72~(1.6)$ & $200 \pm 17~(31)$ & $0.33 \pm 3.3~(6.4)\times10^{-3}$\\
$1000 \pm 1000~(2500)$ & $ 3 \pm 3.9~(10.1)$ & $2000 \pm 110~(230)$ & $0.33 \pm  2.0~(4.3)\times10^{-3}$\\
\hline
\end{tabular}
\caption{Forecast joint $1\sigma$ marginal errors for three benchmark particle DM models. The two leftmost columns are for annihilating DM \low, whilst the rightmost ones for decaying DM. Results are for a $b\bar b$ channel and refer to \Euclid+\Fermissimo. All $\mathcal{A}_i$ parameters encoding the normalisation of the astrophysical components are marginalised over with the addition of the prior mentioned in the text. Tighter errors refer to model A for the astrophysical \g-ray sources, looser constraints (in parenthesis) to model B.}
\label{tab:constraints-dec}
\end{table*}

\section{Conclusions}
\label{sec:conclusions}
In this work we provide realistic prospects for the detection of DM in the
cross-correlation between the extragalactic \g-ray background (EGB) and the weak-lensing signal of cosmic shear. The idea was originally proposed in Paper I \citep{Camera:2012cj}. Here, the concept is further investigated by adopting a tomographic-spectral approach.
We also improve our modelling of the EGB, in particular of its astrophysical components. Unresolved astrophysical sources contribute to the EGB and they represent a background that has to be properly modelled and understood in order to have access to the DM component. Our description of the EGB includes the contributions from unresolved blazars, misaligned AGNs (mAGNs) and star forming galaxies (SFGs), alongside with \g\ rays induced by DM annihilations or decays. For each astrophysical population, we discuss an important ingredient for the computation of their correlation 
with the lensing signal, i.e.\ the relation between the \g-ray luminosity of a 
source and the mass of the host DM halo $M(\mathcal{L})$. We find that such 
a relation is quite unknown but it has only a moderate impact on the prospects
for the detection of DM. 

We discuss how the uncertainties related to the clustering 
of DM at low masses affect the cross-correlation signal. In particular, we 
focus on the dependence of the signal on the value of minimal halo mass 
$M_{\rm min}$ and on the amount of subhaloes. 

Finally, we adopt the Fisher matrix approach to derive forecast for a combination of current and future detectors including, \Fermi-LAT, \DES, \Euclid\ and a future \g-ray detector dubbed \Fermissimo. We determine the minimum annihilation cross section or decay rate that allows us to 
detect a DM particle from the measurement of the cross-correlation 
signal, and estimate the precision in the reconstruction of the DM properties 
(its mass and annihilation or decay rate) that can be reached by analysing the 
data of those experiments through the technique of cross-correlation.

The main conclusions of this paper can be summarised as follows:
\begin{itemize}
\item The \g-ray emission expected from mAGNs and SFGs can contribute
significantly to the cross-correlation signal and they represent an important
background for the detection of DM (see Fig.~\ref{fig:cl_comp}). 
In the most optimistic scenario for annihilating DM (\high), the 
cross-correlation induced by DM is comparable in intensity to that produced 
by astrophysical \g-ray emission (see Fig. \ref{fig:spectra}). However, this
is not enough to compromise the detection of a DM signal. 
We note also that this case would produce a subdominant (negligible) DM contribution to the total EGB (\g-ray auto-correlation APS).  
In the future, the discovery of new SFGs and mAGNs will improve our understanding of 
these populations, facilitating even more the reconstruction of the DM 
component.

\item The intensity of the DM component to the cross-correlation depends on 
the way DM clusters at low masses. Predictions can vary over two orders of 
magnitude, as it can be seen in Figs \ref{fig:cl_DM2} (left panel) and \ref{fig:spectra}. We note that similar uncertainties affect the EGB induced by DM. 

\item In the case of annihilating DM, the measurement of the cross-correlation 
in the data from \DES\ and \Fermitenyr\ has the potential to detect a DM 
particle with an annihilation cross section smaller than the thermal value 
of $3 \times 10^{-26} \,\mathrm{cm^3s^{-1}}$, for a DM mass smaller than 
500 GeV. This result refers to the \high\ subhalo scenario and favorable models of astrophysical \g-ray sources.
The predicted DM signal is reduced by a factor of $\sim$10 for the more conservative
\low\ scenario, and by a further factor of 20 for the more pessimistic case named \ns\ (see Fig. \ref{fig:Detection_Cl_1a_ALL_DESFermi5yr}). We note that this last possibility is quite unlikely but it corresponds to a guaranteed signal that cannot be neglected. In the case of less favorable astrophysical sources scenario, the detection reaches are degraded by a factor of 2.

\item The prospects for detection significantly improve with the inclusion of 
\Euclid\ data. Indeed, for the combination of \Euclid\ and \Fermissimo, a 
thermal annihilation cross section will correspond to a DM detection over 
the whole range of masses considered here (10 GeV to 1 TeV) and for all 
annihilation channels (except for 
annihilations into muons), in 
the \high\ scenario. With this experimental setup, cross sections as low as 
$8 \times 10^{-28} \,\mathrm{cm^3s^{-1}}$ are still able to provide a
detection of DM, at least for masses of the order of 10 GeV.

\item The power of the proposed technique comes from the combination of the 
three independence sources of information: $i)$ the different redshift scaling 
between astrophysical and DM components (see Fig.~\ref{fig:window}); $ii)$ their different energy spectrum (see Fig.~\ref{fig:intensity}); and $iii)$ the fact that the DM signal is dominated by the largest DM haloes (producing a large gravitational lensing effect) whilst astrophysical sources are normally hosted in smaller structures with $M\ll10^{14}M_\odot$. The tomographic-spectral approach provides an excellent sensitivity to DM even when such a component is only subdominant in the total intensity or in the auto-correlation APS.

\item In the absence of a DM detection (i.e.\ the cross-correlation is
found to be compatible with the astrophysical components only), the 
measurement can be used to derive $2\sigma$ limits on the annihilation cross
section (see Fig. \ref{fig:Bounds_ALL_DESFermi5yr}). The upper limits 
already achievable in the near future with \DES+\Fermitenyr\ for the \high\ case can be more stringent than the bounds 
obtained from the observation of the dwarf spheroidal galaxies performed by 
\Fermi-LAT \cite{Ackermann:2013yva} and from the observation of the Galactic 
halo from H.E.S.S. \cite{Abramowski:2011hc} (these limits represent, at the 
moment, the strongest constraints on annihilating DM). This will become the case also for the \low\ subhalo model with the combination of \Euclid+\Fermissimo.
Fig.~\ref{fig:Bounds_ALL_DESFermi5yr} shows how a measurement of the 
cross-correlation signal, potentially available on a timescale of 1--2
years, could become already competitive with long-standing strategies for the 
indirect detection of DM. Even in the unlikely \ns\ scenario, our method 
would be more constraining than what has been achieved to date by Imaging 
Cherenkov Telescopes from observations of dwarf spheroidals \citep[see e.g.][]{Aleksic:2013xea}.

\item When used to reconstruct the properties of the DM particle, the cross-correlation yields very tight constraints. If the DM is a particle with properties similar to a canonical WIMP (i.e., mass around 100 GeV and annihilation rate of the order of $3 \times 10^{-26} \,\mathrm{cm^3s^{-1}}$), $\mdm$ and $\sv$ can be reconstructed at the 20--30\% level for a combination of data from \DES\ and \Fermitenyr\ in the \high\ clustering scenario (see Fig.~\ref{fig:Ellipses_HIGH_DESFermi5yr}) and \Euclid\ and \Fermissimo\ in the \low\ scenario (see Fig.~\ref{fig:Ellipses_LOW} and Table \ref{tab:constraints-dec}).

\item Similar conclusions apply to the decaying DM case.
However, when compared to Galactic probes, the extragalactic signal from decaying DM is less promising than in the case of annihilating DM. Indeed, the latter signal scales with $\rho_{\rm DM}^2$ and can acquire a large boost factor from big structures in the Universe (possibly hosting a large amount of substructures).
This fact can significantly enhance the reach of extragalactic signatures with respect to the Galactic ones, while it does not happen in the decaying DM scenario, which scales linearly with $\rho_{\rm DM}$.
\end{itemize}

In conclusion, the cross-correlation between cosmic shear and diffuse \g-ray emission represents a very powerful mean of investigation for particle DM. A cross-correlation signal robustly interpreted in terms of DM would establish DM as a particle, against alternative interpretations based on modifications of Einstein's general relativity \cite{Camera:2012cj}. Such a measurement would also allow us to reconstruct the properties of the DM particle (e.g.\ its mass and annihilation or decay rate) to a good level of precision for a relevant fraction of its parameter space.

\acknowledgments
SC acknowledges support from the European Research Council under the EC FP7 Grant No. 280127 and from FCT-Portugal under Post-Doctoral Grant No. SFRH/BPD/80274/2011. MF gratefully acknowledges support of the Leverhulme Trust and of the project MultiDark CSD2009-00064. This work is also supported by the research grant {\it Theoretical Astroparticle Physics} number 2012CPPYP7 under the program PRIN 2012 funded by the Ministero dell'Istruzione, Universit\`a e della Ricerca (MIUR), by the research grants {\it TAsP (Theoretical Astroparticle Physics)} and {\it Fermi} funded by the Istituto Nazionale di Fisica Nucleare (INFN), and by the  {\it Strategic Research Grant: Origin and Detection of Galactic and Extragalactic Cosmic Rays} funded by the University of Torino and Compagnia di San Paolo.

\appendix
\section{The $M(\mathcal{L})$ relation for misaligned Active Galactic Nuclei}
\label{sec:appendix_MAGNs}
\begin{figure}
\centering
\hspace{-5mm}
\includegraphics[width=0.48\textwidth]{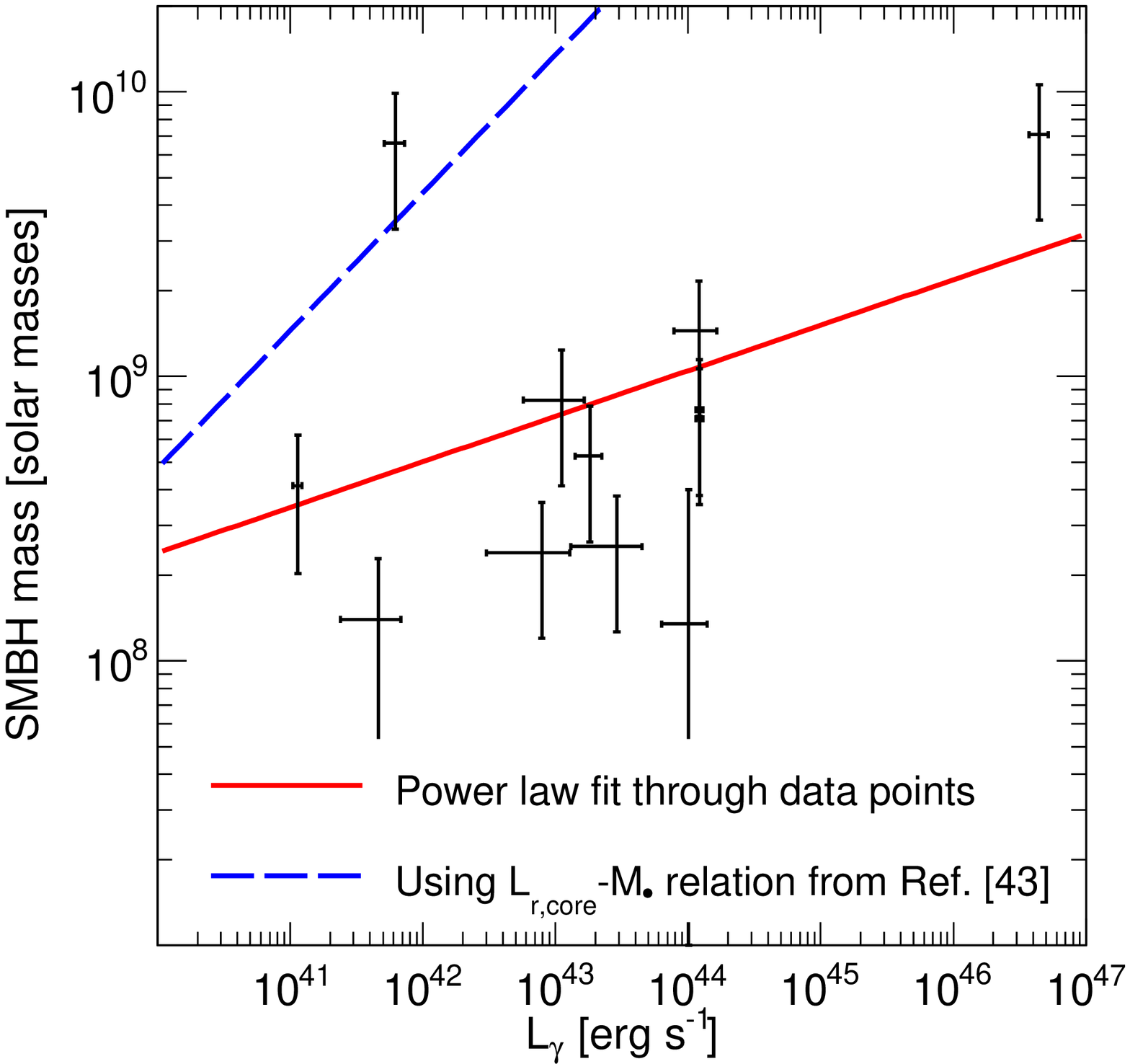}
\hspace{2.5mm}
\includegraphics[width=0.48\textwidth]{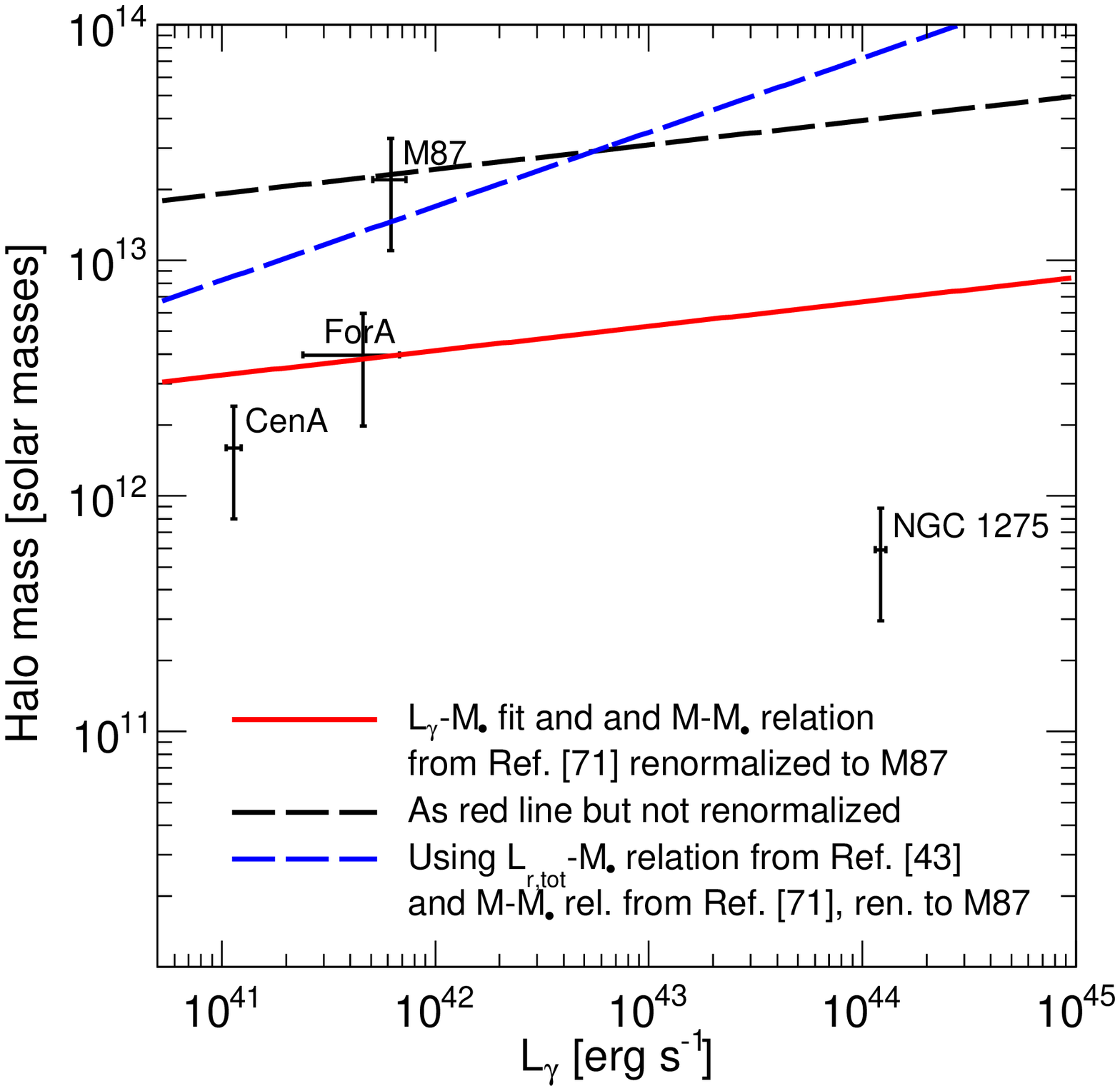}
\caption{Left: SMBH mass as a function of \g-ray luminosity. The 12 data points correspond to the 12 MAGNs from the sample analyzed in Ref.~\cite{DiMauro:2013xta} for which it was possible to find measurement of the mass of their central SMBH $M_\bullet$ (see text for details). The latter is plotted against the \g-ray luminosity $\mathcal{L}$ between 0.1 and 100 GeV taken from Ref.~\cite{DiMauro:2013xta}. The solid red line shows the result of a power law fit to the data. The dashed blue line indicates the $M_{\bullet}-\mathcal{L}$ relation obtained by exploiting the correlation of the two quantites with $L_{r,{\rm core}}$ (following Ref.~\cite{Franceschini:2008tp}). Right: Halo mass as a function of \g-ray luminosity. The solid red line is obtained by combining the fit to the data in the left panel with the $M_{\bullet}-M$ relation from Ref. \cite{Hutsi:2013hwa}, renormalised to reproduce the properties of M 87. The dashed black line shows how the red line would look like without renormalising the $M_{\bullet}-M$ correlation to the case of M 87. The dashed blue line presents the $M(\mathcal{L})$ obtained by using the $L_{r,{\rm core}}-M_{\bullet}$ relation from Ref.~\cite{Franceschini:2008tp} (see blue dashed line in left panel) together with the $M_{\bullet}-M$ relation from Ref. \cite{Hutsi:2013hwa} renormalised to M 87. The 4 data points correspond to the only 4 mAGNs in Ref.~\cite{DiMauro:2013xta} for which is was possible to find estimates of the mass of their DM haloes. See text for details and references.}
\label{fig:luminosity_mass_MAGN}
\end{figure}

In Sec. \ref{sec:cross_correlation_mAGNs}, the \g-ray luminosity of mAGNs has 
been related to the mass $M_\bullet$ of the SMBH at the center of the AGN, 
exploiting the fact that both are correlated to the core radio luminosity 
$L_{r,{\rm core}}$. However, this procedure results in estimates for $M_\bullet$ 
that are too large (see dashed blue line in 
Fig.~\ref{fig:luminosity_mass_MAGN}a). Moreover, the link between $M_\bullet$ 
and $L_{r,{\rm core}}$ is not completely well-established (see 
Refs.~\cite{Franceschini:2008tp,Bettoni:2002rk}). Therefore, we decide to
consider also an alternative approach, based on the information available in
the literature about the mAGNs detected by \Fermi-LAT. Among the 15 mAGNs
discussed in Ref.~\cite{DiMauro:2013xta}, we found measurements for the
central SMBH for 12 of them. More precisely: see Ref.~\cite{Bettoni:2002rk} 
for NGC 1218. Cen A, 3C 120, PKS-0625-35, 3C 84, Ref.~\cite{Woo:2002un} for 
NGC 1218, 3C 120, 3C 380, 3C 84, NGC 1275, Ref.~\cite{Falcke:2003ia} for NGC
1218, NGC 1275, Ref.~\cite{Gebhardt:2011yw} for M 87, 
Ref.~\cite{Neumayer:2010kw} for Cen A, Ref.~\cite{Nowak:2008wu} for For A, 
Ref.~\cite{Chatterjee:2011ix} for 3C 111 and Ref.~\cite{Aleksic:2010xk} for 
IC 310. When more than one measurement is available for the same object, we 
consider the average and we also sum the error in quadrature. If the resulting
error is smaller than 50\% of SMBH mass, we artificially increase it to 50\% 
of $M_\bullet$, in order to be conservative. The masses, estimated in this way, 
are coupled to the \g-ray luminosity taken from Ref.~\cite{DiMauro:2013xta} 
and the 12 data points are plotted in Fig.~\ref{fig:luminosity_mass_MAGN}a. 
The solid red lines shows the result of a fit assuming a power law relation 
between the two quantities.

Finally, the relation between the mass of the SMBH and the host DM halo is 
taken from Ref.~\cite{Hutsi:2013hwa} (which inherits it from 
Ref.~\cite{Bandara:2009sd}). However, we normalise it in order to pass through 
the point ($M_\bullet=6.6 \times 10^9 M_\odot$, $M=2.2 \times 10^{13} M_\odot$) 
\cite{Gebhardt:2009cr,Gebhardt:2011yw}, corresponding to the case of M 87.

This relation combined with the power law fit of $M_\bullet(\mathcal{L})$ (red 
curve in the left panel) gives the solid red line in 
Fig.~\ref{fig:luminosity_mass_MAGN}b. The dashed black line shows how the 
relation would look like without the renormalisation to the case of M 87.

For 4 mAGNs we also find estimates for the mass of ther DM halo. Specifically 
we consider Ref.~\cite{Gebhardt:2009cr} for M 87, Refs~\cite{Woodley:2007zh,
Samurovic:2010} for Cen A, Ref.~\cite{Goudfrooij:2000yc} for For A and
Ref.~\cite{Scharwaechter:2012ag} for NGC 1275\footnote{For NGC 1275, 
Ref.~\cite{Scharwaechter:2012ag} only provides an estimate of the stellar
mass. We estimate the DM halo mass by assuming a value of 0.05 for the ratio
between stellar and DM mass.}. For Cen A, two values are available and we 
consider the average of the two. The error in the DM halo mass is taken to be 
50\% of the estimated mass. These data for these 4 mAGNs are plotted in 
Fig.~\ref{fig:luminosity_mass_MAGN}b, together with their \g-ray 
luminosity from Ref.~\cite{DiMauro:2013xta}. 

Despite having too few points to derive firm conclusions, we note that the 
data are not located far away from the solid red line, while well below 
the dashed blue line. The latter shows the $M(\mathcal{L})$ obtained by 
combining the $M-M_{\bullet}$ relation from Ref. \cite{Hutsi:2013hwa} 
(renormalised to reproduce the properties of M 87) and the 
$L_{r,{\rm core}}-M_{\bullet}$ relation from Ref.~\cite{Franceschini:2008tp} (see blue dashed line in the left panel). This represents our benchmark 
$M(\mathcal{L})$ relation employed in the rest of the paper. We note 
that it may over-estimate the halo mass, although it should be kept in mind 
that the four data-points considered might be not representative. A large value for the mass of the host DM halo would correspond to a generous 1-halo 
term in the 3-dimensional power spectrum and, thus, a conservative estimate
of how the cross-correlation of mAGNs may compete with a DM signal.

\section{The $M(\mathcal{L})$ relation for star-forming galaxies}
\label{sec:appendix_SFGs}
In Sec.~\ref{sec:cross_correlation_SFGs}, the mass of the DM halo hosting
a SFG is derived from its \g-ray luminosity via the determination of its SFR. 
Here, we discuss how to derive the $M(\mathcal{L})$ empirically, based on the
few SFGs detected by \Fermi-LAT for which we could find measurements for the 
mass of their DM halo. In particular, we consider the MW from 
Ref.~\cite{Catena:2009mf} and  M31 from Ref.~\cite{Geehan:2005aq}. For SMC, 
NGC 253, NGC 1068 and NGC 4945 we refer to the rotation curves in 
Refs~\cite{Bekki:2008db,Sofue:1999jy}. These are fitted by a simple model
with a DM halo and a disk, leaving the normalisations of the two components 
free, as well as the scale radius of the DM halo and the scale radius of the 
disk (4 parameters). The fit is performed excluding the first radial bins 
(corresponding to the inner region) in order to be independent on possible 
structures (e.g. bulges or bars) in the very center of the galaxies.

\begin{figure}
\centering
\includegraphics[width=0.6\textwidth]{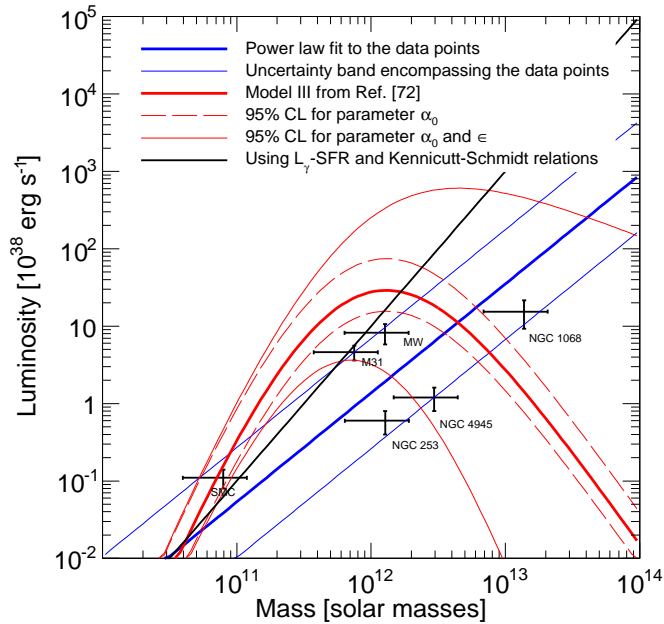}
\caption{The relation between \g-ray luminosity and the host DM mass for SFGs. The data points are taken from Ref.~\cite{Ackermann:2012vca}, whilst the measurements of the DM halo mass are derived from literature or from rotation curves (see text for details). The thick solid blue lines indicates the result of a power law fit of the 6 data points. The thin solid blue line show a reasonable uncertainty band, encompassing the data. The red lines show the results implementing the relation between SFR and DM halo mass obtained in Ref.~\cite{Lu:2013aoa} (Model III). See text for details. The solid black line represents the fiducial model adopted in this paper and it is obtained by combining the $\mathcal{L}$-SFR relation from Ref. \cite{Ackermann:2012vca} and the Kennicutt-Schmidt relation.}
\label{fig:Mass_luminosity_SFGs}
\end{figure}

Fig. \ref{fig:Mass_luminosity_SFGs} present the estimated masses for the DM
haloes. The solid thick blue line also shows the result of a power law fit to 
the data and it gives the desired $M(\mathcal{L})$ relation. The thin solid 
blue lines indicates a reasonable uncertainty band that encompass the data 
points considered. We also show the predictions obtained implementing the 
$M-$SFR from Ref.~\cite{Lu:2013aoa} (Model III) and relating the SFR to 
$\mathcal{L}$ as found in Ref.~\cite{Ackermann:2012vca}. The solid red line 
is obtained with the parameters of Model III fixed to their average value. To 
estimate the uncertainty band, we left the two main parameters of their 
description (i.e $\alpha_0$ and $\epsilon$, see Ref.~\cite{Lu:2013aoa} for 
their definition) free to vary. For the thin dashed red lines we take 
$\epsilon$ within its 95\% confidence level region, leaving $\alpha_0$ to its 
average value. For the thin red solid lines we also vary $\alpha_0$ to the 
values corresponding to its 95\% confidence level.

\bibliographystyle{JHEP}
\bibliography{../bibliography}

\end{document}